\documentclass[12pt]{article}
\usepackage{graphicx,amsmath,amssymb,cite}

\newlength{\dinwidth}
\newlength{\dinmargin}
\renewcommand{\vec}[1]{\boldsymbol{#1}}
\newcommand{\dif}{\mathrm{d}}

\newcommand{\xB}{x_{\scriptscriptstyle{B}}}
\newcommand{\chisq}{\chi^2/\mathrm{d.o.f.}}
\newcommand{\Pom}{{\mathbb{P}}}

\begin{document}
\titlepage
\begin{flushright}
  DESY 06-095      \\
  29th August 2006 \\
\end{flushright}

\vspace*{0.5cm}

\begin{center}
  
  {\Large \bf Exclusive diffractive processes at HERA \\[1ex] within the dipole picture}

  \vspace*{1cm}

  \textsc{H. Kowalski$^a$, L. Motyka$^{a,b}$ and G. Watt$^{a,c}$} \\

  \vspace*{0.5cm}

  $^a$ Deutsches Elektronen-Synchrotron DESY, 22607 Hamburg, Germany \\
  $^b$ Institute of Physics, Jagellonian University, 30-059 Krak\'ow, Poland \\
  $^c$ Department of Physics \& Astronomy, University College London, WC1E 6BT, UK

\end{center}

\vspace*{0.5cm}

\begin{abstract}
  We present a simultaneous analysis, within an impact parameter dependent saturated dipole model, of exclusive diffractive vector meson ($J/\psi$, $\phi$ and $\rho$) production, deeply virtual Compton scattering and the total $\gamma^*p$ cross section data measured at HERA.  Various cross sections measured as a function of the kinematic variables $Q^2$, $W$ and $t$ are well described, with little sensitivity to the details of the vector meson wave functions.  We determine the properties of the gluon density in the proton in both longitudinal and transverse dimensions, including the impact parameter dependent saturation scale.  The overall success of the description indicates universality of the emerging gluon distribution and proton shape.
\end{abstract}

\section{Introduction} \label{sec:introduction}
Exclusive diffractive processes at HERA, such as exclusive vector meson production or deeply virtual Compton scattering (DVCS), are excellent probes of the proton shape in the perturbative regime.  Several investigations have already shown that these processes can be well described within a QCD dipole approach with the vector meson wave functions determined by educated guesses and the photon wave function computed within QED; see, for example, Refs.~\cite{Kowalski:2003hm,Nemchik:1994fp,Gotsman:1995bn,Nemchik:1996cw,Dosch:1996ss,McDermott:1999fa,Munier:2001nr,Caldwell:2001ky,Forshaw:2003ki,Frankfurt:2005mc,Forshaw:2006np}.  It was also pointed out some time ago that the exclusive vector meson and DVCS processes provide severe constraints on the gluon density at low-$x$ \cite{Ryskin:1992ui,Brodsky:1994kf,Frankfurt:1995jw,Martin:1996bp,Collins:1996fb,Frankfurt:1997fj,Frankfurt:1997at,Martin:1999wb,Gotsman:2001ic,Gotsman:2003ww}.

The vector meson and DVCS processes are measured at HERA \cite{Breitweg:2000yn,Chekanov:2001qu,Adloff:2000qk,Chekanov:2002xi,Chekanov:2004mw,Aktas:2005xu,Chekanov:2005cq,Adloff:1999kg,Aktas:2005ty,Chekanov:2003ya} in the small-$x$ regime where the behaviour of the inclusive deep-inelastic scattering (DIS) cross section, or the structure function $F_2$, is driven by the gluon density.  The dipole model allows these processes to be calculated, through the optical theorem, from the gluon density determined by a fit to the total inclusive DIS cross sections.  Usually, it is assumed that the evolution of the gluon density is independent of the proton shape in the transverse plane.  The investigation of Kowalski and Teaney~(KT)~\cite{Kowalski:2003hm} has shown that the Gaussian form of the proton shape, implied by the data, has implications on the emerging pattern of QCD evolution and saturation effects.  The interplay of saturation and evolution effects was first investigated by Bartels, Golec-Biernat and Kowalski~\cite{Bartels:2002cj}, where it was found that the total inclusive DIS cross sections, or $F_2$, can be described either by strong saturation and weak evolution or by strong evolution and weak saturation effects.  The investigation of Ref.~\cite{Kowalski:2003hm}, which took into account also the proton shape in the transverse plane, concluded that saturation effects are substantial in the proton centre, but that the Gaussian form implies that a large contribution to the cross section has to come from the outskirts of the proton, where the gluon density is diluted.  Hence the evolution effects have to be strong and play an important role.  An alternative approach to determining the impact parameter dependent gluon distribution, based on a two-Pomeron model, is discussed in Refs.~\cite{Shoshi:2002in,Shoshi:2002fq}.

Another important result of dipole model investigations is that a wide variety of DIS data can be described with only a few assumptions.  The investigations of Refs.~\cite{Golec-Biernat:1998js,Golec-Biernat:1999qd,Bartels:2002cj,Iancu:2003ge,Forshaw:2004vv} show that the inclusive DIS cross section can be described together with the inclusive diffractive DIS cross section.  Moreover, in Ref.~\cite{Kowalski:2003hm} it was shown that the inclusive DIS process can be described together with inclusive charm production and exclusive diffractive $J/\psi$ photoproduction.  This description preserves also the main properties of the inclusive diffractive DIS cross section~\cite{Kowalski:2005}.

In this paper we will extend the analysis of Ref.~\cite{Kowalski:2003hm} and show that the same minimal set of assumptions allows the description of a much wider set of recently measured data on exclusive $J/\psi$, $\phi$ and $\rho$ photo- and electroproduction and also the DVCS process.  The cross sections for these processes have been measured as a function of the photon virtuality, $Q^2$, the $\gamma^* p$ centre-of-mass energy, $W$, and the squared momentum transfer, $t$.  In addition, for vector mesons the ratios of the cross sections for longitudinally and transversely polarised incoming photons have been determined as a function of $Q^2$.

To perform the analysis we use an impact parameter dependent saturated dipole model in which the gluon density is determined by a DGLAP fit to the total inclusive DIS cross sections.  The wave function of the virtual photon is known from QED and the proton and vector meson wave functions are assumed to have a Gaussian shape.  The parameters of these Gaussian distributions are easily determined from data.  The results are compared to numerous data distributions provided by the HERA experiments.  In this framework the $W$ distributions are mainly sensitive to the square of the gluon density and the $Q^2$ distributions and $\sigma_L/\sigma_T$ ratios to the properties of the vector meson wave functions.  The proper choice of the wave functions is also confirmed by the agreement of the predicted size of the cross sections with data.  In the dipole model the absolute normalisation of the vector meson cross sections follows from the optical theorem.

The $t$-distributions determine the area size of the interaction region, $B_D$.  The parameter $B_D$ is obtained by making a fit to the $t$-distributions of the form $\dif\sigma/\dif t\propto \exp(-B_D|t|)$.  For scattering of very small dipoles $B_D$ is connected to the proton radius $R_p$ via $B_D=R_p^2/3$.  However, for larger dipoles the size of the interaction area depends not only on the proton radius but also on the size of the produced vector meson or real photon, which we take into account following the work of Bartels, Golec-Biernat and Peters (BGBP)~\cite{Bartels:2003yj}.  This allows the data for all vector mesons and DVCS to be described using a unique Gaussian proton shape, independent of the produced final state.

\section{The dipole model}
\begin{figure}
  \centering
  \includegraphics[width=0.6\textwidth,clip]{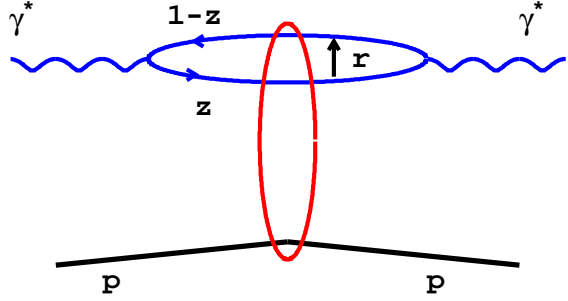}
  \caption{The elastic scattering of a virtual photon on a proton in the dipole representation.}
  \label{fig:fgraqqtot}
\end{figure}
In the dipole model, deep inelastic scattering is viewed as the interaction of a colour dipole, that is, mostly a quark--antiquark pair, with the proton.  The transverse size of the pair is denoted by $\vec{r}$ and a quark carries a fraction $z$ of the photon's light-cone momentum.  In the proton rest frame, the dipole lifetime is much longer than the lifetime of its interaction with the target proton.  Therefore, the elastic $\gamma^* p$ scattering is assumed to proceed in three stages: first the incoming virtual photon fluctuates into a quark--antiquark pair, then the $q\bar{q}$ pair scatters elastically on the proton, and finally the $q\bar{q}$ pair recombines to form a virtual photon.  This is shown schematically in Fig.~\ref{fig:fgraqqtot}.

The amplitude for the elastic process $\gamma^* p\rightarrow \gamma^* p$, $\mathcal{A}^{\gamma^*p}(x,Q,\Delta)$, is simply the product of amplitudes of these three subprocesses integrated over the dipole variables $\vec{r}$ and $z$:
\begin{eqnarray}
  \mathcal{A}^{\gamma^*p}(x,Q,\Delta) = \sum_f \sum_{h,\bar h} \int\!\dif^2\vec{r}\,\int_0^1\!\frac{\dif{z}}{4\pi}\,\Psi^*_{h\bar h}(r,z,Q)\,\mathcal{A}_{q\bar q}(x,r,\Delta)\,\Psi_{h\bar h}(r,z,Q),
  \label{eq:elamp}
\end{eqnarray}
where $\Psi_{h\bar h}(r,z,Q)$ denotes the amplitude for the incoming virtual photon to fluctuate into a quark--antiquark dipole with helicities $h$ and $\bar h$ and flavour $f$.  We suppress here references to the photon helicities for simplicity.  $\mathcal{A}_{q\bar q}(x,r,\Delta)$ is the elementary amplitude for the scattering of a dipole of size $\vec{r}$ on the proton, $\vec{\Delta}$ denotes the transverse momentum lost by the outgoing proton, and $x$ is the Bjorken variable.  Note that, following Ref.~\cite{Kowalski:2003hm}, we choose a slightly different convention from that commonly used, in that we include a factor of $1/(4\pi)$ in the integration measure; this convention is reflected in the normalisation of the photon and vector meson wave functions.

The elementary elastic amplitude $\mathcal{A}_{q\bar q}$ is defined such that the elastic differential cross section for the $q\bar{q}$ pair scattering on the proton is
\begin{eqnarray}
  \frac{\dif\sigma_{q\bar q}}{\dif t} = \frac{1}{16\pi}\left\lvert\mathcal{A}_{q\bar q}(x,r,\Delta)\right\rvert^2,
  \label{eq:elsiggp}
\end{eqnarray}
where $t=-\Delta^2$.  It can be related to the $S$-matrix element $S(x,r,b)$ for the scattering of a dipole of size $\vec{r}$ at impact parameter $\vec{b}$:
\begin{eqnarray}
  \mathcal{A}_{q\bar q}(x,r,\Delta) = \int\!\dif^2\vec{b}\;\mathrm{e}^{-\mathrm{i}\vec{b}\cdot\vec{\Delta}}\,\mathcal{A}_{q\bar q}(x,r,b)
  = \mathrm{i}\,\int \dif^2\vec{b}\;\mathrm{e}^{-\mathrm{i}\vec{b}\cdot\vec{\Delta}}\,2\left[1-S(x,r,b)\right]. 
  \label{eq:smatrix}
\end{eqnarray}
This corresponds to the intuitive notion of impact parameter when the dipole size is small compared to the size of the proton.  The optical theorem then connects the total cross section for the $q\bar q$ pair scattering on the proton to the imaginary part of the forward scattering amplitude:
\begin{eqnarray}
  \sigma_{q\bar q}(x,r)=\mathrm{Im}\,\mathcal{A}_{q\bar q}(x,r,\Delta=0)=\int \dif^2 \vec{b}\; 2[1-\mathrm{Re}\,S(x,r,b)].
  \label{eq:totsiggp}
\end{eqnarray}
The integration over $\vec{b}$ of the S-matrix element motivates the definition of the $q\bar{q}$--$p$ differential cross section as
\begin{eqnarray}
  \frac{\dif\sigma_{q\bar q}}{\dif^2 \vec{b}}= 2[1-\mathrm{Re}\,S(x,r,b)].
  \label{eq:difsiggp}
\end{eqnarray}

\begin{figure}
  \centering
  \begin{minipage}{0.5\textwidth}
    \centering
    \includegraphics[height=4.5cm,clip]{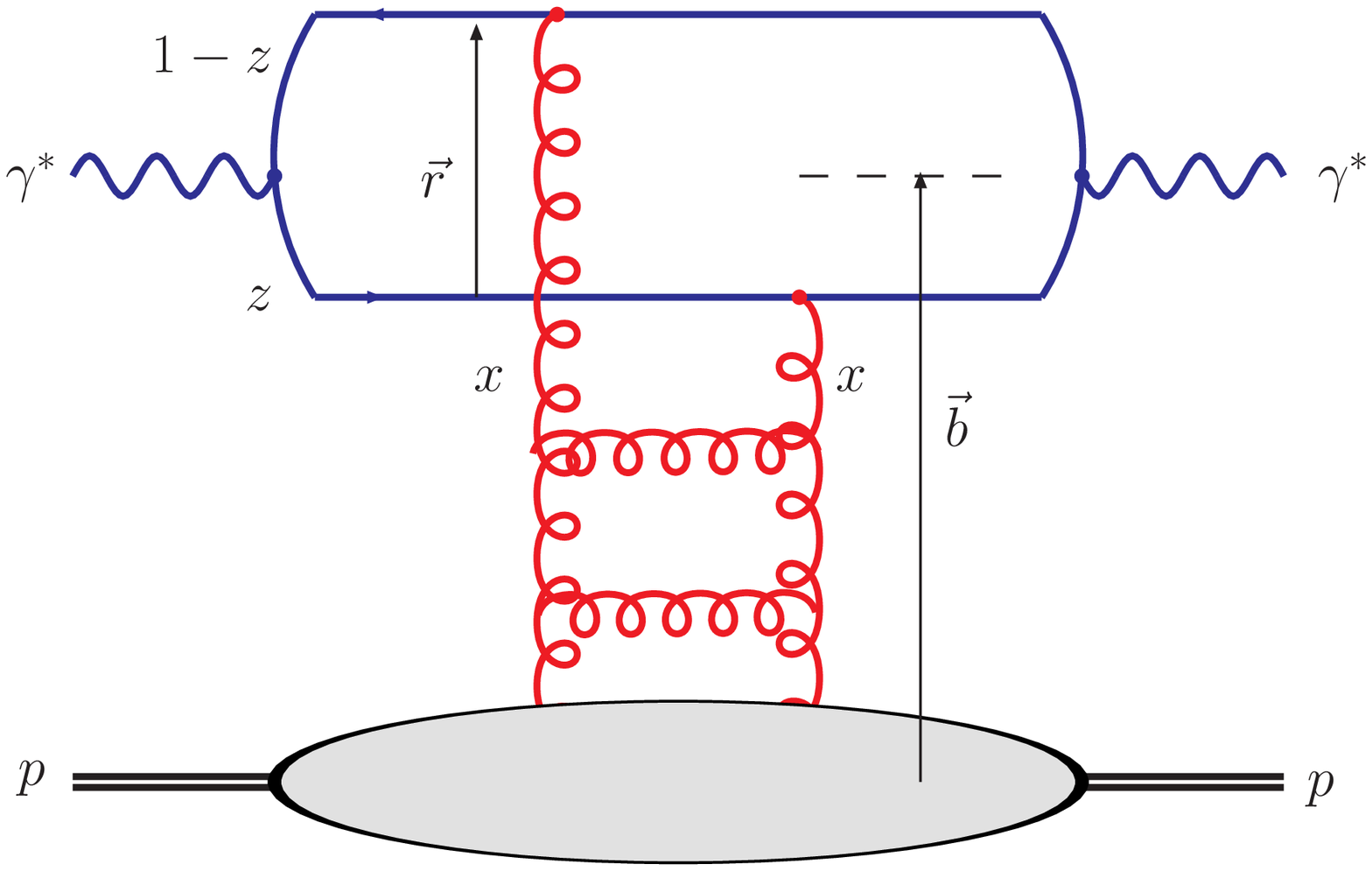}
  \end{minipage}%
  \begin{minipage}{0.5\textwidth}
    \centering
    \includegraphics[height=4.5cm,clip]{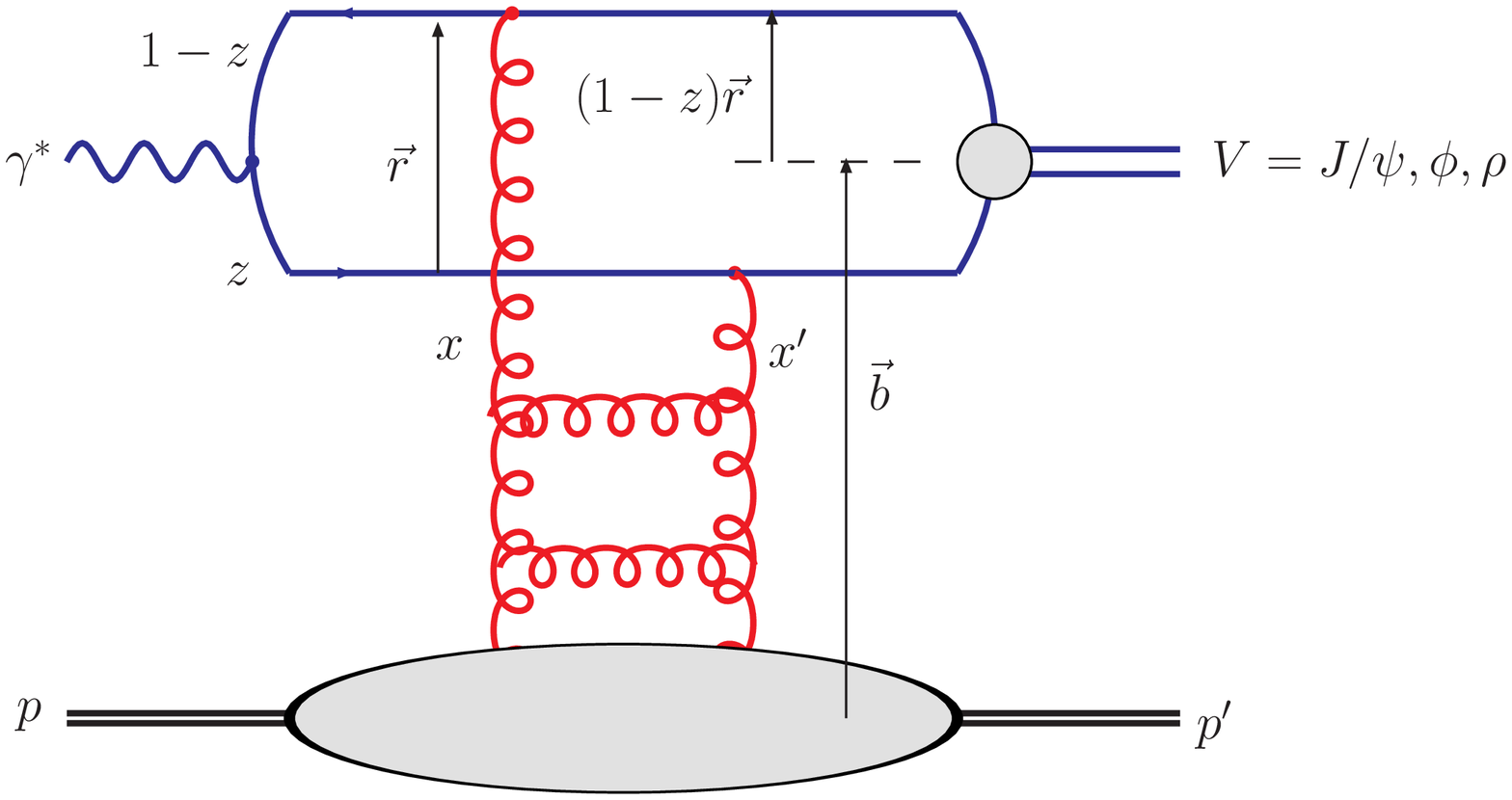}
  \end{minipage}
  \caption{The elastic scattering amplitude for inclusive DIS (left) and vector meson production (right).  For DVCS, the outgoing vector meson in the right-hand diagram is replaced by a \emph{real} photon.}
  \label{fig:diag}
\end{figure}

The total cross section for $\gamma^* p$ scattering, or equivalently $F_2$, is obtained, using \eqref{eq:elamp} and \eqref{eq:totsiggp}, by integrating the dipole cross section with the photon wave functions: 
\begin{eqnarray}
  \sigma^{\gamma^* p}_{T,L}(x,Q) 
  =  \mathrm{Im}\,\mathcal{A}^{\gamma^* p}_{T,L}(x,Q,\Delta=0) 
  = \sum_f \int\!\dif^2\vec{r} \int_0^1\!\frac{\dif z}{4\pi}
  (\Psi^{*}\Psi)_{T,L}^f
  \, \sigma_{q\bar q}(x,r),
  \label{eq:siggp}
\end{eqnarray}
with the overlap of the photon wave functions $(\Psi^{*}\Psi)_{T,L}^f$ defined as
\begin{align}
  (\Psi^{*}\Psi)_{T}^f &\equiv \frac{1}{2} \sum_{h, \bar h}
  \left[\Psi_{h \bar h,\lambda=+1}^* 
    \Psi_{h \bar h,\lambda=+1} 
    +
    \Psi_{h \bar h,\lambda=-1}^* 
    \Psi_{h \bar h,\lambda=-1}\right], 
  \label{eq:overlap2}\\
  (\Psi^*\Psi)_{L}^f &\equiv \sum_{h, \bar h}
  \Psi_{h \bar h,\lambda=0}^* 
  \Psi_{h \bar h,\lambda=0},
  \label{eq:overlap1}
\end{align}
where $\lambda $ denotes the photon helicity and $f$ the flavour of the $q\bar{q}$ pair.  The dependence on the quark flavour $f$ is specified below in Sect.~\ref{sec:wavefunctions}.  In the perturbative region, that is, for small dipole sizes $r$, the dipole cross section corresponds to exchange of a gluon ladder; see Fig.~\ref{fig:diag} (left).  The same diagram applies for exclusive final state production if the wave function of the outgoing virtual photon is replaced by the wave function of a specific final state; see Fig.~\ref{fig:diag} (right).

The amplitude for production of an exclusive final state $E$, such as a vector meson ($E=V$) or a real photon in DVCS ($E=\gamma$), is given by
\begin{eqnarray}
  \mathcal{A}^{\gamma^* p\rightarrow Ep}_{T,L}(x,Q,\Delta) &=& \int\!\dif^2\vec{r}\int_0^1\!\frac{\dif{z}}{4\pi}\;(\Psi_{E}^{*}\Psi)_{T,L}\;\mathcal{A}_{q\bar q}(x,r,\Delta)\\
  &=& \mathrm{i}\,\int\!\dif^2\vec{r}\int_0^1\!\frac{\dif{z}}{4\pi}\int\!\dif^2\vec{b}\;(\Psi_{E}^{*}\Psi)_{T,L}\;\mathrm{e}^{-\mathrm{i}\vec{b}\cdot\vec{\Delta}}\;2 [1-S(x,r,b)],
  \label{eq:ampvecm}
\end{eqnarray}
where $(\Psi_E^*\Psi)_{T,L}$ denotes the overlap of the photon and exclusive final state wave functions.  For DVCS, the amplitude involves a sum over quark flavours.  This expression, used in the analysis of exclusive $J/\psi$ photoproduction by Kowalski and Teaney~\cite{Kowalski:2003hm}, is derived under the assumption that the size of the quark--antiquark pair is much smaller than the size of the proton. The explicit perturbative QCD calculation of Bartels, Golec-Biernat and Peters \cite{Bartels:2003yj} shows that the non-forward wave functions can be written as the usual forward wave functions multiplied by exponential factors $\exp[\pm\mathrm{i}(1-z)\vec{r}\cdot\vec{\Delta}/2]$.  Effectively, the momentum transfer $\vec{\Delta}$ should conjugate to $\vec{b}+(1-z)\vec{r}$, the transverse distance from the centre of the proton to one of the two quarks of the dipole, rather than to $\vec{b}$, the transverse distance from the centre of the proton to the centre-of-mass of the quark dipole; see the right-hand diagram of Fig.~\ref{fig:diag}.

Assuming that the S-matrix element is predominantly real we may substitute $2[1-S(x,r,b)]$ in \eqref{eq:ampvecm} with $\dif\sigma_{q\bar q}/\dif^2\vec{b}$.

These two changes lead to
\begin{eqnarray}
  \mathcal{A}^{\gamma^* p\rightarrow Ep}_{T,L}(x,Q,\Delta) = \mathrm{i}\,\int\!\dif^2\vec{r}\int_0^1\!\frac{\dif{z}}{4\pi}\int\!\dif^2\vec{b}\;(\Psi_{E}^{*}\Psi)_{T,L}\; 
  \mathrm{e}^{-\mathrm{i}[\vec{b}-(1-z)\vec{r}]\cdot\vec{\Delta}}
  \;\frac {\dif\sigma_{q\bar q}}{\dif^2\vec{b}}.
  \label{eq:newampvecm}
\end{eqnarray}
The elastic diffractive cross section is then given by
\begin{eqnarray}
  \frac{\dif\sigma^{\gamma^* p\rightarrow Ep}_{T,L}}{\dif t}
  =\frac{1}{16\pi}\left\lvert\mathcal{A}^{\gamma^* p\rightarrow Ep}_{T,L}\right\rvert^2
  =\frac{1}{16\pi}
  \left\lvert
  \int\!\dif^2\vec{r}\int_0^1\!\frac{\dif{z}}{4\pi}\int\!\dif^2\vec{b}\;(\Psi_{E}^{*}\Psi)_{T,L}\; 
  \mathrm{e}^{-\mathrm{i}[\vec{b}-(1-z)\vec{r}]\cdot\vec{\Delta}}
  \;\frac {\dif\sigma_{q\bar q}}{\dif^2\vec{b}}
  \right\rvert^2.
  \label{eq:xvecm1}
\end{eqnarray}
This is the basic equation for the simultaneous analysis of different exclusive processes performed in this paper.

\subsection{Forward photon wave functions} \label{sec:wavefunctions}
The forward photon wave functions were perturbatively calculated in QCD by many authors; see, for example, Refs.~\cite{Dosch:1996ss,Lepage:1980fj}.  The normalised photon wave function for the longitudinal photon polarisation ($\lambda = 0$) is given by~\cite{Forshaw:2003ki} 
\begin{equation}
  \Psi_{h\bar{h},\lambda=0}(r,z,Q) =  e_f e \, \sqrt{N_c}\, 
  \delta_{h,-\bar h} \, 2Qz(1-z)\, \frac{K_0(\epsilon r)}{2\pi}, 
  \label{lspinphot}
\end{equation}
and for the transverse photon polarisations ($\lambda = \pm 1$) by
\begin{equation}
  \Psi_{h\bar{h},\lambda=\pm 1}(r,z,Q) =
  \pm e_f e \, \sqrt{2N_c}\,
  \left\{
  \mathrm{i}e^{\pm \mathrm{i}\theta_r}[
    z\delta_{h,\pm}\delta_{\bar h,\mp} - 
    (1-z)\delta_{h,\mp}\delta_{\bar h,\pm}] \partial_r \, + \, 
  m_f \delta_{h,\pm}\delta_{\bar h,\pm}
  \right\}\, \frac{K_0(\epsilon r)}{2\pi},
  \label{tspinphot}
\end{equation}
where $e=\sqrt{4\pi\alpha_{\mathrm{em}}}$, the subscripts $h$ and $\bar h$ are the helicities of the quark and the antiquark respectively and $\theta_r$ is the azimuthal angle between the vector $\vec{r}$ and the $x$-axis in the transverse plane.  $K_0$ is a modified Bessel function of the second kind, $\epsilon^2 \equiv z(1-z)Q^2+m_f^2$ and $N_c=3$ is the number of colours.  The flavour $f$ dependence enters through the values of the quark charge $e_f$ and mass $m_f$, and $\partial_r K_0(\epsilon r) = -\epsilon K_1(\epsilon r)$.

\subsubsection{Total DIS cross sections} \label{sec:wavefunction0}
In the case of the total DIS cross section $\sigma^{\gamma^* p}$, which is obtained from the elastic $\gamma^* p \to \gamma^* p$ amplitude via the optical theorem, the squared photon wave functions summed over the quark helicities for a given photon polarisation and quark flavour are given by the tree-level QED expressions:
\begin{align}
  (\Psi^*\Psi)_{T}^f &\equiv \frac{1}{2} \sum_{\substack{h,\bar{h}=\pm\frac{1}{2}\\\lambda=\pm1}}\Psi_{h\bar{h},\lambda}^*\Psi_{h\bar{h},\lambda}
  = \frac{2N_c}{\pi}\alpha_{\mathrm{em}}e_f^2\left\{\left[z^2+(1-z)^2\right]\epsilon^2 K_1^2(\epsilon r) + m_f^2 K_0^2(\epsilon r)\right\},
  \label{eq:overgg}
  \\
  (\Psi^*\Psi)_{L}^f &\equiv  \sum_{h,\bar{h}=\pm\frac{1}{2}}\Psi_{h\bar{h},\lambda=0}^*\Psi_{h\bar{h},\lambda=0}
  = \frac{8N_c}{\pi}\alpha_{\mathrm{em}}e_f^2 Q^2 z^2(1-z)^2 K_0^2(\epsilon r).
\label{eq:overgg1}
\end{align}
At small dipole sizes these expressions are well motivated since they can be derived from the LO $k_t$-factorisation formulae.  At large dipole sizes the wave functions are suppressed, since for large values of the argument the modified Bessel functions behave as $K_0(\epsilon r), K_1(\epsilon r) \sim \sqrt{\pi/(2\epsilon r)}\, \exp(-\epsilon r)$.  At larger $Q^2$ values the wave functions are suppressed for large $r$ unless $z$ is close to the end-point values of zero or one.\footnote{This is the origin of the statement that the transverse cross section is more inherently non-perturbative than the longitudinal cross section, since the contribution from the end-points is suppressed for the longitudinal but not the transverse case, see \eqref{eq:overgg} and \eqref{eq:overgg1}.} Near the end-points or at small $Q^2$  the wave functions are sensitive to the non-zero quark masses $m_f$, which prevent the integrals over $r$ of the modified Bessel functions from diverging.  Of course, near the end-points or at small $Q^2$ the expressions \eqref{eq:overgg} and \eqref{eq:overgg1} should be considered as a model in which the value of the light quark masses provides a cut-off scale which should be related to the physical cut-off scale generated by confinement effects.  It is therefore customary in dipole models to identify the light quark masses with the pion mass.

\subsubsection{Deeply virtual Compton scattering} \label{sec:wavefunction1}
In addition to the total DIS cross section $\sigma^{\gamma^* p}$, the photon wave functions determine also the DVCS process, $\gamma^* p \to \gamma p$. Here the outgoing photon is real and therefore the process is directly observed at HERA. For real photons, only the transversely polarised overlap function contributes to the cross section.  Summed over the quark helicities, for a given quark flavour $f$ it is given by
\begin{equation}
  (\Psi_\gamma^*\Psi)_{T}^f = \frac{2N_c}{\pi}\alpha_{\mathrm{em}}e_f^2\left\{\left[z^2+(1-z)^2\right]\epsilon K_1(\epsilon r) m_f K_1(m_f r)+ m_f^2 K_0(\epsilon r) K_0(m_f r)\right\}.
  \label{eq:overlap_dvcs}
\end{equation}

\subsection{Forward vector meson wave functions} \label{sec:wavefunction2}
Various conventions are used in the literature for the forward vector meson wave functions.  Recently, Forshaw, Sandapen and Shaw (FSS) \cite{Forshaw:2003ki} suggested some guidelines for bringing order into this problem.  We will adopt their prescription in this section, apart from the overall normalisation factor of $1/(4\pi)$ discussed previously, which in our case appears in the integration measure.

The simplest approach to modelling the vector meson wave function is to assume, following Refs.~\cite{Forshaw:2003ki,Dosch:1996ss,Kowalski:2003hm}, that the vector meson is predominantly a quark--antiquark state and that the spin and polarisation structure is the same as in the photon case.  In complete analogy to the transversely polarised photon wave function \eqref{tspinphot}, the transversely polarised vector meson wave function is
\begin{equation}
  \Psi^V_{h\bar{h},\lambda=\pm 1}(r,z) =
  \pm\sqrt{2N_c}\, \frac{1}{z(1-z)} \, 
  \left\{
  \mathrm{i}e^{\pm \mathrm{i}\theta_r}[
    z\delta_{h,\pm}\delta_{\bar h,\mp} - 
    (1-z)\delta_{h,\mp}\delta_{\bar h,\pm}] \partial_r \, + \, 
  m_f \delta_{h,\pm}\delta_{\bar h,\pm}
  \right\}\, \phi_T(r,z).
  \label{tspinvm}
\end{equation}
The longitudinally polarised wave function is slightly more complicated due to the fact that the coupling of the quarks to the meson is non-local, contrary to the photon case~\cite{Forshaw:2003ki}.  It is given by
\begin{equation}
  \Psi^V_{h\bar{h},\lambda=0}(r,z) = \sqrt{N_c}\,
  \delta_{h,-\bar h} \,
  \left[ M_V\,+ \, \delta \, \frac{m_f^2 - \nabla_r^2}{M_Vz(1-z)}\,  
    \right] \, \phi_L(r,z),
  \label{lspinvm}
\end{equation}
where $\nabla_r^2 \equiv (1/r)\partial_r + \partial_r^2$ and $M_V$ is the meson mass.  The difference in the structure of the longitudinal wave function is due to the non-local term proportional to $\delta$, which was first introduced by Nemchik, Nikolaev, Predazzi and Zakharov (NNPZ) \cite{Nemchik:1994fp,Nemchik:1996cw}.

Formulae \eqref{tspinvm} and \eqref{lspinvm} uniquely define the scalar part of the vector meson wave function $\phi_{T,L}(r,z)$, which is obtained from the photon wave function by the replacement
\begin{equation}
  e_f e\, z(1-z) \, \frac{K_0(\epsilon r)}{2\pi}\qquad\longrightarrow \qquad 
  \phi_{T,L}(r,z),
  \label{eq:wvident} 
\end{equation}
with the prefactor $2Q \to M_V$ for the case of the longitudinal polarisation.  Note that this definition of $\phi_{T,L}(r,z)|_{r=0}$ matches, up to a constant factor, the definition of the distribution amplitude in QCD.

The overlaps between the photon and the vector meson wave functions read then:
\begin{align}
  (\Psi_V^*\Psi)_{T} &= \hat{e}_f e\, \frac{N_c}{\pi z(1-z)} \,
  \left\{m_f^2 K_0(\epsilon r)\phi_T(r,z) - \left[z^2+(1-z)^2\right]\epsilon K_1(\epsilon r) \partial_r \phi_T(r,z)\right\},
  \label{eq:overt}
  \\
  (\Psi_V^*\Psi)_{L} &=  \, \hat{e}_f e \, \frac{N_c}{\pi}\,
  2Qz(1-z)\,K_0(\epsilon r)\,
  \left[M_V\phi_L(r,z)+ \delta\,\frac{m_f^2 - \nabla_r^2}{M_Vz(1-z)}
    \phi_L(r,z)\right],
  \label{eq:overl}
\end{align}
where the effective charge $\hat{e}_f=2/3$, $1/3$, or $1/\sqrt{2}$, for $J/\psi$, $\phi$, or $\rho$ mesons respectively.  Although it seems to be more natural to set $\delta=1$ as it was done in Refs.~\cite{Nemchik:1994fp,Nemchik:1996cw,Forshaw:2003ki}, we shall also use the value $\delta=0$ in order to match the assumptions of other models \cite{Dosch:1996ss,Kowalski:2003hm}.  Note that the additional factor of $1/[z(1-z)]$ in \eqref{eq:overt} and \eqref{eq:overl} as compared to the photon overlap functions \eqref{eq:overgg} and \eqref{eq:overgg1} is due to the identification \eqref{eq:wvident}.

The usual assumption that the quantum numbers of the meson are saturated by the quark--antiquark pair, that is, that the possible contributions of gluon or sea-quark states to the wave function may be neglected, allows the normalisation of the vector meson wave functions to unity:
\begin{equation}
  1 = \sum_{h,\bar h} \int \dif^2 \vec{r} 
  \int_0 ^1 \frac{\dif z}{4\pi}\, \left\lvert\Psi^V _{h\bar h, \lambda}(\vec{r},z)\right\rvert^2.
  \label{fullnorm}
\end{equation}
Thus, in the scheme presented here the normalisation conditions for the scalar parts of the wave functions are
\begin{gather}
  \label{eq:nnz_normt}
  1 = \frac{N_c}{2\pi}\int_0^1\!\frac{\dif{z}}{z^2(1-z)^2}\int\!\dif^2\vec{r}\;
  \left\{m_f^2\phi_T^2+\left[z^2+(1-z)^2\right]
  \left(\partial_r\phi_T\right)^2\right\},\\
  \label{eq:nnz_norml}
  1 = \frac{N_c}{2\pi} \int_0^1\!
  \dif{z}\,
  \int\!\dif^2\vec{r}\;
  \left[
    M_V\phi_L+
    \delta\,
    \frac{m_f^2-\nabla_r^2}{M_V z(1-z)}\,\phi_L\right]^2.
\end{gather}
Another important constraint on the vector meson wave functions is obtained from the decay width.  It is commonly assumed that the decay width can be described in a factorised way; the perturbative matrix element $q \bar{q} \to \gamma^* \to l^+ l^-$ factorises out from the details of the wave function, which contributes only through its properties at the origin.\footnote{Usually, one assumes that the factorisation holds and that the perturbative QCD corrections are similar for the process of vector meson production $\gamma^*(Q^2) + 2g \to V$ and for the vector meson decay $V\to \gamma^* \to l^+ l^-$, thus the corrections can be absorbed into the wave function.}  The decay widths are then given by
\begin{gather}
  \label{eq:nnz_fvt}
  f_{V,T} = \hat{e}_f\, \left.\frac{N_c}{2\pi M_V}
  \int_0^1\!\frac{\dif{z}}{z^2(1-z)^2}
  \left\{m_f^2-\left[z^2+(1-z)^2\right]\nabla_r^2\right\}\phi_T(r,z)\right\rvert_{r=0},\\
  \label{eq:nnz_fvl}
  f_{V,L} = \hat{e}_f\, \left.\frac{N_c}{\pi}
  \int_0^1\!
  \dif{z}\,
  \left[
    M_V +\delta\, \frac{m_f^2-\nabla_r^2}{M_Vz(1-z)}\right]
  \phi_L(r,z)\right\rvert_{r=0}.
\end{gather} 
The coupling of the meson to the electromagnetic current, $f_V$, is obtained from the measured electronic decay width by
\begin{equation}
  \Gamma_{V\to e^+e^-} = \frac{4\pi\alpha_{\rm em}^2f_V^2}{3M_V}.
\end{equation}

In order to complete the model of the vector meson wave function the scalar parts of the wave functions $\phi_{T,L}(r,z)$ should be specified.  In the photon case the scalar part is given by modified Bessel functions, whereas for vector mesons various quark models tell us that a hadron at rest can be modelled by Gaussian fluctuations in transverse separation.  The \emph{proton} wave function is also directly seen to have a Gaussian form from the $t$-distributions of vector mesons at HERA; see the discussion of the proton shape below.  After assuming a Gaussian form the modelling freedom reduces to the choice of a fluctuating variable.

Dosch, Gousset, Kulzinger and Pirner (DGKP) \cite{Dosch:1996ss} made the simplest assumption that the longitudinal momentum fraction $z$ fluctuates independently of the transverse quark momentum $\vec{k}$, where $\vec{k}$ is the Fourier conjugate variable to the dipole vector $\vec{r}$.  In what follows, this type of scalar wave function will be called the factorised wave function.  In the DGKP model the parameter $\delta=0$ in \eqref{eq:overl}, \eqref{eq:nnz_norml} and \eqref{eq:nnz_fvl}.  The DGKP model was further simplified by Kowalski and Teaney~\cite{Kowalski:2003hm}, who assumed that the $z$ dependence of the wave function for the longitudinally polarised meson is given by the short-distance limit of $z(1-z)$ \cite{Frankfurt:1997fj}. For the transversely polarised meson they set $\phi_T(r,z) \propto [z(1-z)]^2$ in order to suppress the contribution from the end-points ($z\to 0,1$).  This leads to the ``Gaus-LC'' \cite{Kowalski:2003hm} wave functions given by\footnote{Kowalski and Teaney \cite{Kowalski:2003hm} used a somewhat different convention; see the appendix for more details.}
\begin{align}
  \phi_{T}(r,z) &= N_{T} [z(1-z)]^2\exp(-r^2/2R_{T}^2),\label{eq:Gaus-LC-T}\\
  \phi_{L}(r,z) &= N_{L} z(1-z)\exp(-r^2/2R_{L}^2).
  \label{eq:Gaus-LC-L}
\end{align}
The values of the constants $N_{T,L}$ and $R_{T,L}$ in \eqref{eq:Gaus-LC-T} and \eqref{eq:Gaus-LC-L}, determined by requiring the correct normalisation and by the condition $f_V=f_{V,T}=f_{V,L}$, are given in Table \ref{tab:GLCparams}. 
\begin{table}
  \centering
  \begin{tabular}{cccc|cccc}
    \hline\hline
    Meson & $M_V$/GeV & $f_V$ & $m_f$/GeV & $N_T$ & $R_T^2$/GeV$^{-2}$ & $N_L$ & $R_L^2$/GeV$^{-2}$ \\ \hline
    $J/\psi$ & 3.097 & 0.274 & 1.4 & 1.23 & 6.5 & 0.83 & 3.0  \\
    $\phi$ & 1.019 & 0.076 & 0.14 & 4.75 & 16.0 & 1.41 & 9.7  \\
    $\rho$ & 0.776 & 0.156 & 0.14 & 4.47 & 21.9 & 1.79 & 10.4 \\
    \hline\hline
  \end{tabular}
  \caption{Parameters of the ``Gaus-LC'' vector meson wave functions.}
  \label{tab:GLCparams}
\end{table}

The main advantage of the factorised wave functions is their simplicity.  Probably a more realistic approach starts from the observation of Brodsky, Huang and Lepage \cite{Brodsky:1980vj} that the fluctuation of the quark three-momentum $\vec{p}$ in the rest frame of the meson could be described in a boost-invariant form.  In the meson rest frame, the momentum $\vec{p}$ is connected to the $q\bar q$ invariant mass by $M^2=4(p^2+m_f^2)$.  In the light-cone frame, the $q\bar q$ invariant mass is given by $M^2=(k^2+m_f^2)/[z(1-z)]$.  This leads to 
\begin{equation}
  p^2= \frac{k^2+m_f^2}{4z(1-z)} - m_f^2,
\end{equation}
and a simple ansatz for the scalar wave function in momentum space of
\begin{align}
  \tilde{\phi}_{T,L}(k,z) \propto \exp\left[-\frac{\mathcal{R}^2}{8} \left(\frac{k^2+m_f^2}{z(1-z)}-4m_f^2\right)\right].
  \label{eq:Boosted-Gaus}
\end{align}
This is the basis for the ``boosted Gaussian'' wave function of FSS \cite{Forshaw:2003ki}, which was first proposed by NNPZ \cite{Nemchik:1994fp,Nemchik:1996cw}.\footnote{Following FSS~\cite{Forshaw:2003ki} we set the Coulombic part of the NNPZ wave function \cite{Nemchik:1994fp,Nemchik:1996cw} to zero to avoid singular behaviour at the origin.  This should be reasonable for $\rho$ and $\phi$ mesons, but has less justification for $J/\psi$ mesons.}  In the configuration space these wave functions are given by the Fourier transform of \eqref{eq:Boosted-Gaus}:
\begin{equation}
  \phi_{T,L}(r,z) = \mathcal{N}_{T,L} z(1-z)
  \exp\left(-\frac{m_f^2 \mathcal{R}^2}{8z(1-z)} - \frac{2z(1-z)r^2}{\mathcal{R}^2} + \frac{m_f^2\mathcal{R}^2}{2}\right).
\end{equation}
Note that the ``boosted Gaussian'' wave function has the proper short-distance limit, $\sim z(1-z)$, for $m_f \to 0$.  Following the authors of the model we set $\delta=1$ in equations \eqref{eq:overl}, \eqref{eq:nnz_norml} and \eqref{eq:nnz_fvl}, defining the longitudinally polarised overlap, the normalisation and the decay constant respectively.  We choose the ``radius'' parameter $\mathcal{R}$ to reproduce the experimentally measured leptonic decay width of the vector meson for the longitudinally polarised case.  This means that the calculated decay width for the transversely polarised case will be slightly different.  The parameters $\mathcal{R}$ and $\mathcal{N}_{T,L}$ are determined by the normalisation conditions \eqref{eq:nnz_normt} and \eqref{eq:nnz_norml} and the decay width condition \eqref{eq:nnz_fvl}.

The parameters of the ``boosted Gaussian'' wave function are given in Table~\ref{tab:bGparams}, where we also show the value of $f_{V,T}$ \eqref{eq:nnz_fvt} computed using the given values of $\mathcal{R}$ and $\mathcal{N}_T$.  (Recall that we require that $f_{V,L}=f_V$.)
\begin{table}
  \centering
  \begin{tabular}{cccc|cccc}
    \hline\hline
    Meson & $M_V$/GeV & $f_V$ & $m_f$/GeV & $\mathcal{N}_T$& $\mathcal{N}_L$& $\mathcal{R}^2$/GeV$^{-2}$ & $f_{V,T}$ \\ \hline
     $J/\psi$ & 3.097 & 0.274 & 1.4 & 0.578  & 0.575  & 2.3 & 0.307 \\
     $\phi$ & 1.019 & 0.076 & 0.14 &  0.919  & 0.825  & 11.2 & 0.075 \\
     $\rho$ & 0.776 & 0.156 & 0.14 & 0.911 & 0.853 & 12.9 & 0.182 \\
     \hline\hline
  \end{tabular}
  \caption{ Parameters of the ``boosted Gaussian'' vector meson wave functions.}
  \label{tab:bGparams}
\end{table}

The ``boosted Gaussian'' wave function is very similar to the ``Gaus-RF'' wave function used in the KT investigation \cite{Kowalski:2003hm}, except for the Jacobian of the transformation from the rest frame variables to the light-cone variables.  We focus here on the ``boosted Gaussian'' version because of the proper short distance limit of the $z$ dependence.  The ``CORNELL'' wave function used in Ref.~\cite{Kowalski:2003hm} cannot be used for light vector mesons since it was obtained within the nonrelativistic bound-state model.

Comparing the values of the radius parameters given in Tables \ref{tab:GLCparams} and \ref{tab:bGparams} we note that the meson description with the ``boosted Gaussian'' wave function is more self-consistent; the values of the radius parameters $R_T$ and $R_L$ for the ``Gaus-LC'' wave functions are very different indicating that there are large dynamical corrections to at least one of the meson polarisation states.  For the ``boosted Gaussian'' there is only one radius parameter $\mathcal{R}$, since the description of the meson is assumed to be boost-invariant between the meson rest frame and the light-cone frame.  The shortcoming of this approach is that the predicted decay constant $f_V$ differs slightly between the transverse and the longitudinal polarisation components.  However, the differences between the decay constants of the ``boosted Gaussian'' wave function are relatively small compared to the differences between the radii of the ``Gaus-LC'' wave function.  To quantify this effect we fix the parameter $R_T$ of the ``Gaus-LC'' wave function to the same value as $R_L$, then we predict the value of the decay constant $f_{V,T}$ (allowing for $N_T$ to be determined from the normalisation constraint).  The resulting values of $f_{V,T}$ were $0.44$, $0.13$ and $0.33$ for $J/\psi$, $\phi$ and $\rho$ mesons respectively, to be compared with the experimental values of $f_V$ ($=f_{V,L}$) of $0.27$, $0.08$ and $0.16$.  That is, the differences between $f_{V,T}$ and $f_{V,L}$ for the ``Gaus-LC'' wave function are much larger than the equivalent differences for the ``boosted Gaussian'' wave function; see Table \ref{tab:bGparams}.

The agreement between the decay constants for the longitudinal and transverse polarisation with the ``boosted Gaussian'' wave function is particularly good for the $\phi$ meson wave function.  We note, {\it en passant}, that the difference between the two decay constants $f_{V,T}$ and $f_{V,L}$ depends on the assumed quark mass; for the $\phi$ meson the difference is minimal for the strange quark mass of 0.14 GeV, for the $J/\psi$ meson it is minimal for the charm quark mass of 1.15 GeV, and for the $\rho$ meson it decreases slightly with decreasing quark mass but there is still a significant difference even when the quark mass is set to zero.\footnote{For the $\phi$ meson, the relative difference of decay constants $f_{V,T}$ and $f_{V,L}$ is 11\% for $m_{s}=0.3$ GeV and 3\% for $m_{s}=0.05$ GeV.  For the $\rho$ meson, the relative difference of decay constants is 36\% for $m_{u,d}=0.3$ GeV and 14\% for $m_{u,d}=0.05$ GeV.}

\begin{figure}
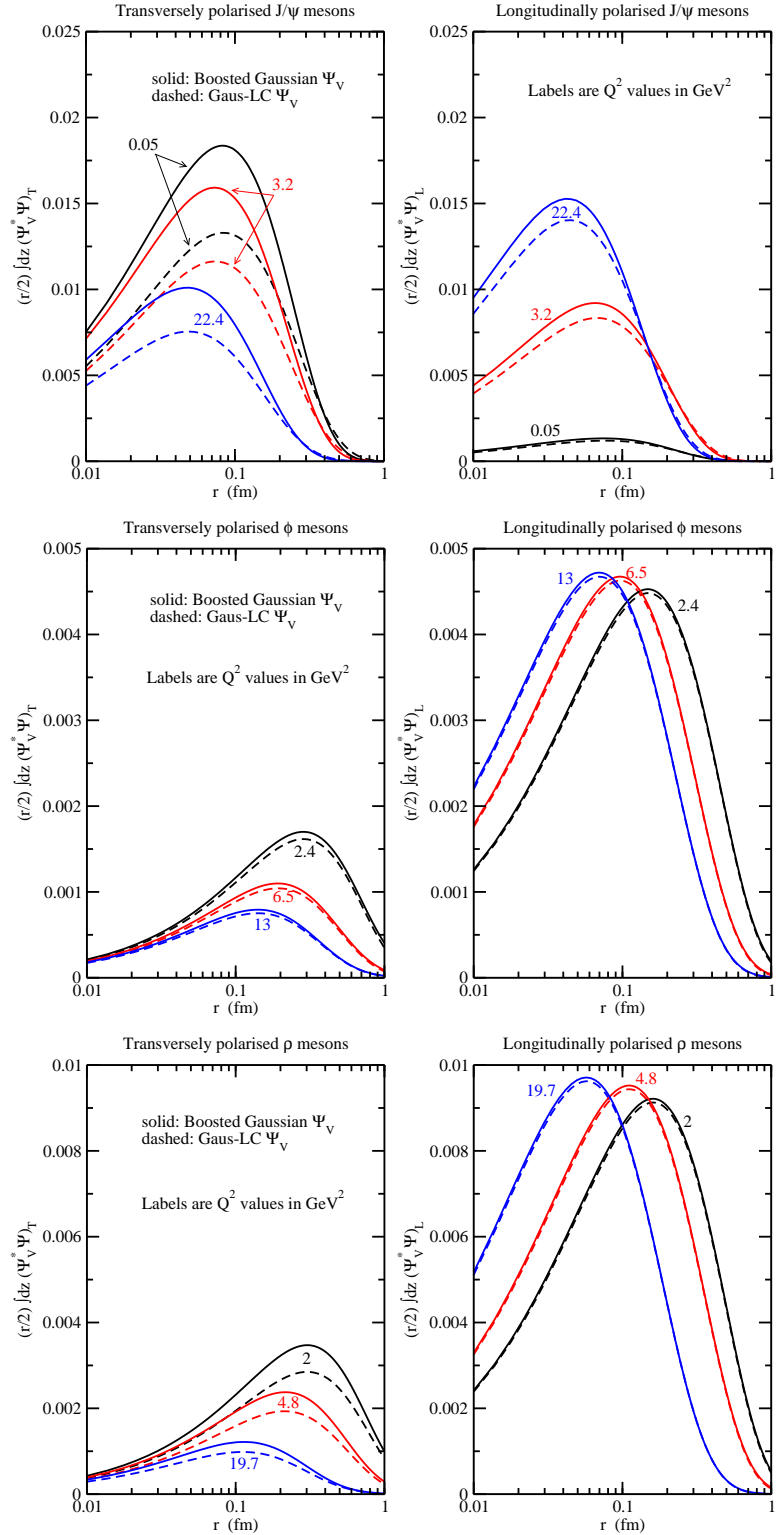

  \centering
  \includegraphics[width=0.6\textwidth,clip]{overlap_jpsi.eps}\\[2mm]
  \includegraphics[width=0.6\textwidth,clip]{overlap_phi.eps}\\[2mm]
  \includegraphics[width=0.6\textwidth,clip]{overlap_rho.eps}
  \caption{Overlap functions \eqref{eq:overt} and \eqref{eq:overl} between the photon and vector meson wave functions integrated over $z$ for the three different vector mesons at $Q^2$ values representative of the data.}
  \label{fig:overlap}
\end{figure}
In Fig.~\ref{fig:overlap} we show the overlap functions between the photon and vector meson wave functions integrated over $z$ for the three different vector mesons at $Q^2$ values representative of the data discussed later in Sect.~\ref{sec:vmdesc}.  To be precise, we plot the quantity
\begin{equation}
  2\pi r\int_0^1\!\frac{\dif{z}}{4\pi}\;(\Psi_V^*\Psi)_{T,L}.
\end{equation}
The plots show that the longitudinal overlap functions for the ``Gaus-LC'' and ``boosted Gaussian'' cases are more similar than the transverse overlap functions for all three vector mesons.  For the $\phi$ meson there is also a good agreement for the transverse overlap function.  This indicates that observable quantities for $\phi$ mesons computed with either the ``Gaus-LC'' or ``boosted Gaussian'' wave functions should be very similar, in spite of the sizable disagreement between $R_T^2$ and $R_L^2$ for the ``Gaus-LC'' wave function.

\subsection{Dipole cross sections}
\subsubsection{Review of dipole cross sections}
The dipole model became an important tool in investigations of deep-inelastic scattering due to the initial observation of Golec-Biernat and W\"usthoff (GBW)~\cite{Golec-Biernat:1998js,Golec-Biernat:1999qd} that a simple ansatz for the dipole cross section integrated over the impact parameter $\vec{b}$, $\sigma_{q\bar q}$, was able to describe simultaneously the total inclusive and diffractive DIS cross sections:
\begin{equation}
  \sigma_{q\bar{q}}^{\rm GBW}(x,r) = \sigma_0\left(1-\mathrm{e}^{-r^2Q_s^2(x)/4}\right),
  \label{eq:sigGBW}
\end{equation}
where $\sigma_0$ is a constant and $Q_s(x)$ denotes the $x$ dependent saturation scale, $Q_s^2(x) = (x_0/x)^{\lambda_{\rm GBW}}$ GeV$^2$.  The parameters $\sigma_0=23$ mb, $\lambda_{\rm GBW}=0.29$ and $x_0=3\times 10^{-4}$ were determined from a fit to the $F_2$ data without including charm quarks.  After inclusion of the charm quark contribution with mass $m_c=1.5$~GeV into the fit, the parameters of the GBW model changed to $\sigma_0=29$ mb, $\lambda_{\rm GBW}=0.28$ and $x_0=4\times 10^{-5}$.  Although the dipole model is theoretically well justified for small-size dipoles only, the GBW model provided a good description of data from medium $Q^2$ values ($\sim$ 30 GeV$^2$) down to low $Q^2$ ($\sim$ 0.1 GeV$^2$).  The saturation scale $Q_s^2$ is intimately related to the gluon density in the transverse plane.  The exponent $\lambda_{\rm GBW}$ determines therefore the growth of the total and diffractive cross sections with decreasing $x$.  For dipole sizes which are large in comparison to $1/Q_s$ the dipole cross section saturates by approaching a constant value $\sigma_0$, which becomes independent of $\lambda_{\rm GBW}$.  It is a characteristic feature of the GBW model that a good description of data is due to large saturation effects, that is, the strong growth due to the factor $x^{-\lambda_{\rm GBW}}$ is, for large dipoles, significantly flattened by the exponentiation in \eqref{eq:sigGBW}.
 
The assumption of dipole saturation provided an attractive theoretical background for investigation of the transition from the perturbative to non-perturbative regimes in the HERA data.  Despite the appealing simplicity and success of the GBW model it suffers from clear shortcomings.  In particular it does not include scaling violations, that is, at large $Q^2$ it does not match with QCD (DGLAP) evolution.  Therefore, Bartels, Golec-Biernat and Kowalski (BGBK)~\cite{Bartels:2002cj} proposed a modification of the original ansatz of \eqref{eq:sigGBW} by replacing $Q_s^2$ by a gluon density with explicit DGLAP evolution:
\begin{eqnarray} 
  \sigma_{q\bar q}^{\rm BGBK}(x,r) = \sigma_0 \left\{1-\exp\left[-\pi^2r^{2}\alpha_s(\mu^2)xg(x,\mu^2)/(3\sigma_0)\right]\right\}.
  \label{eq:sigBGBK}
\end{eqnarray}
The scale of the gluon density, $\mu^2$, was assumed to be $\mu^2 = C/r^2 + \mu_0^2$, and the gluon density was evolved according to the leading-order (LO) DGLAP equation without quarks.

The BGBK form of the dipole cross section led to significantly better fits to the HERA $F_2$ data than the original GBW model, especially in the region of larger $Q^2$. The good agreement of the original model with the DIS diffractive HERA data was also preserved.  However, the contribution from charm quarks was omitted in the BGBK analysis.

The BGBK analysis found, surprisingly, that there exist two distinct solutions, both giving a very good description of the HERA data, depending on the quark mass in the photon wave function.  The first solution was obtained assuming $m_{u,d,s} =0.14$ GeV and led to the initial gluon density, $xg(x,\mu_0^2)\propto x^{-\lambda_g}$, with the value of exponent $\lambda_g=0.28$ at $\mu_0^2=0.52$ GeV$^2$, which is very similar to the $\lambda_{\rm GBW}$.  As in the original model, the good agreement with data was due to substantial saturation effects.  In the second solution, which took $m_{u,d,s} = 0$, the value of the exponent was very different, $\lambda_g=-0.41$ at a fixed $\mu_0^2=1$ GeV$^2$.  The initial gluon density no longer rose at small $x$; it was valence-like, and QCD evolution played a much more significant role than in the solution with $m_{u,d,s} =0.14$ GeV.

The DGLAP evolution, which is generally used in the analysis of HERA data, may not be appropriate when $x$ approaches the saturation region.  Therefore, Iancu, Itakura and Munier \cite{Iancu:2003ge} proposed a new saturation model, the Colour Glass Condensate (CGC) model, in which gluon saturation effects are incorporated via an approximate solution of the Balitsky--Kovchegov equation \cite{Balitsky:1995ub,Kovchegov:1999yj,Kovchegov:1999ua}.  The CGC dipole cross section is
\begin{equation} \label{eq:cgc}
  \sigma_{q\bar{q}}^{\rm CGC}(x,r) = \sigma_0\times
  \begin{cases}
    \mathcal{N}_0\left(\frac{rQ_s}{2}\right)^{2\left(\gamma_s+\frac{1}{\kappa\lambda Y}\ln\frac{2}{rQ_s}\right)} & :\quad rQ_s\le 2\\
    1-\mathrm{e}^{-A\ln^2(BrQ_s)} & :\quad rQ_s>2
  \end{cases},
\end{equation}
where $Y=\ln(1/x)$, $\gamma_s=0.63$, $\kappa=9.9$ and $Q_s\equiv Q_s(x)=(x_0/x)^{\lambda/2}$.  The free parameters $\sigma_0$, $\mathcal{N}_0$, $\lambda$ and $x_0$ were determined by a fit to HERA $F_2$ data.  The coefficients $A$ and $B$ in the second line of \eqref{eq:cgc} are determined uniquely from the condition that $\sigma_{q\bar{q}}^{\rm CGC}$, and its derivative with respect to $rQ_s$, are continuous at $rQ_s=2$:
\begin{equation}
  A = -\frac{\mathcal{N}_0^2\gamma_s^2}{(1-\mathcal{N}_0)^2\ln(1-\mathcal{N}_0)}, \qquad B = \frac{1}{2}\left(1-\mathcal{N}_0\right)^{-\frac{(1-\mathcal{N}_0)}{\mathcal{N}_0\gamma_s}}.
\end{equation}

Later, also Forshaw and Shaw (FS) \cite{Forshaw:2004vv} proposed a Regge-type model with saturation effects.  The CGC and FS models provide a description of HERA inclusive and diffractive DIS data which is better than the original GBW model and comparable in quality to the BGBK analysis. Both models find strong saturation effects in HERA data comparable to the GBW model and the solution of the BGBK model with $m_{u,d,s}=0.14$ GeV.  

All approaches to dipole saturation discussed so far ignored a possible impact parameter dependence of the dipole cross section.  This dependence was introduced in this context by KT~\cite{Kowalski:2003hm}, who assumed that the dipole cross section is a function of the opacity $\Omega$, following for instance Ref.~\cite{Gotsman:1995bn}:
\begin{eqnarray}
  \frac{\dif\sigma_{qq}}{\dif^2\vec{b}} = 2\, \left(1-\mathrm{e}^{-\frac{\Omega}{2}}\right). 
  \label{eq:xdip}
\end{eqnarray}
At small $x$ the opacity $\Omega$ can be directly related to the gluon density, $xg(x,\mu^2)$, and the transverse profile of the proton, $T(b)$:
\begin{eqnarray}
  \Omega = \frac{\pi^2}{N_c}\, r^2\, \alpha_S(\mu^2)\, xg(x,\mu^2)\, T(b).
  \label{eq:omega}
\end{eqnarray}
The formulae of \eqref{eq:xdip} and \eqref{eq:omega} are called the Glauber--Mueller dipole cross section.  The diffractive cross section of this type was used around 50 years ago to study the diffractive dissociation of deuterons by Glauber \cite{Glauber:1955} and reintroduced by Mueller \cite{Mueller:1989st} to describe dipole scattering in deep-inelastic processes.

\subsubsection{Applied dipole cross sections}
Since the description of exclusive vector meson production is the focus of this investigation we concentrate here on impact parameter dependent dipole cross sections.  First, we use the same form of the differential dipole cross section as in the KT investigation \cite{Kowalski:2003hm}:
\begin{equation} \label{eq:dsigmad2b}
  \frac{\dif\sigma_{q\bar{q}}}{\dif^2\vec{b}} = 2\left[1-\exp\left(-\frac{\pi^2}{2N_c}r^2\alpha_S(\mu^2)xg(x,\mu^2)T(b)\right)\right].
\end{equation}
Here, the scale $\mu^2$ is related to the dipole size $r$ by $\mu^2=4/r^2+\mu_0^2$.  The gluon density, $xg(x,\mu^2)$, is evolved from a scale $\mu_0^2$ up to $\mu^2$ using LO DGLAP evolution without quarks:
\begin{equation}
  \frac{\partial xg(x,\mu^2)}{\partial\ln\mu^2} = \frac{\alpha_S(\mu^2)}{2\pi}\int_x^1\!\dif{z}\;P_{gg}(z)\frac{x}{z}g\left(\frac{x}{z},\mu^2\right).
\end{equation}
The initial gluon density at the scale $\mu_0^2$ is taken in the form
\begin{equation} \label{eq:inputgluon}
  xg(x,\mu_0^2) = A_g\,x^{-\lambda_g}\,(1-x)^{5.6}.
\end{equation}
The values of the parameters $\mu_0^2$, $A_g$, and $\lambda_g$ are determined from a fit to $F_2$ data.  For the light quarks, the gluon density is evaluated at $x=\xB$ (Bjorken-$x$), while for charm quarks, $x=\xB(1+4m_c^2/Q^2)$.  The contribution from beauty quarks is neglected.  For vector meson production, the gluon density is evaluated at $x=\xB(1+M_V^2/Q^2)$.  The LO formula for the running strong coupling $\alpha_S(\mu^2)$ is used, with three fixed flavours and $\Lambda_{\mathrm{QCD}}=0.2$ GeV.

The proton shape function $T(b)$ is normalised so that
\begin{equation} \label{eq:TGnorm}
  \int\!\dif^2\vec{b}\;T(b) = 1.
\end{equation}
We consider first a Gaussian form for $T(b)$, that is,
\begin{equation} \label{eq:GaussianTb}
  T_G(b) = \frac{1}{2\pi B_G}\mathrm{e}^{-\frac{b^2}{2B_G}},
\end{equation}
where $B_G$ is a free parameter which is fixed by the fit to the differential cross sections $\dif\sigma/\dif t$ for exclusive vector meson production.  This distribution yields the average squared transverse radius of the proton,
\begin{equation} \label{eq:BB}
  \langle  b^2 \rangle = 2B_G.
\end{equation}
Assuming that the Gaussian distribution given by \eqref{eq:GaussianTb} holds also in three dimensions (with a different normalisation factor) we obtain the relationship between the parameter $B_G$ and the Hofstadter radius of the proton $R_p$, namely $R_p^2 = 3B_G$.  Note that the Hofstadter experiment \cite{Mcallister:1956ng} measured the electromagnetic radius whereas we probe the gluonic distribution of the proton.

The two-dimensional Fourier transform of \eqref{eq:GaussianTb} has the exponential form which is supported by the data:\footnote{Note that for exclusive diffractive processes at large values of $(M_V^2 + Q^2)$ the typical dipole size $r$ is small, and the $t$-dependence of the cross section is determined entirely by the proton transverse profile.}
\begin{eqnarray}
  \frac{\dif\sigma^{\gamma^* p\to V p}}{\dif t} \propto \mathrm{e}^{-B_G |t|}.
  \label{eq:DiVM}
\end{eqnarray}

Alternatively, we assume that the gluonic density in the proton is evenly distributed over a certain area within a sharp boundary, and is zero beyond this boundary.  That is, we assume a step function, again normalised as in \eqref{eq:TGnorm}:
\begin{equation} \label{eq:StepTb}
  T_S(b) = \frac{1}{\pi b_S^2}\Theta\left(b_S-b\right),
\end{equation}
where $b_S$ is a free parameter, for which the average squared transverse radius of the proton is
\begin{equation} \label{eq:BBstep}
  \langle b^2 \rangle = \frac{b_S^2}{2}.
\end{equation}
This is the form of $T(b)$ implicitly used in all $b$-\emph{independent} parameterisations of the dipole cross section.  That is, it is usually assumed that
\begin{equation}
  \frac{\dif\sigma_{q\bar{q}}}{\dif^2\vec{b}} \equiv 2[1-\mathrm{Re}\,S(x,r,b)] \equiv 2\,\mathcal{N}(x,r,b) = 2\,\mathcal{N}(x,r)\,\Theta\left(b_S-b\right),
\end{equation}
so that integration over $\vec{b}$ gives
\begin{equation}
  \sigma_{q\bar{q}}(x,r) = \sigma_0\,\mathcal{N}(x,r),
\end{equation}
where the parameter $\sigma_0\equiv 2\pi b_S^2$ is usually obtained by fitting to the $F_2$ data.  This is the form assumed in the GBW model \eqref{eq:sigGBW}, the BGBK model \eqref{eq:sigBGBK}, and the CGC model \eqref{eq:cgc}.  Note that the scattering amplitudes $\mathcal{N}(x,r,b)$ or $\mathcal{N}(x,r)$ can vary between zero and one, where $\mathcal{N}=1$ is the unitarity limit.

To introduce the impact parameter dependence into the CGC model \cite{Iancu:2003ge}, we modify \eqref{eq:cgc} to obtain the ``b-CGC'' model:
\begin{equation} \label{eq:bcgc}
  \frac{\dif\sigma_{q\bar{q}}}{\dif^2\vec{b}} \equiv 2\,\mathcal{N}(x,r,b)=2\times\begin{cases}
  \mathcal{N}_0\left(\frac{rQ_s}{2}\right)^{2\left(\gamma_s+\frac{1}{\kappa\lambda Y}\ln\frac{2}{rQ_s}\right)} & :\quad rQ_s\le 2\\
  1-\mathrm{e}^{-A\ln^2(BrQ_s)} & :\quad rQ_s>2
  \end{cases},
\end{equation}
where now the parameter $Q_s$ depends on the impact parameter:
\begin{equation} \label{eq:bcgc1}
  Q_s\equiv Q_s(x,b)=\left(\frac{x_0}{x}\right)^{\frac{\lambda}{2}}\;\left[\exp\left(-\frac{b^2}{2B_{\rm CGC}}\right)\right]^{\frac{1}{2\gamma_s}}.
\end{equation}
Note that, in contrast to the parameter $B_G$ in the KT approach, a straightforward interpretation of $B_{\rm CGC}$ in terms of the proton size is not possible due to the $r$~and~$Y$ dependence of the exponent $2\left(\gamma_s+\frac{1}{\kappa\lambda Y}\ln\frac{2}{rQ_s}\right)$ in \eqref{eq:bcgc}.

Following KT \cite{Kowalski:2003hm} we define the saturation scale $Q_S^2\equiv 2/r_S^2$, where the saturation radius $r_S$ is the dipole size where the scattering amplitude $\mathcal{N}$ has a value of $1-\exp(-1/2)\simeq 0.4$, that is, $r_S$ is defined by solving
\begin{equation} \label{eq:satdef}
  \mathcal{N}(x,r_S,b) = 1 - \mathrm{e}^{-\frac{1}{2}},
\end{equation}
with the same condition for the $b$-independent dipole models.  For the GBW model \eqref{eq:sigGBW}, the saturation scale $Q_S^2=2/r_S^2$ defined by \eqref{eq:satdef} coincides with $Q_s^2(x) \equiv (x_0/x)^{\lambda_{\rm GBW}}$ GeV$^2$.  However, for the CGC \eqref{eq:cgc} and b-CGC \eqref{eq:bcgc} models, the saturation scale $Q_S$ defined by \eqref{eq:satdef} differs from the parameter $Q_s$.  Note that we use upper-case $S$ and lower-case $s$ to distinguish between these two scales.  The saturation scale $Q_S$ is the quantity we shall later compute and compare for the different dipole models in Sect.~\ref{sec:Sat}.

\subsubsection{Phenomenological corrections for exclusive processes} \label{sec:pheno-improvements}
After performing the angular integrations, \eqref{eq:newampvecm} reduces to
\begin{equation} \label{eq:ampVM2}
  \mathcal{A}_{T,L}^{\gamma^* p\rightarrow Ep} = \mathrm{i}\,\int_0^\infty\!\dif{r}\,(2\pi r)\int_0^1\!\frac{\dif{z}}{4\pi}\int_0^\infty\!\dif{b}\,(2\pi b)\;(\Psi_E^*\Psi)_{T,L}\;J_0(b\Delta)\;J_0\left([1-z]r\Delta\right)\;\frac{\dif\sigma_{q\bar{q}}}{\dif^2\vec{b}},
\end{equation}
where $J_0$ is the Bessel function of the first kind and $E=V,\gamma$ denotes either the exclusive vector meson or DVCS final state.
The derivation of the expression for the exclusive vector meson production or DVCS amplitude, \eqref{eq:newampvecm}, relies on the assumption that the $S$-matrix is purely real and therefore the exclusive amplitude $\mathcal{A}$ is purely imaginary.  The real part of the amplitude can be accounted for by multiplying the differential cross section for vector meson production or DVCS, \eqref{eq:xvecm1}, by a factor $(1+\beta^2)$, where $\beta$ is the ratio of real to imaginary parts of the scattering amplitude $\mathcal{A}$ and is calculated using
\begin{equation}
  \beta = \tan(\pi\lambda/2), \quad\text{with}\quad \lambda \equiv \frac{\partial\ln\left(\mathcal{A}_{T,L}^{\gamma^* p\rightarrow Ep}\right)}{\partial\ln(1/x)}.
\end{equation}
This procedure (or similar) is adopted in other descriptions of vector meson production to account for the real part of the amplitude; see, for example, Refs.~\cite{Forshaw:2003ki,Nemchik:1996cw,Martin:1999wb}.

For vector meson production or DVCS, we should use the off-diagonal (or generalised) gluon distribution, since the two gluons in the right-hand diagram of Fig.~\ref{fig:diag} carry different fractions $x$ and $x^\prime$ of the proton's (light-cone) momentum.  In the leading $\ln(1/x)$ limit, the skewed effect vanishes.  However, the skewed effect can be accounted for, in the limit that $x^\prime \ll x \ll 1$, by multiplying the gluon distribution $xg(x,\mu^2)$ in \eqref{eq:dsigmad2b} by a factor $R_g$, given by \cite{Shuvaev:1999ce}
\begin{equation} \label{eq:Rg}
  R_g(\lambda) = \frac{2^{2\lambda+3}}{\sqrt{\pi}}\frac{\Gamma(\lambda+5/2)}{\Gamma(\lambda+4)}, \quad\text{with}\quad \lambda \equiv \frac{\partial\ln\left[xg(x,\mu^2)\right]}{\partial\ln(1/x)}.
\end{equation}
This skewedness effect is also accounted for in the calculation of vector meson production by Martin, Ryskin and Teubner (MRT) \cite{Martin:1999wb}, but is neglected in most other dipole model descriptions.

\section{Description of HERA data with the ``b-Sat'' model} \label{sec:vmdesc}
In this section we describe HERA data within the generalised impact parameter dipole saturation (``b-Sat'') model in which the dipole cross section is given by \eqref{eq:dsigmad2b} and the proton shape function $T(b)$ is assumed to be purely Gaussian \eqref{eq:GaussianTb}.  The total DIS cross section is given by \eqref{eq:siggp} and the photon overlap functions by \eqref{eq:overgg} and \eqref{eq:overgg1}.  For exclusive processes, the differential cross sections are given by \eqref{eq:xvecm1} with the phenomenological improvements described in Sect.~\ref{sec:pheno-improvements}.  For vector mesons the overlaps of wave functions are given by \eqref{eq:overt} and \eqref{eq:overl}, and for the DVCS process by \eqref{eq:overlap_dvcs}.

\begin{figure}
  \centering
  \includegraphics[width=0.6\textwidth,clip]{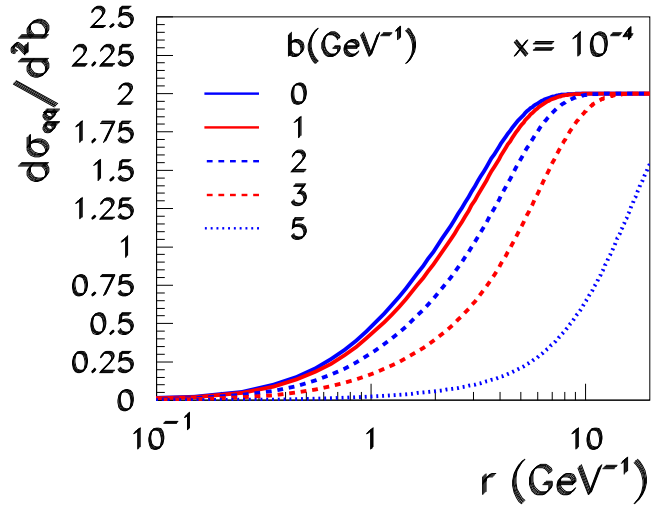}
  \caption{Dipole cross section at various impact parameters, as determined in the b-Sat model.}
  \label{fig:dsigb}
\end{figure}

The light quark masses are taken to be $m_{u,d,s}=0.14$ GeV, the value of the pion mass, which ensures the proper exponential cut-off of the photon wave functions \eqref{eq:overgg} and \eqref{eq:overgg1} at large distances.  The value of the charm mass was chosen to be $m_c=1.4$~GeV, but other choices for the charm and light quark masses are also discussed below.  The free parameters of the model are $\mu_0^2$, $A_g$ and $\lambda_g$ of the initial gluon distribution, $xg(x,\mu_0^2) = A_g\,x^{-\lambda_g}\,(1-x)^{5.6}$, and the proton width $B_G$.  The aim of the model is to describe with these four parameters the total DIS cross section for $\xB\le 0.01$ and all total and differential cross sections for $J/\psi$, $\phi$ and $\rho$ meson production, as well as DVCS.  The dipole cross section as determined in the b-Sat model is shown at various impact parameters in Fig.~\ref{fig:dsigb}.

\subsection{Total $\gamma^* p$ cross section} \label{sec:totalx}
\begin{table}
  \centering
  \begin{tabular}{ccccc|ccc|c}
    \hline\hline
    Model & $T(b)$ & $Q^2$/GeV$^2$ & $m_{u,d,s}$/GeV & $m_c$/GeV & $\mu_0^2/\mathrm{GeV}^2$ & $A_g$ & $\lambda_g$ & $\chisq$ \\ \hline
    b-Sat & Gaussian & [0.25,650] & $0.14$ & $1.4$ & $1.17$ & $2.55$ & $0.020$ & $193.0/160=1.21$ \\
    b-Sat & Gaussian & [0.25,650] & $0.14$ & $1.35$ & $1.20$ & $2.51$ & $0.024$ & $190.2/160=1.19$ \\
b-Sat & Gaussian & [0.25,650] & $0.14$ & $1.5$ & $1.11$ & $2.64$ & $0.011$ & $198.1/160=1.24$ \\
    b-Sat & Gaussian & [0.25,650] & $0.05$ & $1.4$ & $0.77$ & $3.61$ & $-0.118$ & $144.7/160=0.90$ \\
    b-Sat & Step & [0.25,650] & $0.14$ & $1.4$ & $1.50$ & $2.20$ & 0.071 & $199.6/160=1.25$ \\ \hline\hline
  \end{tabular}
  \caption{Parameters of the initial gluon distribution \eqref{eq:inputgluon} determined from fits to $F_2$ data \cite{Breitweg:2000yn,Chekanov:2001qu}.  All predictions using the b-Sat model in this paper are evaluated with the set of parameters given in the first line unless explicitly stated otherwise.}
  \label{tab:bSat}
\end{table}
\begin{figure}
  \centering
  \includegraphics[width=0.53\textwidth,clip]{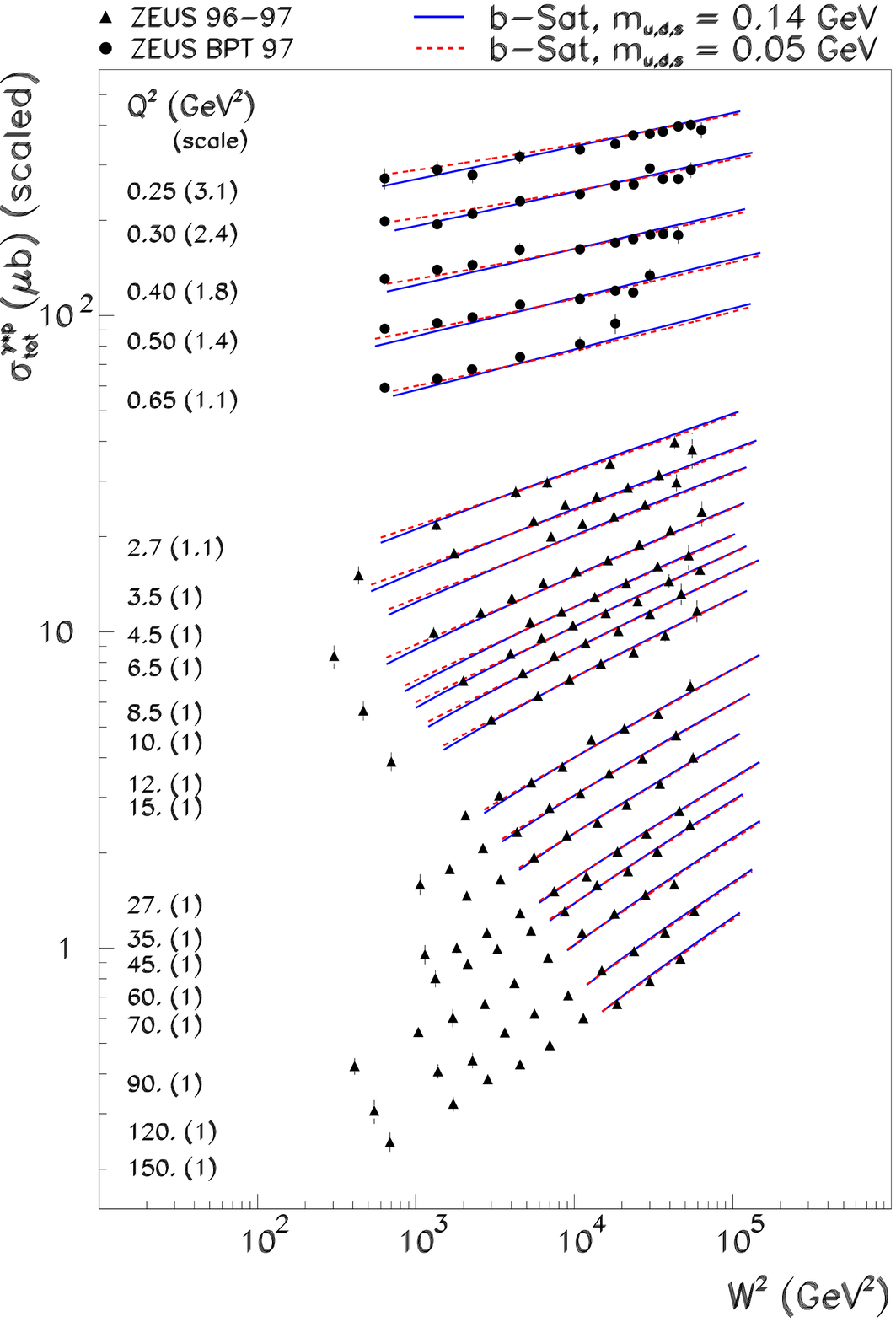}\\
  \includegraphics[width=0.53\textwidth,clip]{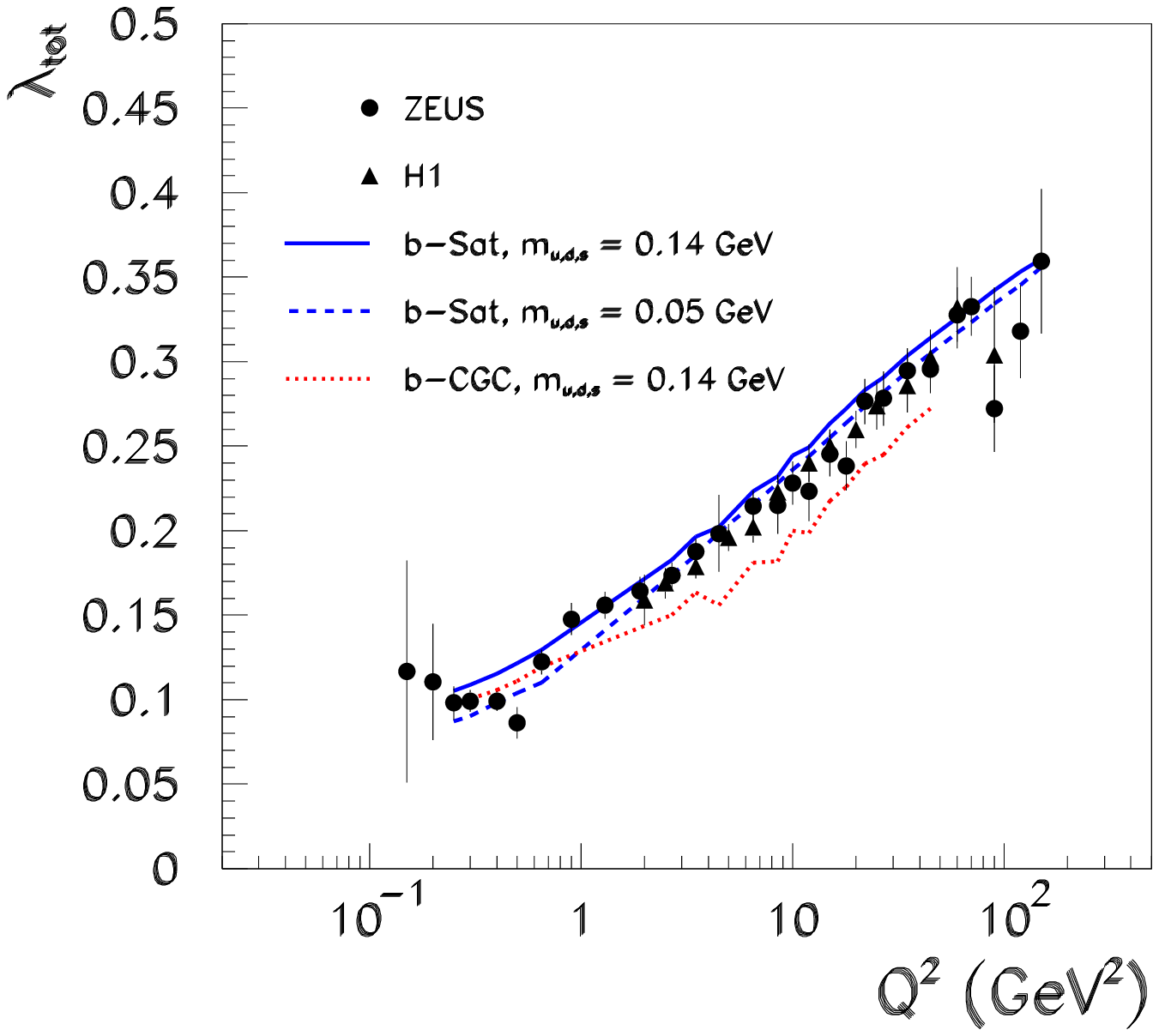}
  \caption{{\it Top:} The total DIS cross section $\sigma_{\rm tot}^{\gamma^*p}$ vs.~$W^2$ for different $Q^2$.  The data points plotted are from ZEUS \cite{Breitweg:2000yn,Chekanov:2001qu}.
    {\it Bottom:} 
    The $\lambda_{\rm tot}$ parameter for inclusive DIS defined by $\sigma^{\gamma^*p}_{\rm tot}\propto (1/x)^{\lambda_{\rm tot}}$.  The data points plotted are from ZEUS \cite{Breitweg:2000yn,Chekanov:2001qu} and H1 \cite{Adloff:2000qk}.}
  \label{fig:sigtot}
\end{figure}
The parameters in the initial gluon distribution \eqref{eq:inputgluon} are determined by fitting the ZEUS $F_2$ data \cite{Breitweg:2000yn,Chekanov:2001qu} with $\xB\le 0.01$ and $Q^2\in[0.25,650]$ GeV$^2$.  They are obtained in a quickly converging iterative procedure in which the $F_2$ data are fitted alternately with the $t$-distributions of the vector meson data (see Sect.~\ref{sec:tdistr}) which determine the parameter $B_G=4$ GeV$^{-2}$.  As well as our main fit with $m_{u,d,s}=0.14$ GeV and $m_c=1.4$ GeV, shown in the first line of Table \ref{tab:bSat}, we also make alternative fits with different quark masses.  As in Ref.~\cite{Kowalski:2003hm}, the best fit to $F_2$ is obtained with very low light quark masses, $m_{u,d,s}=0.05$ GeV.  The quark mass of 0.14 GeV, which is more appropriate as a cut-off mass for vector meson bound states, gives a fit to $F_2$ of still acceptable quality; see Table~\ref{tab:bSat}.  The last line of Table~\ref{tab:bSat} shows also the fit results performed with the step-like proton shape defined by \eqref{eq:StepTb} with the parameter $b_S=4$ GeV$^{-1}$, which we discuss further in Sect.~\ref{sec:Sat}.

To compare with the fits obtained by global analysis using the NLO DGLAP formalism, we evaluated the $\chi^2$ for a subset of the ZEUS $F_2$ data \cite{Chekanov:2001qu} with $\xB\le 0.01$ and $Q^2\ge 2$ GeV$^2$ comprising of 116 data points.  The main b-Sat fit shown in the first line of Table \ref{tab:bSat} gave a $\chi^2$ of 114, while the most recent NLO DGLAP fit by the MRST group \cite{Thorne:2006zu} gave a $\chi^2$ of 96 for the 116 data points.

In Fig.~\ref{fig:sigtot} we show the comparison of the main b-Sat fit results with measurements of the total DIS cross section $\sigma_{\rm tot}^{\gamma^*p}$.  In the same figure we also show the comparison for the rate of rise of the total DIS cross section, $\lambda_{\rm tot}$, defined by $\sigma^{\gamma^*p}_{\rm tot}\propto (1/x)^{\lambda_{\rm tot}}$.  Both comparisons show a very good agreement between data and the b-Sat model results.

\begin{figure}
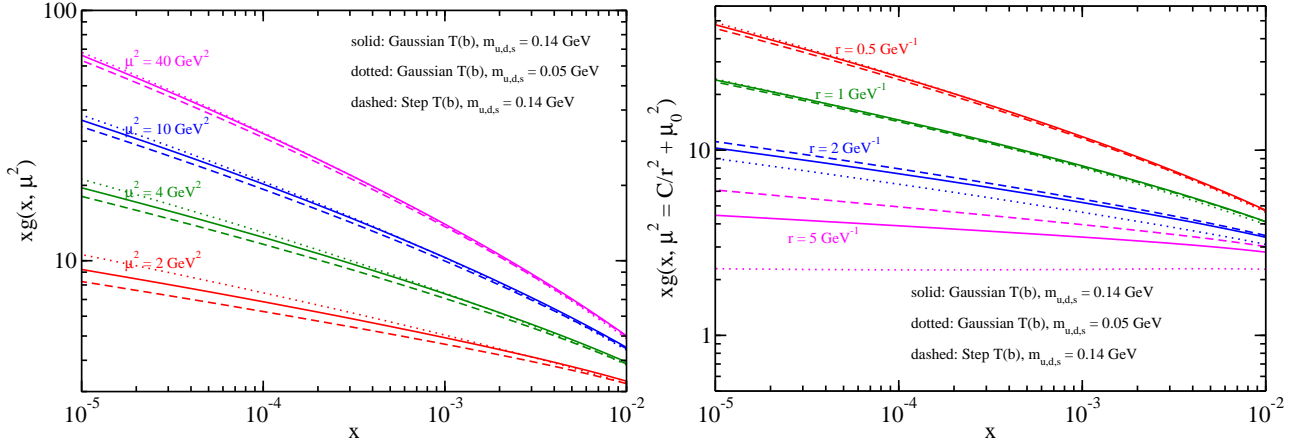

  \centering
  \includegraphics[width=0.5\textwidth,clip]{xgluq.eps}%
  \includegraphics[width=0.5\textwidth,clip]{xglur.eps}
  \caption{The gluon distribution $xg(x,\mu^2)$ for different fixed $\mu^2$ (left) or fixed $r$ (right).}
  \label{fig:gluon}
\end{figure}

Let us make some general remarks about the sensitivity of the fit to the assumed quark masses and the proton shape.  Table~\ref{tab:bSat} shows that the variation of the charm quark mass does not sizably change the fit parameters.  On the other hand, the choice of the light quark mass influences the value of the $\lambda_g$ parameter and consequently the evolution of the gluon density.  In Fig.~\ref{fig:gluon} we show the gluon distribution for different scales $\mu^2$ or dipole sizes $r$.
The correlation between the assumed value of the light quark mass and the $\lambda_g$ and $\mu_0^2$ parameters was investigated in detail in Ref.~\cite{Kowalski:2003hm}.  Consequently, in the b-Sat model the description of the change of the parameter $\lambda_{\rm tot}$ with $Q^2$ is mainly due to evolution effects and not to saturation effects as in, for example, the GBW model \cite{Golec-Biernat:1998js,Golec-Biernat:1999qd}.

\subsection{Vector meson total cross sections}
We now compare our predictions for exclusive vector meson production with recent published HERA data for $J/\psi$ \cite{Chekanov:2002xi,Chekanov:2004mw,Aktas:2005xu}, $\phi$ \cite{Chekanov:2005cq} and $\rho$ \cite{Adloff:1999kg} meson production.\footnote{The ZEUS $\gamma^*p$ cross sections \cite{Chekanov:2002xi,Chekanov:2004mw,Chekanov:2005cq} are given as $\sigma=\sigma_T+\sigma_L$, while H1 \cite{Aktas:2005xu,Adloff:1999kg} give $\sigma=\sigma_T+\varepsilon\sigma_L$, where $\varepsilon=(1-y)/(1-y+y^2/2)$ and $\langle\varepsilon\rangle\approx 0.99$.  We use the ZEUS definition in our calculations.}  The H1 $J/\psi$ cross sections \cite{Aktas:2005xu} are measured in the range $|t|<1.2$ GeV$^2$ while ZEUS measure $|t|<1$ GeV$^2$ for electroproduction \cite{Chekanov:2004mw} and $|t|<1.8$ GeV$^2$ ($J/\psi\to\mu^+\mu^-$) or $|t|<1.25$ GeV$^2$ ($J/\psi\to e^+e^-$) for photoproduction \cite{Chekanov:2002xi}.  The ZEUS $\phi$ data \cite{Chekanov:2005cq} have $|t|<0.6$ GeV$^2$, while the H1 $\rho$ data \cite{Adloff:1999kg} have $|t|<0.5$ GeV$^2$.

\begin{figure}
  \centering
  \includegraphics[width=0.33\textwidth]{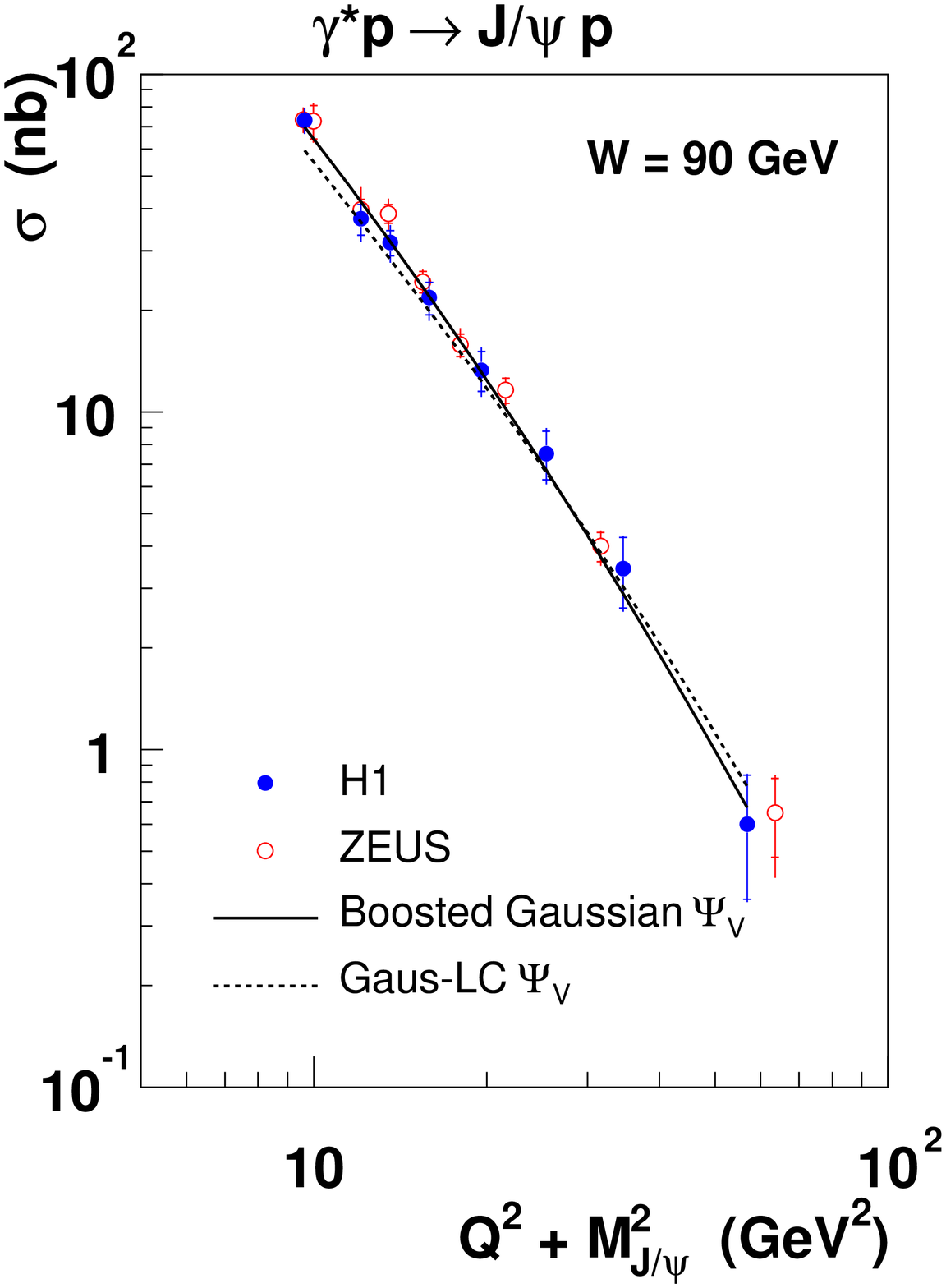}%
  \includegraphics[width=0.33\textwidth]{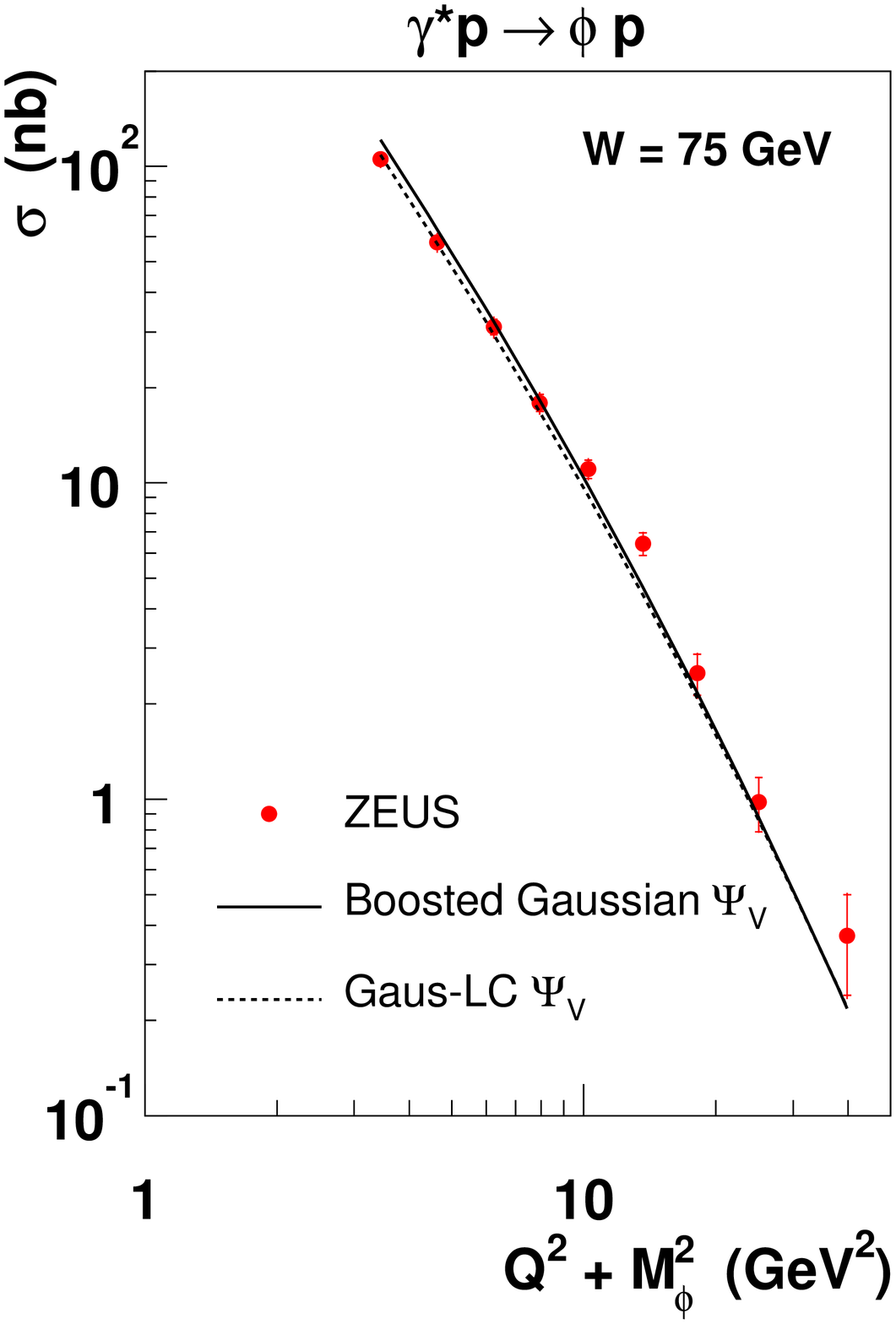}%
  \includegraphics[width=0.33\textwidth]{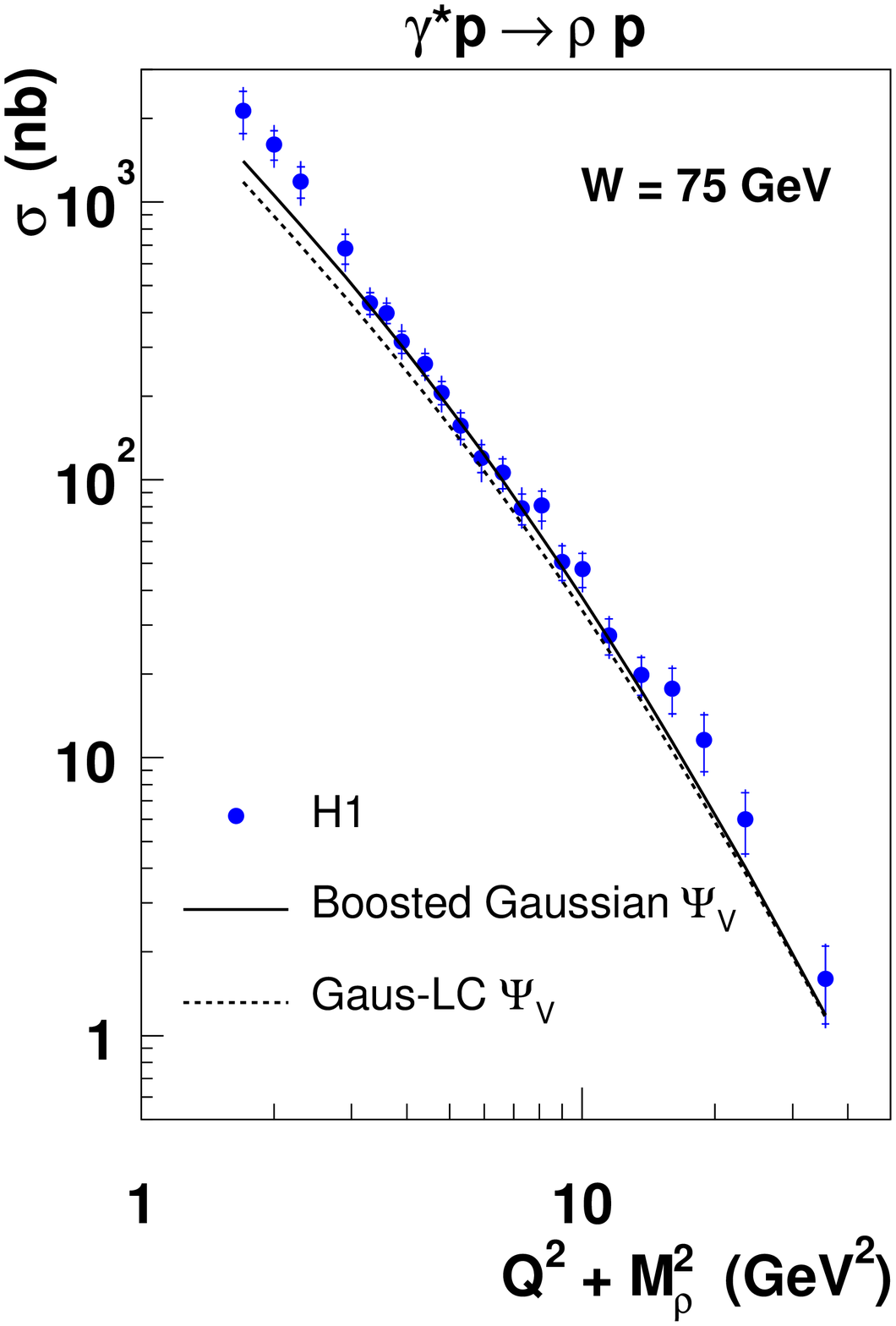}
  \caption{Total vector meson cross section $\sigma$ vs.~$(Q^2+M_V^2)$ compared to predictions from the b-Sat model using two different vector meson wave functions.  The ZEUS $J/\psi$ photoproduction point is taken from Table 1 of Ref.~\cite{Chekanov:2002xi}, from the muon decay channel with $W = 90$--$110$ GeV.}
  \label{fig:crossq}
\end{figure}
In Fig.~\ref{fig:crossq} we show the $(Q^2+M_V^2)$ dependence of the total cross section $\sigma$ for all three vector mesons at a fixed value of $W$.  The inner error bars indicate the statistical uncertainties only, while the outer error bars include the systematic uncertainties added in quadrature.  The predictions are given integrated over the appropriate $t$ range.  For the $J/\psi$ data, the predictions shown correspond to the H1 $t$ range.  The predictions of the model are in good agreement with data for both vector meson wave functions. The model reproduces the $Q^2$ dependence as well as the absolute magnitude of the data. The prediction for the absolute normalisation is determined mainly by the gluon density obtained from the fit to the total DIS cross section (or $F_2$) and the shapes of the ``Gaus-LC'' and ``boosted Gaussian'' wave functions, discussed in Sect.~\ref{sec:wavefunction2}.  Although these two vector meson wave functions are quite different, they lead to similar predictions using the constraints from the normalisation and vector meson decay width conditions given in \eqref{eq:nnz_normt}, \eqref{eq:nnz_norml}, \eqref{eq:nnz_fvt} and \eqref{eq:nnz_fvl}.  Note that, unlike the MRT calculations \cite{Martin:1999wb} compared to the H1 $J/\psi$ data in \cite{Aktas:2005xu}, we do \emph{not} require an additional normalisation factor $\sim2$ to achieve agreement with the data.  Note also that the MRT calculations \cite{Martin:1999wb}, based on $k_t$-factorisation using an unintegrated gluon distribution, take as input the gluon density determined from the global analyses using collinear factorisation.  There is no \emph{a priori} reason why the fitted parameters in the two gluon distributions determined in these two calculational frameworks should be identical.  The dipole approach is self-consistent in that the gluon density is determined from the inclusive process and applied to exclusive processes within the \emph{same} calculational framework.

\begin{figure}
  \centering
  \includegraphics[width=0.33\textwidth]{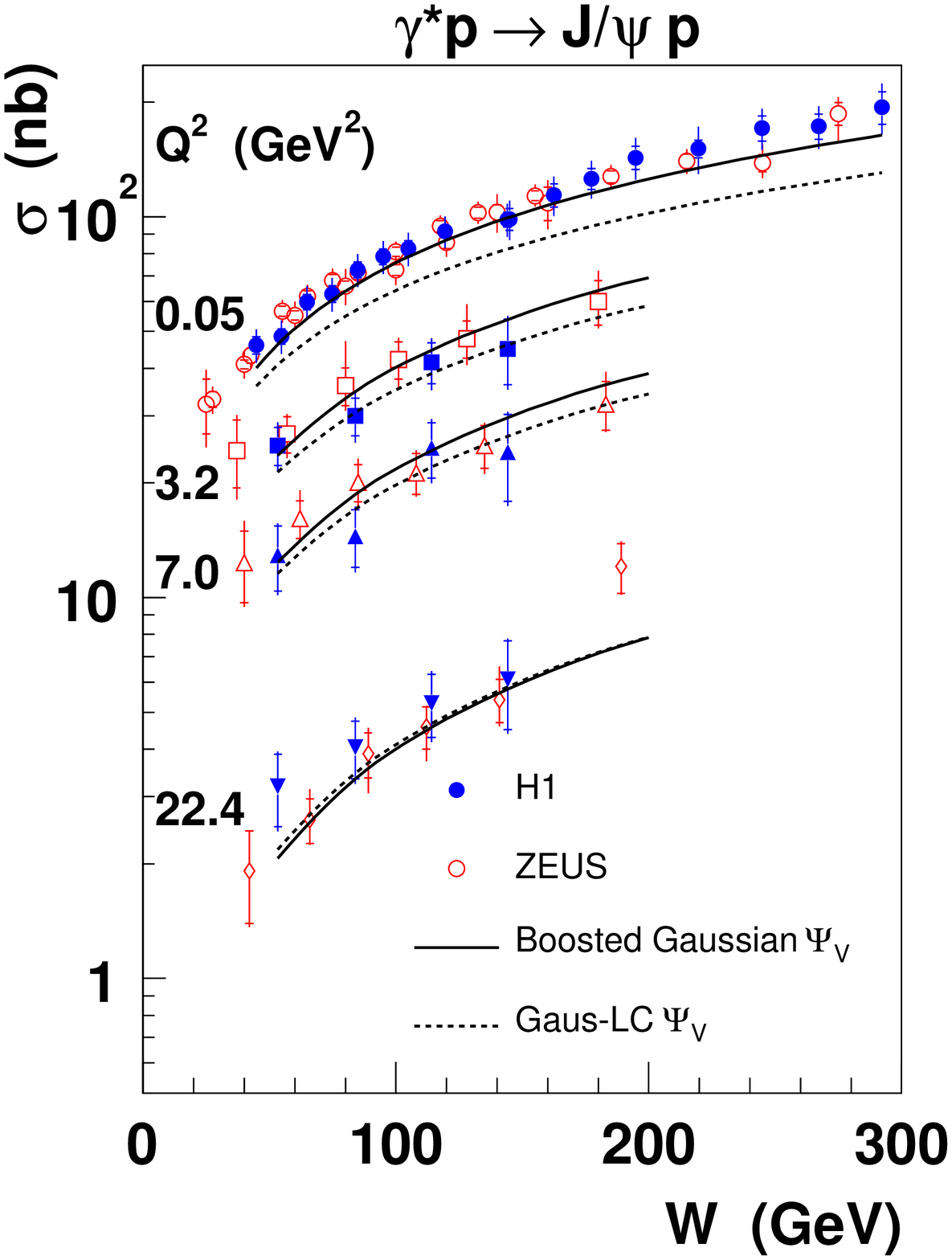}%
  \includegraphics[width=0.33\textwidth]{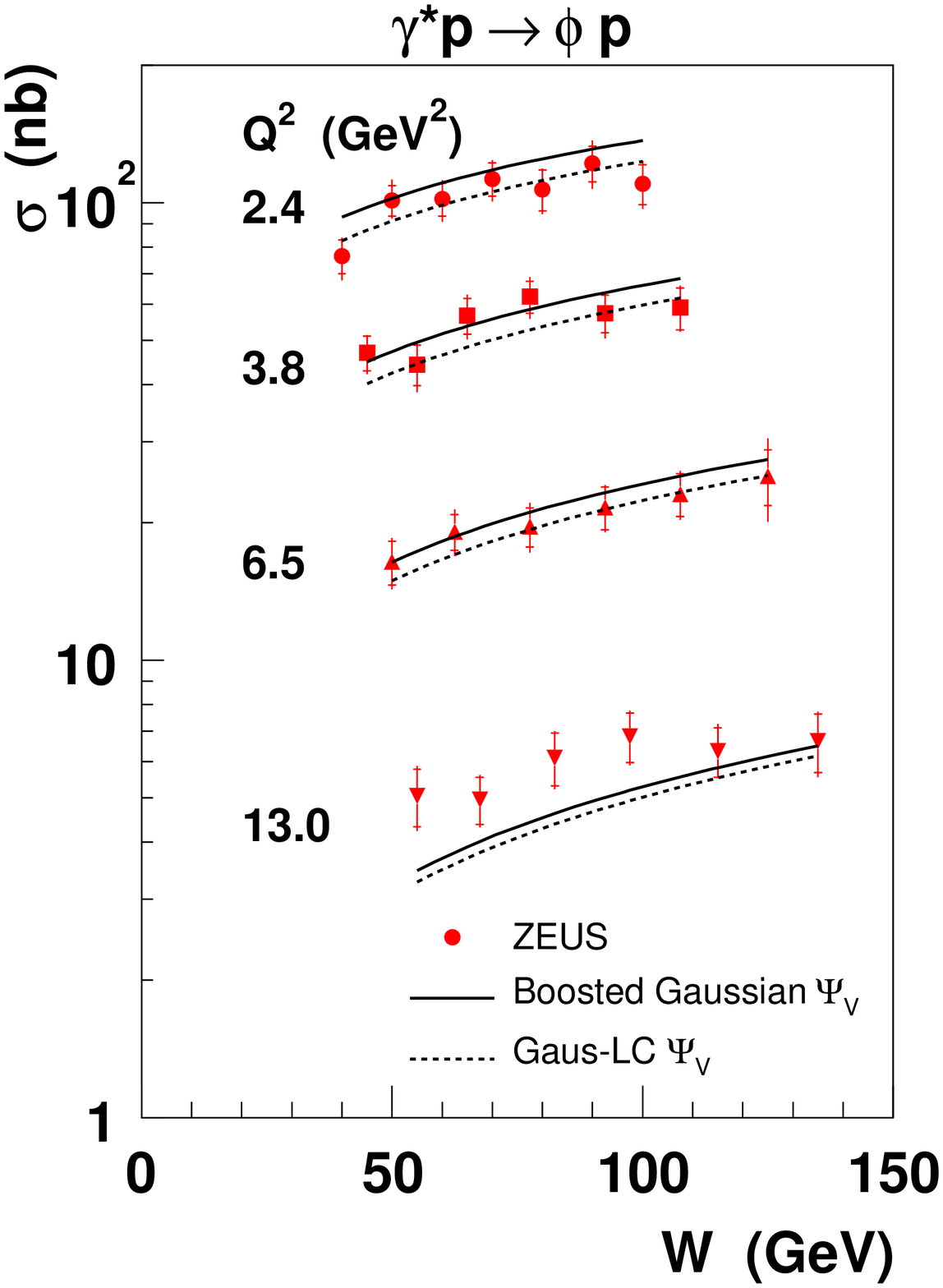}%
  \includegraphics[width=0.33\textwidth]{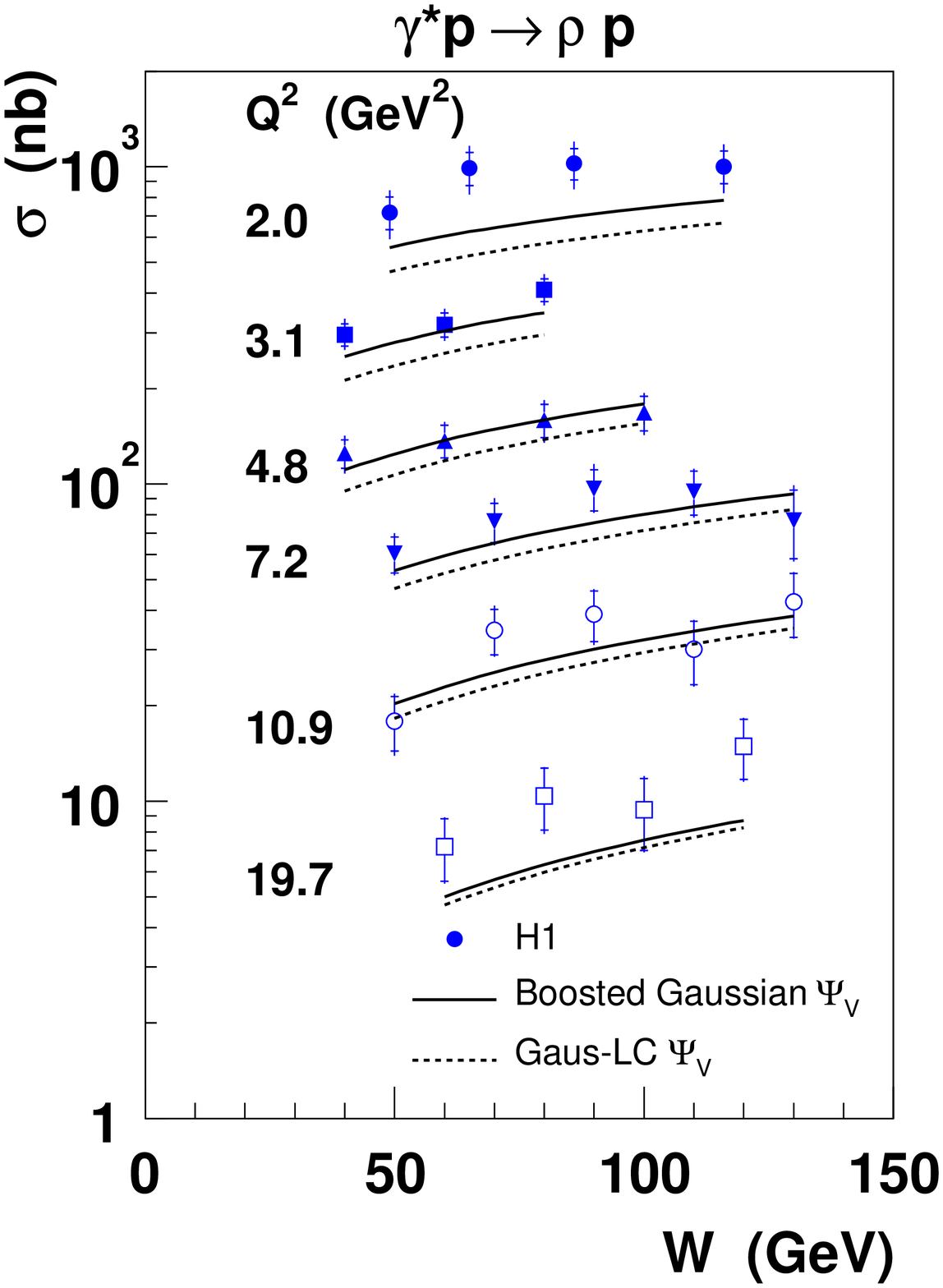}
  \caption{Total vector meson cross section $\sigma$ vs.~$W$ compared to predictions from the b-Sat model using two different vector meson wave functions.  The ZEUS $J/\psi$ data points \cite{Chekanov:2002xi,Chekanov:2004mw} have been scaled to the H1 $Q^2$ values \cite{Aktas:2005xu} using the $Q^2$ dependence measured by ZEUS of the form $\sigma\propto (Q^2+M_V^2)^{-2.44}$ \cite{Chekanov:2004mw}.}
  \label{fig:crossw}
\end{figure}
In Fig.~\ref{fig:crossw} we show the $W$ dependence of the total cross section $\sigma$ for fixed values of $Q^2$.  Here, the  ``boosted Gaussian'' vector meson wave function gives a slightly better description of the data.
\begin{figure}
  \centering
  \includegraphics[width=0.33\textwidth]{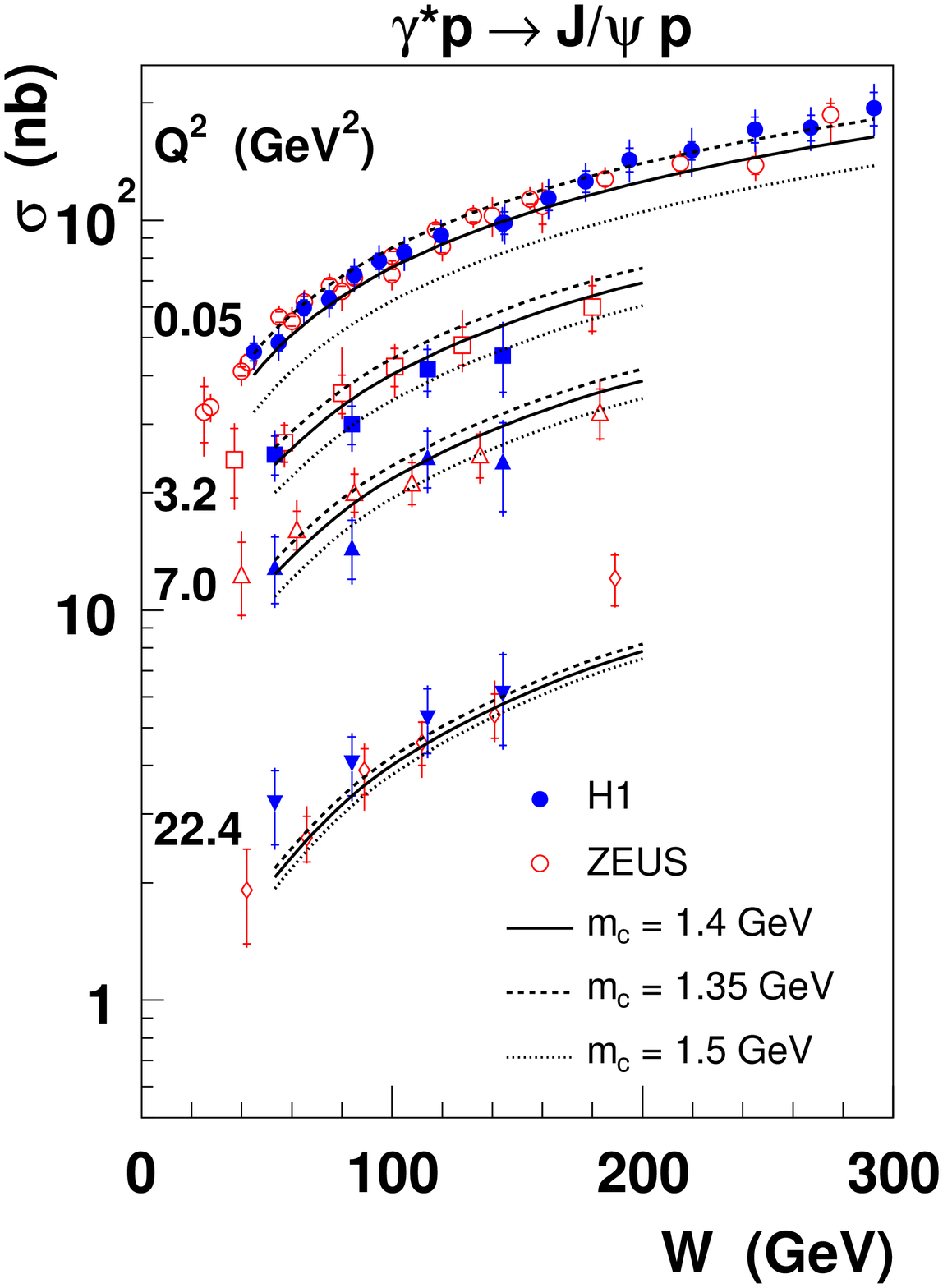}%
  \includegraphics[width=0.33\textwidth]{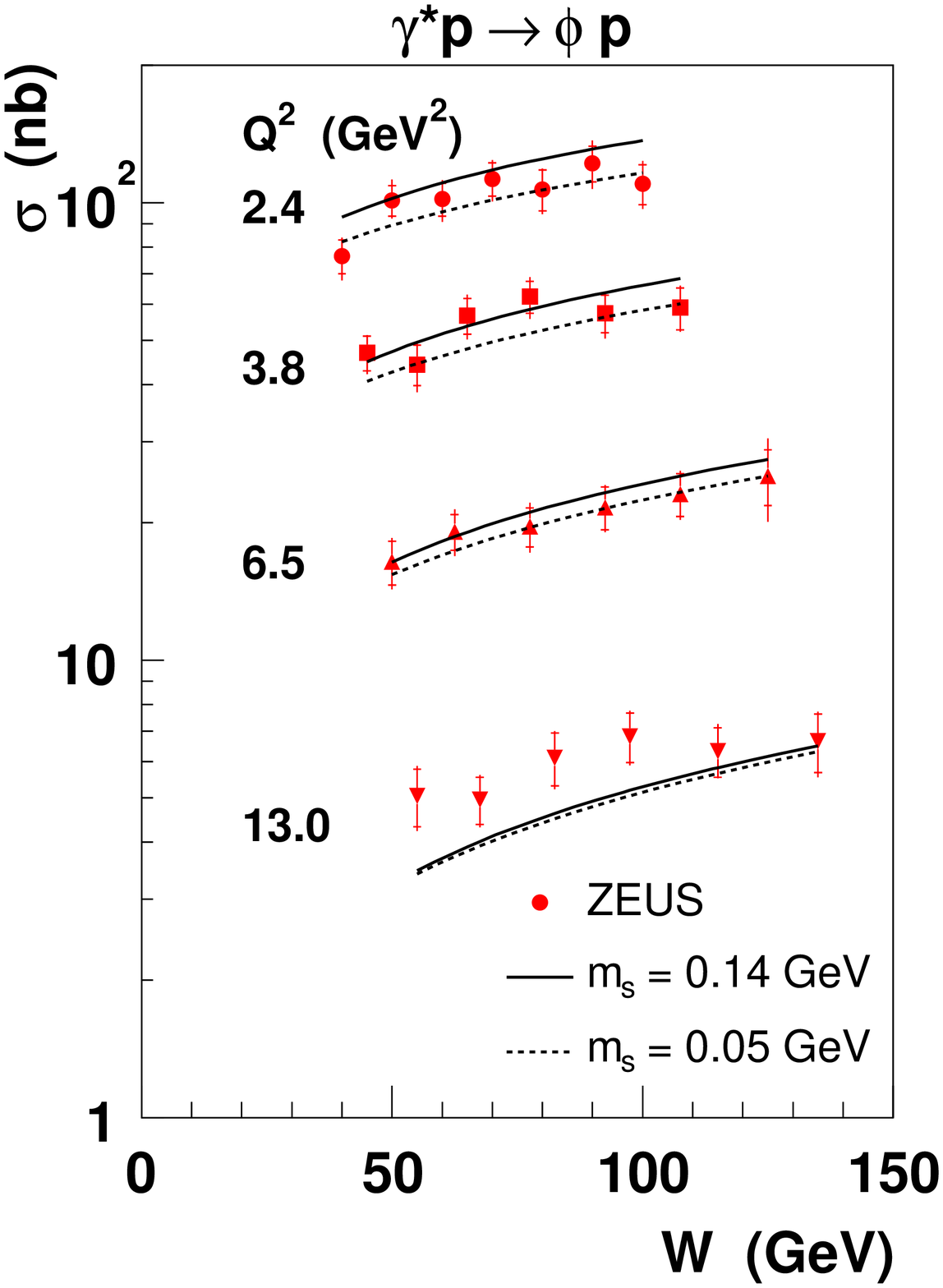}%
  \includegraphics[width=0.33\textwidth]{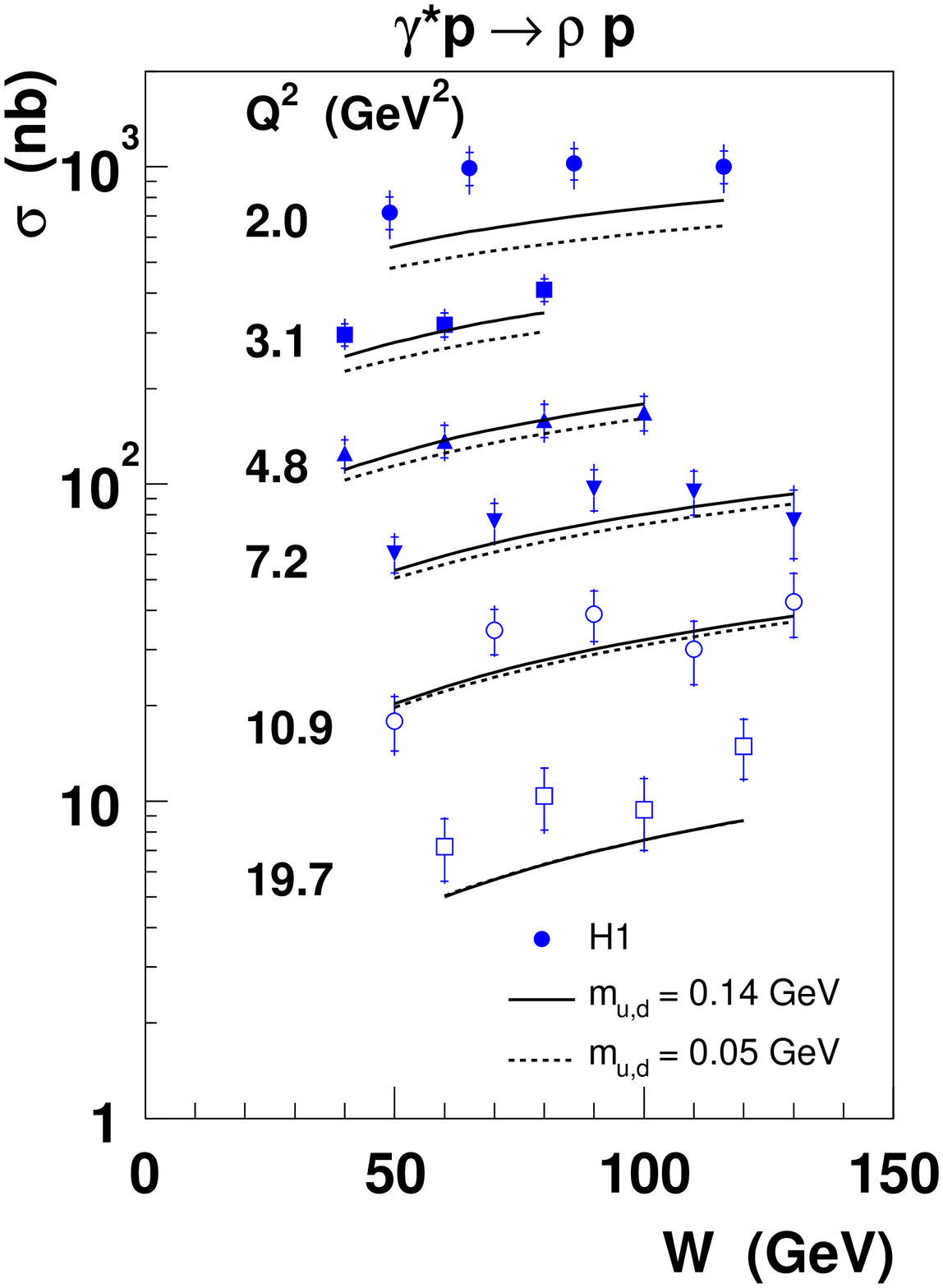}
  \caption{Total vector meson cross section $\sigma$ vs.~$W$ compared to predictions from the b-Sat model using the ``boosted Gaussian'' vector meson wave function for different quark masses.}
  \label{fig:mqdepcros}
\end{figure}
In Fig.~\ref{fig:mqdepcros} we show the effect of changing the charm quark mass from the default value of 1.4 GeV to 1.35 GeV or 1.5 GeV.  We also show the effect of changing the light quark masses from 0.14 GeV to 0.05 GeV.  In each case, we refit the $F_2$ data to determine the gluon distribution with parameters given in Table~\ref{tab:bSat}.  The absolute magnitude of the $J/\psi$ cross sections is strongly dependent on the choice of the charm quark mass, particularly at small $Q^2$ values.  The cross sections for the $\phi$ and $\rho$ vector mesons are only weakly dependent on the choice of the light quark masses.  This is because, in the $Q^2$ range considered in this paper, the scale for light vector meson production, given by $\epsilon^2=z(1-z)Q^2+m_f^2$, is predominantly given by $Q^2$ whereas for $J/\psi$ mesons the scale $\epsilon^2$ is dominated by the square of the charm quark mass.  Note also that for all vector mesons the sensitivity of the cross section to the quark mass decreases with increasing $Q^2$.

\begin{figure}
  \centering
  \includegraphics[width=0.33\textwidth]{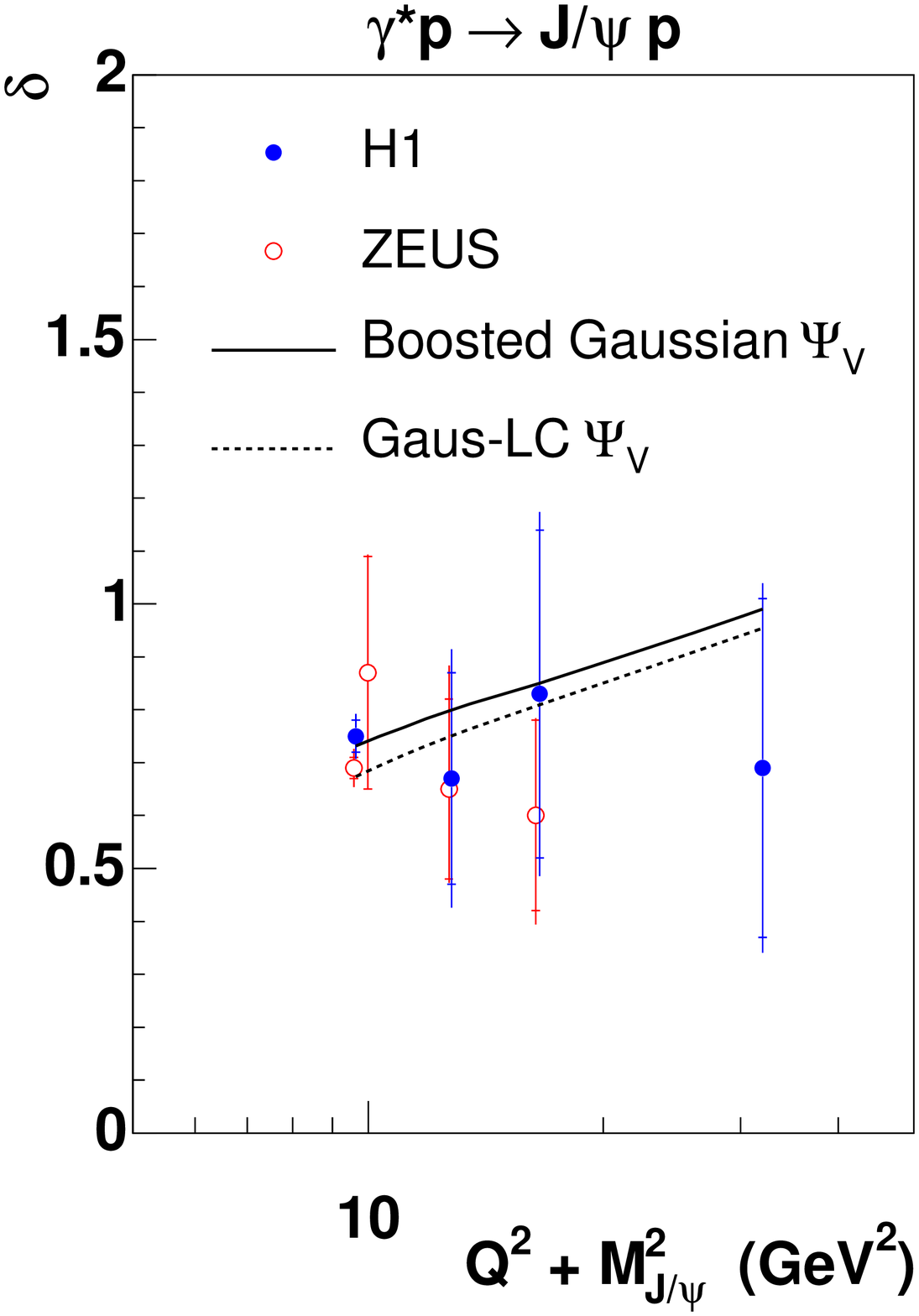}%
  \includegraphics[width=0.33\textwidth]{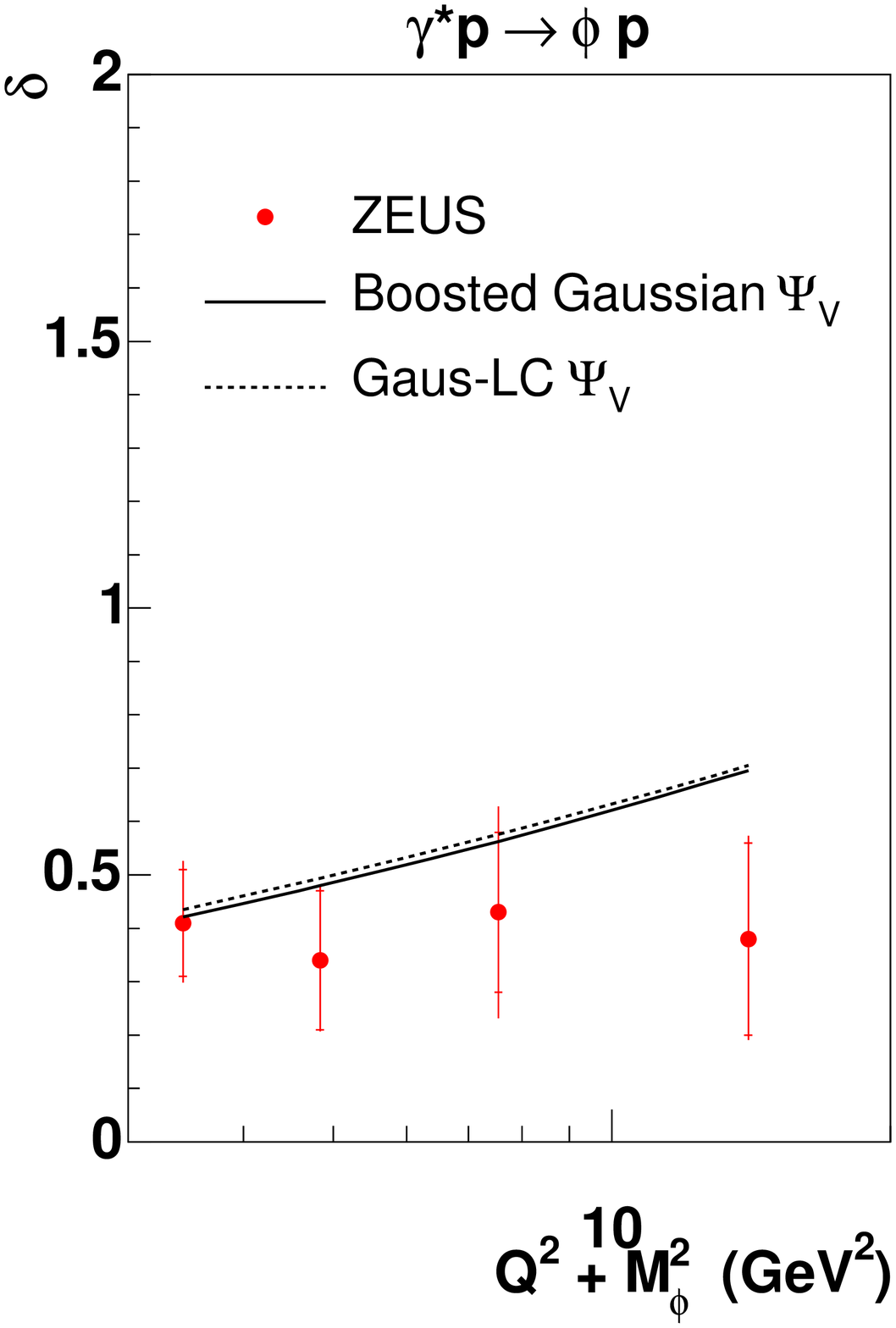}%
  \includegraphics[width=0.33\textwidth]{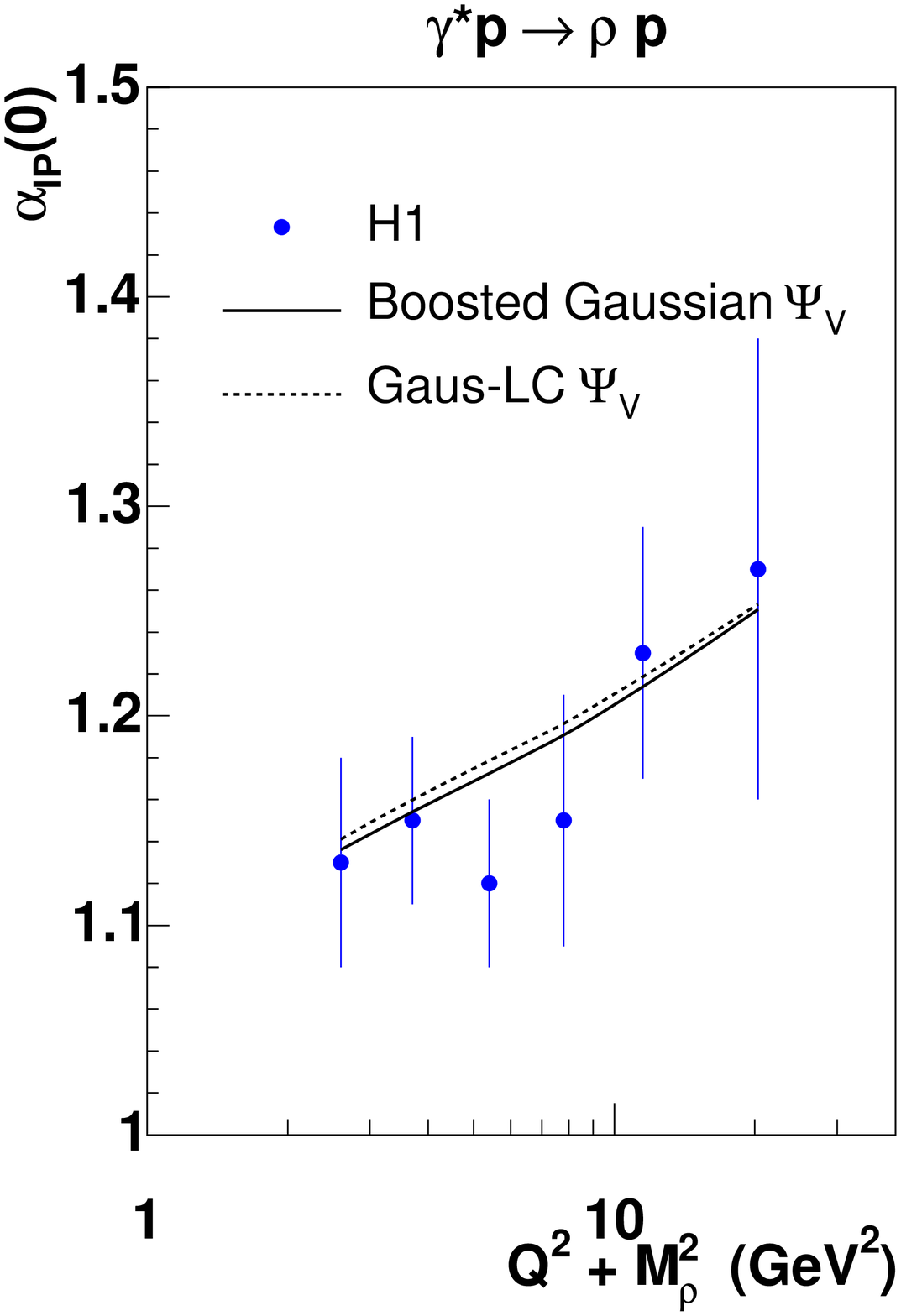}
  \caption{The power $\delta$ vs.~$(Q^2+M_V^2)$, where $\delta$ is defined by fitting $\sigma\propto W^\delta$, compared to predictions from the b-Sat model using two different vector meson wave functions.  For $\rho$ mesons, we instead show $\alpha_\Pom(0)$ defined in the main text; the error bars represent the statistical and non-correlated systematic uncertainties only \cite{Adloff:1999kg}.}
  \label{fig:delta}
\end{figure}
We then perform a fit to the theory predictions shown in Fig.~\ref{fig:crossw} of the form $\sigma\propto W^\delta$ and compare the values of $\delta$ obtained to the experimental values; see Fig.~\ref{fig:delta}.  For $\rho$ production, we instead show $\alpha_\Pom(0)$ calculated from $\delta = 4[\alpha_\Pom(\langle t\rangle)-1]$, where $\alpha_\Pom(\langle t\rangle) = \alpha_\Pom(0) + \alpha_\Pom^\prime \langle t\rangle$, $\langle t\rangle = -1/B_D$, $B_D$ is the theoretical prediction (see Fig.~\ref{fig:bd}), and $\alpha_\Pom^\prime=0.25$ GeV$^{-2}$.  We observe again a reasonable agreement of the model results with data.

\begin{figure}
 \centering
  \includegraphics[width=0.33\textwidth]{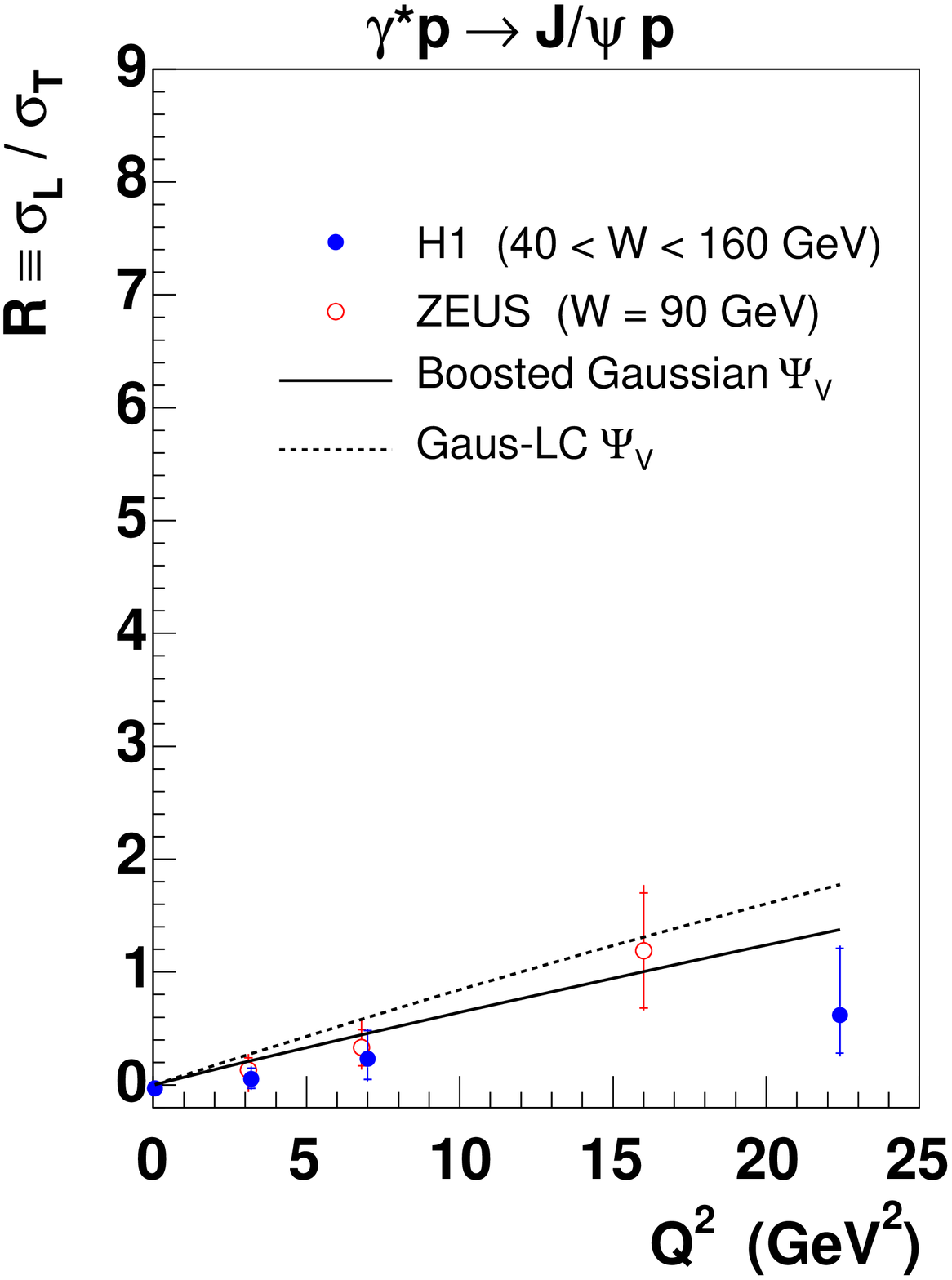}%
  \includegraphics[width=0.33\textwidth]{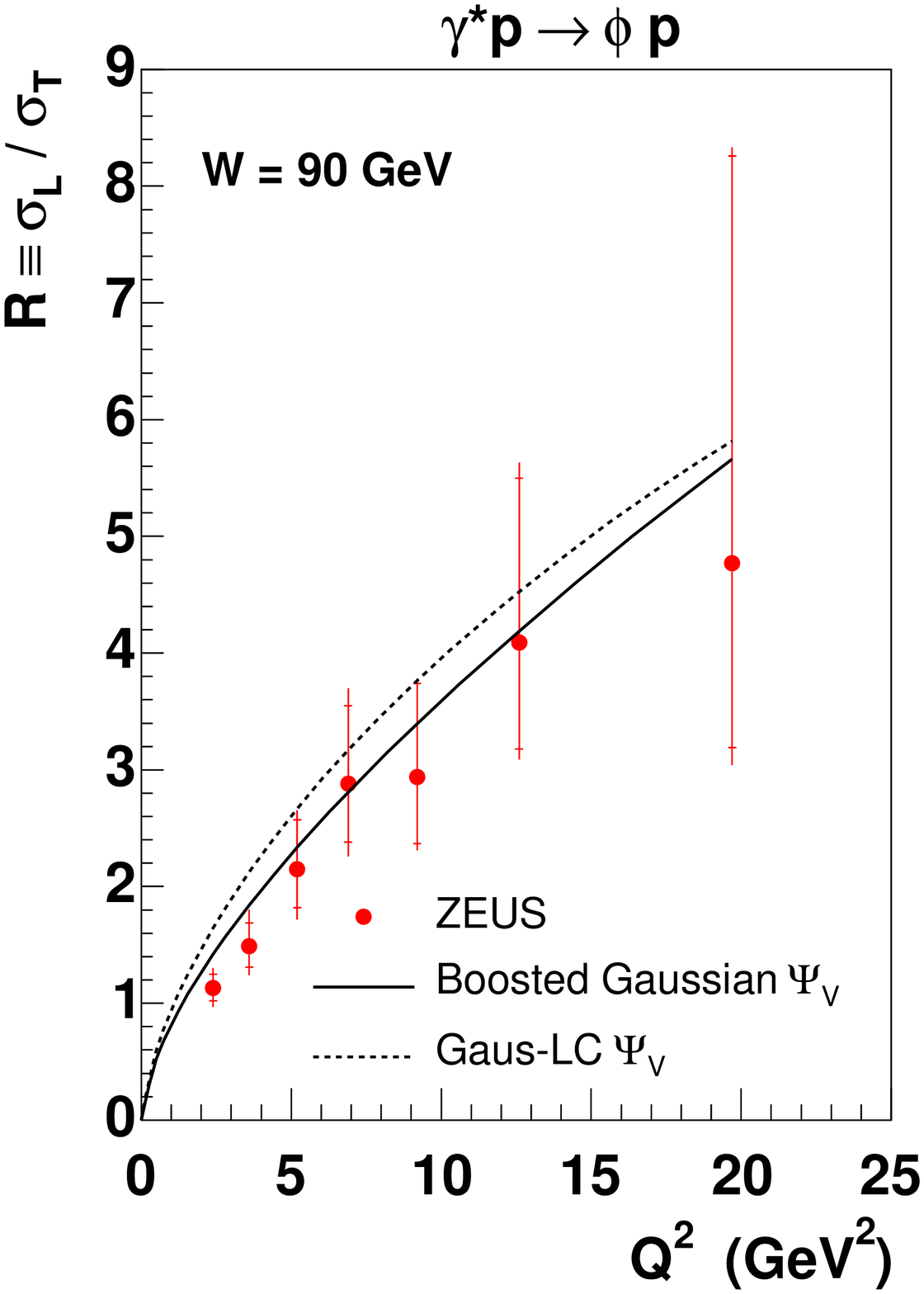}%
  \includegraphics[width=0.33\textwidth]{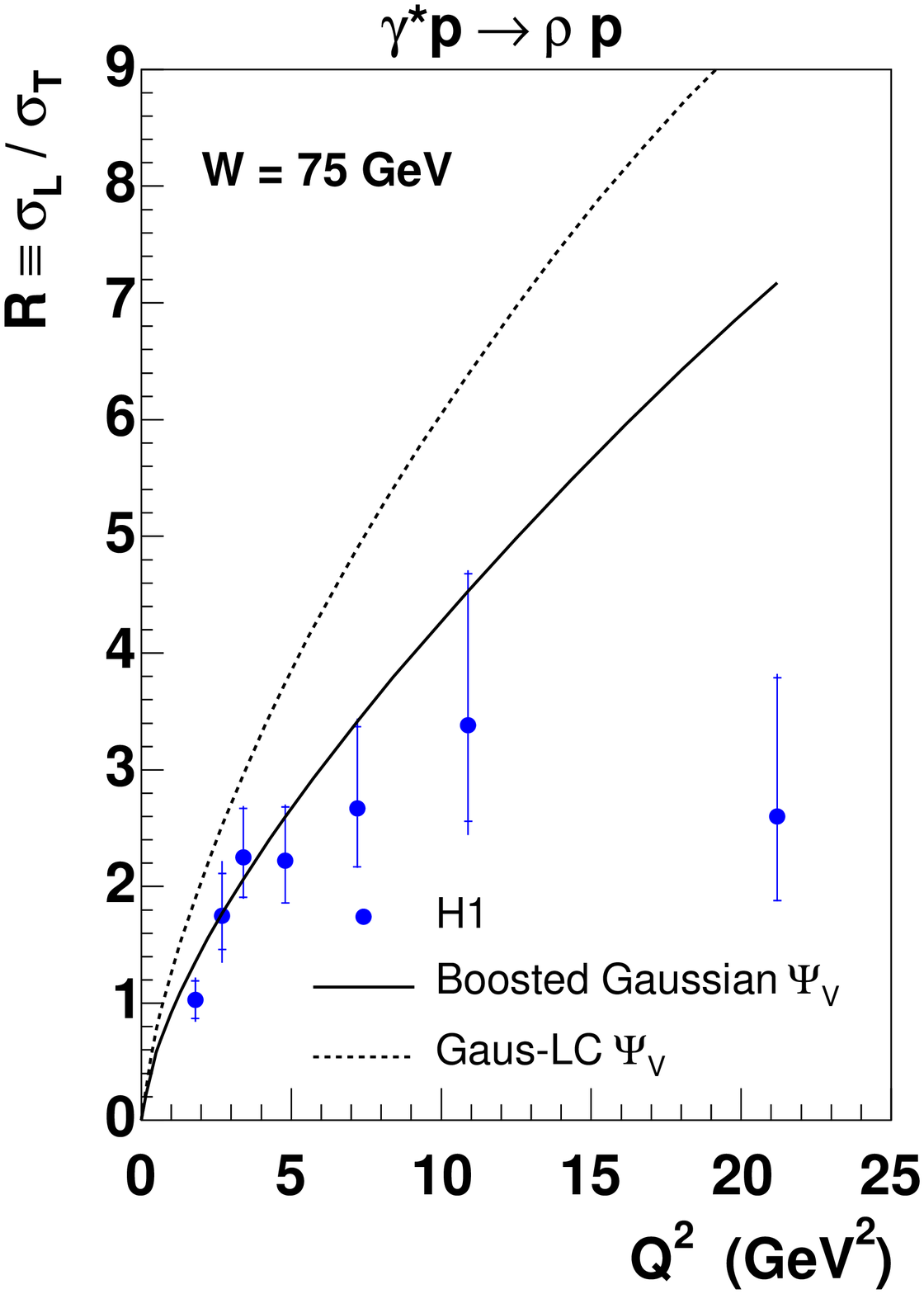}
  \caption{The ratio $R\equiv\sigma_L/\sigma_T$ vs.~$Q^2$ compared to predictions from the b-Sat model using two different vector meson wave functions.}
  \label{fig:r}
\end{figure}
A variable which is more sensitive to the details of the wave function is the ratio of the longitudinal to the transverse cross sections, $R\equiv \sigma_L/\sigma_T$, shown in Fig.~\ref{fig:r}.  This is due to the fact that the ratio $\sigma_L/\sigma_T$ probes the behaviour of the transversely polarised vector meson wave function close to the end-points ($z\to 0,1$).  At large values of $Q^2$, the contributions from the intermediate values of $z\simeq 1/2$ follow the simple, perturbative scaling that leads to $\sigma_L/\sigma_T \,\sim\, Q^2$.  This simple scaling is affected by the $Q^2$ evolution of the anomalous dimension of the gluon distribution \cite{Martin:1996bp,Martin:1999wb}, and by the contributions from the end-points to the transverse cross section, which are different for the ``Gaus-LC'' and ``boosted Gaussian'' vector meson wave functions.  Fig.~\ref{fig:r} shows that the ``boosted Gaussian'' wave function is favoured by the $\rho$ meson data, where the ``Gaus-LC'' wave function leads to a value of $\sigma_L/\sigma_T$ which rises too rapidly with increasing $Q^2$.  For $J/\psi$ and $\phi$ mesons, both vector meson wave functions lead to a similar behaviour.
\begin{figure}
  \centering
  \includegraphics[width=0.33\textwidth]{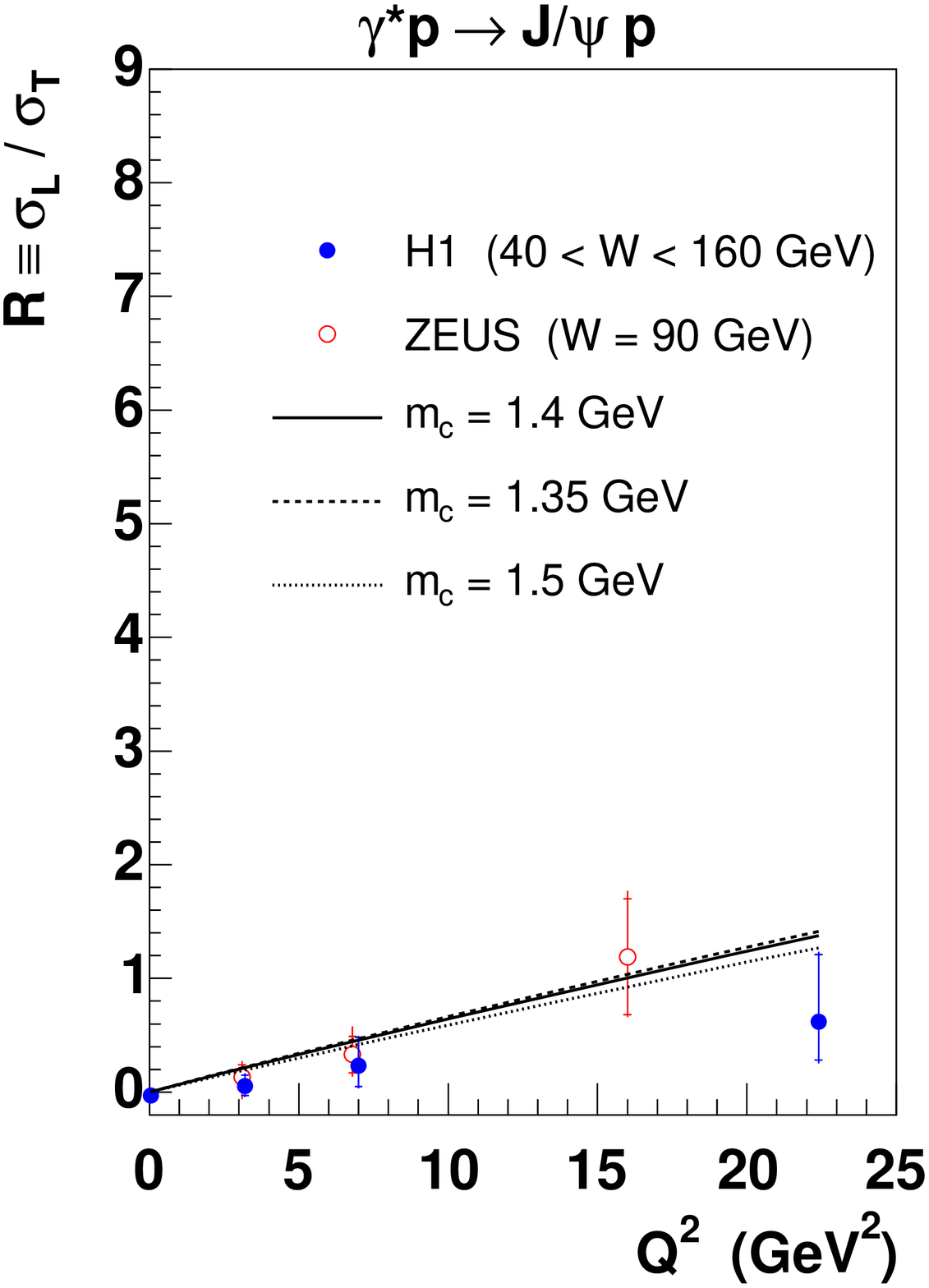}%
  \includegraphics[width=0.33\textwidth]{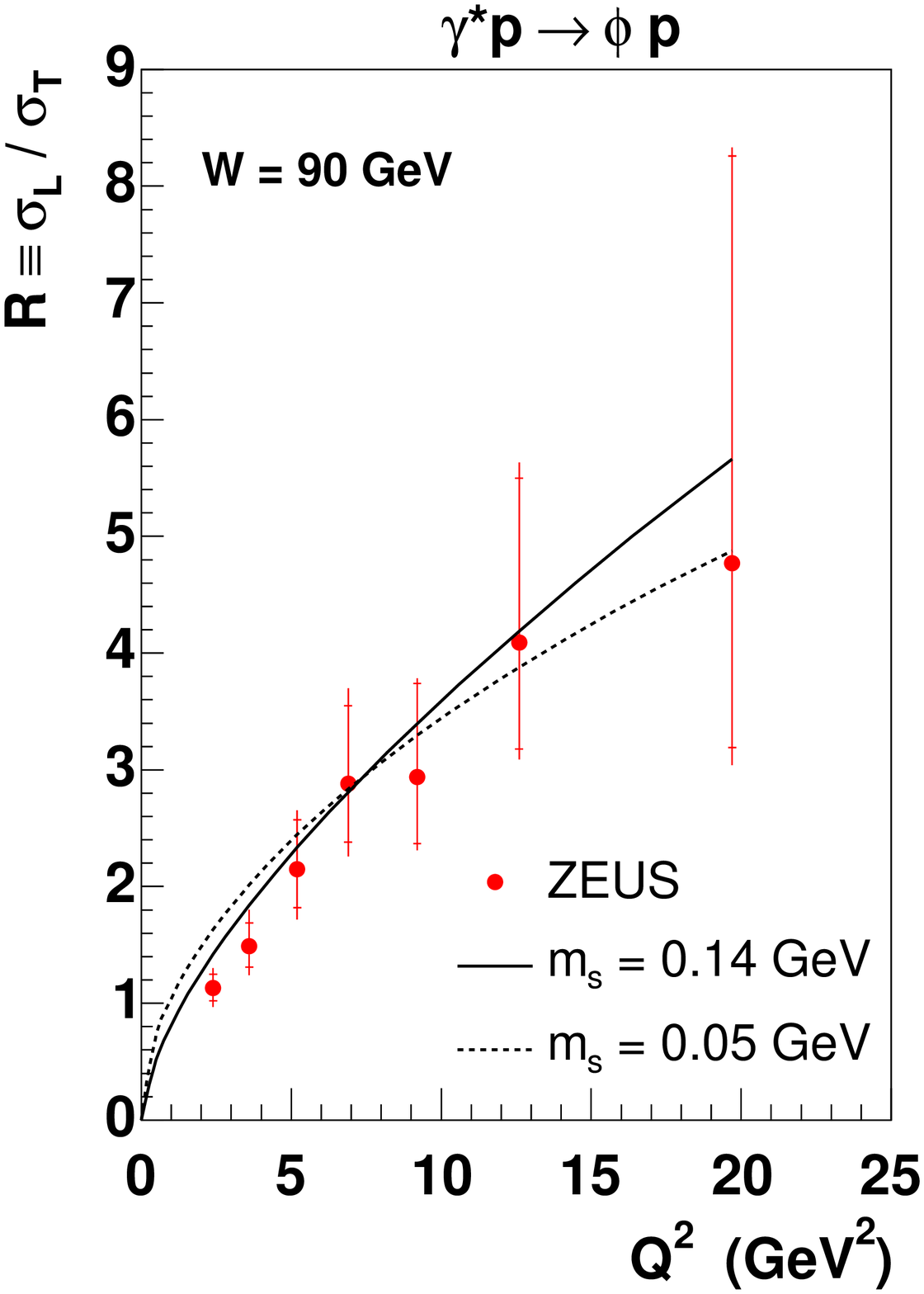}%
  \includegraphics[width=0.33\textwidth]{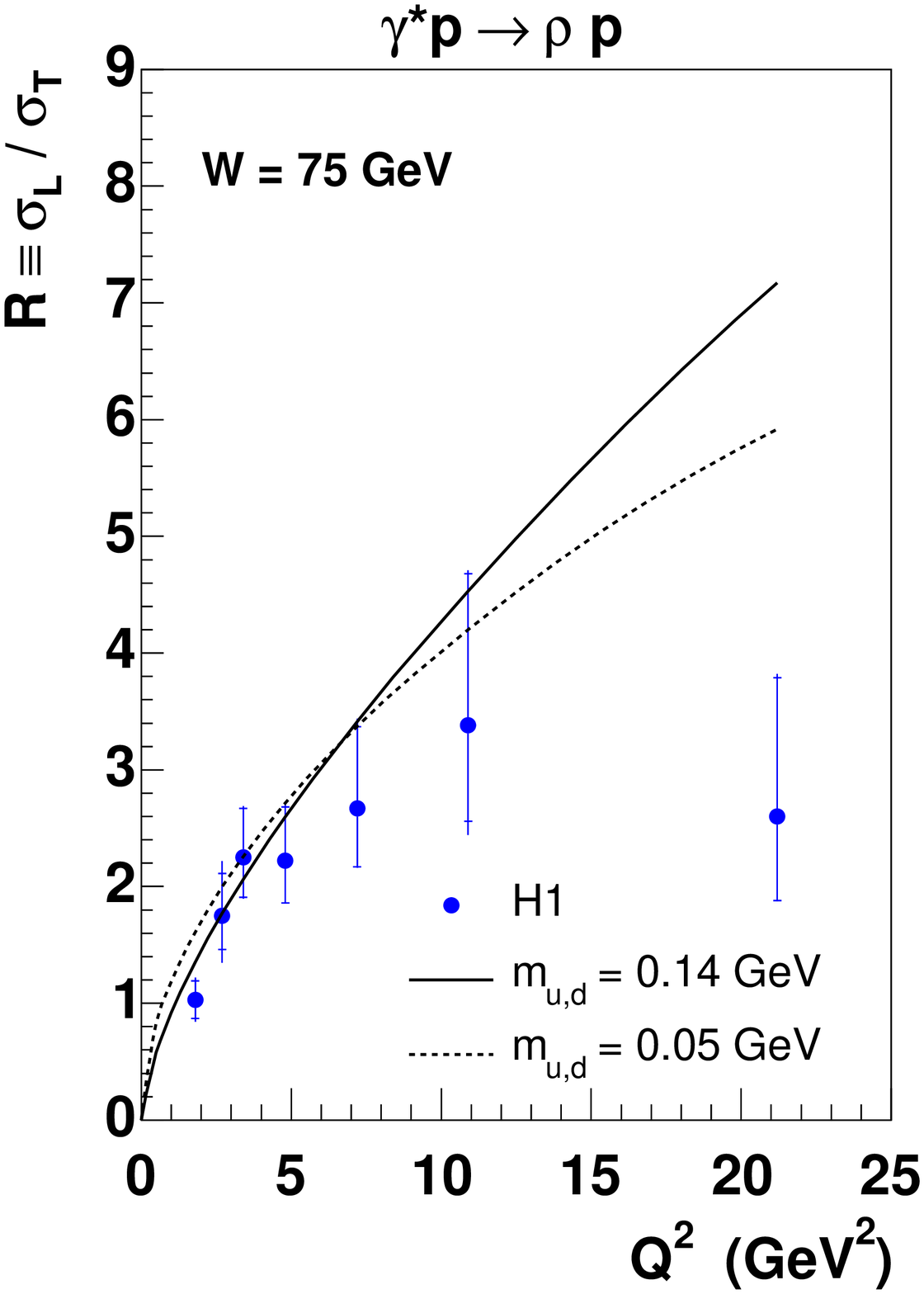}\\
  \caption{The ratio $R\equiv\sigma_L/\sigma_T$ vs.~$Q^2$ compared to predictions from the b-Sat model using the ``boosted Gaussian'' vector meson wave function for different quark masses.}
  \label{fig:mqdeprat}
\end{figure}
In Fig.~\ref{fig:mqdeprat} we show the effect of changing the quark masses when using the ``boosted Gaussian'' wave function.  For $\rho$ mesons, the ratio $\sigma_L/\sigma_T$ shows a strong dependence on the quark mass.  A more precise analysis, which goes beyond the scope of this paper, shows that the ratio $\sigma_L/\sigma_T$ is very sensitive to the behaviour of the wave functions at the end-points ($z\to 0,1$).

\subsection{Vector meson $t$-distributions} \label{sec:tdistr}
The observed $t$-distributions of the vector meson processes are an important source of information on the shape of the proton in the low-$x$ region.
\begin{figure}
  \centering
  \includegraphics[width=0.33\textwidth]{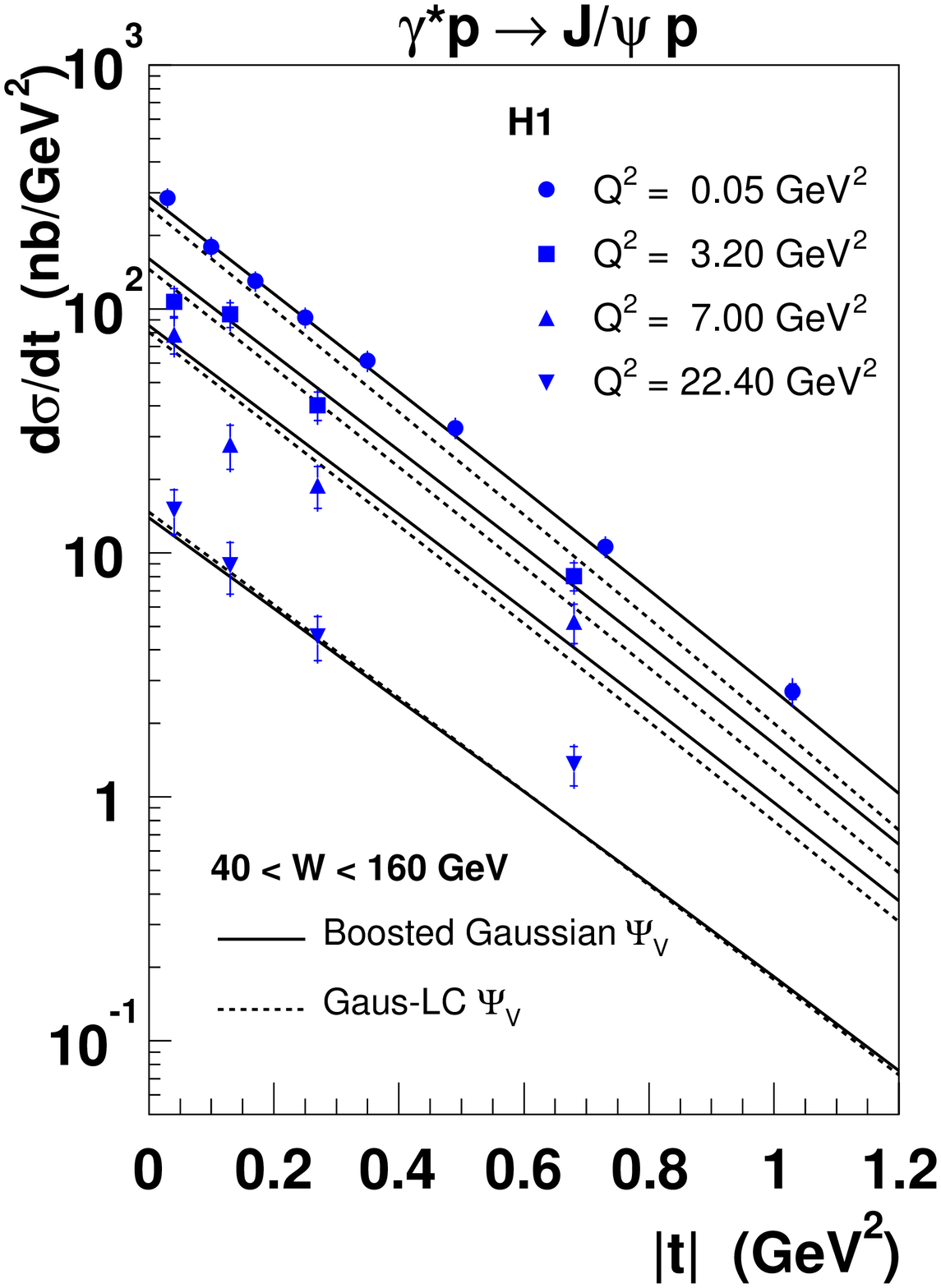}%
  \includegraphics[width=0.33\textwidth]{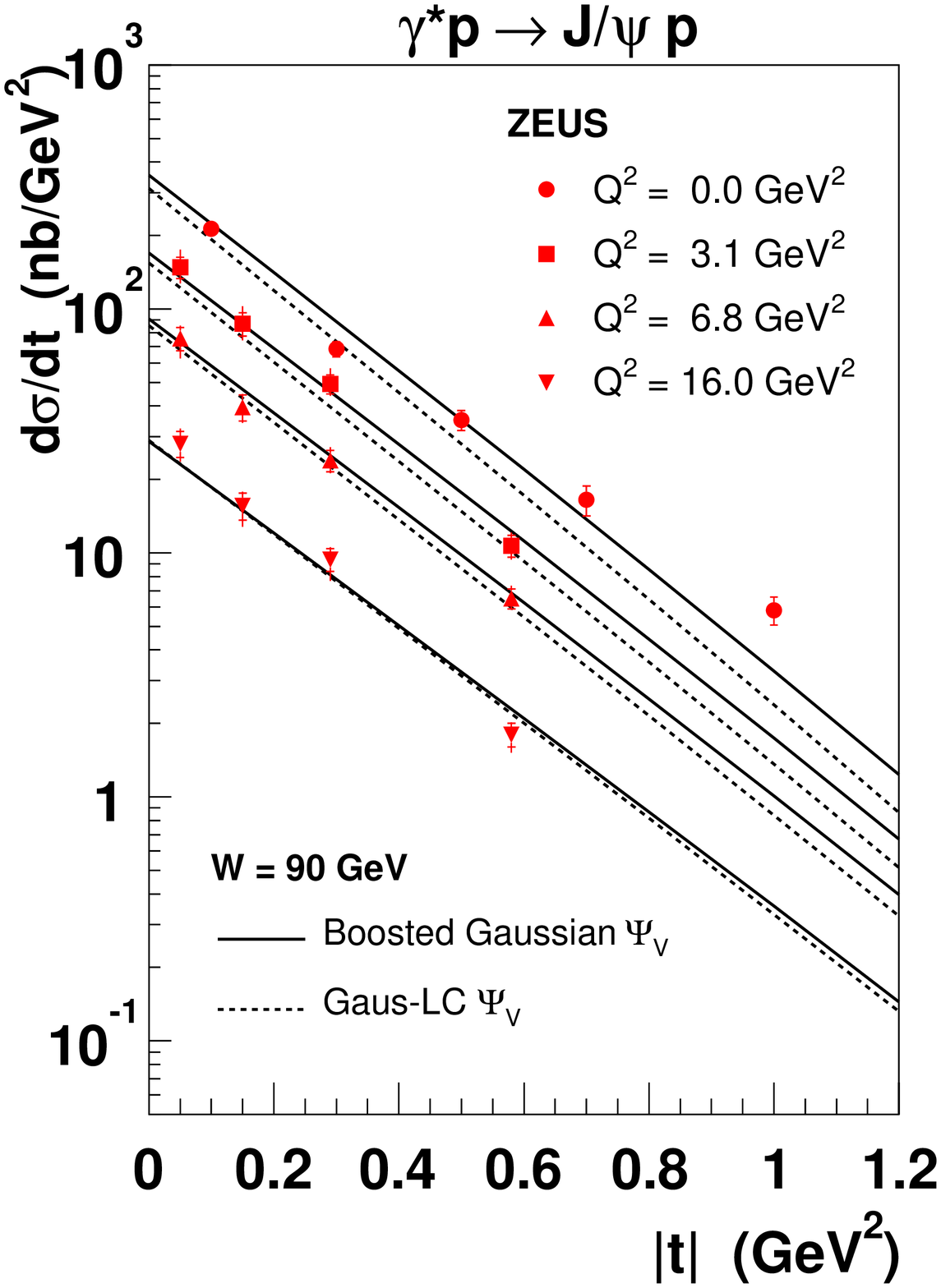}%
  \includegraphics[width=0.33\textwidth]{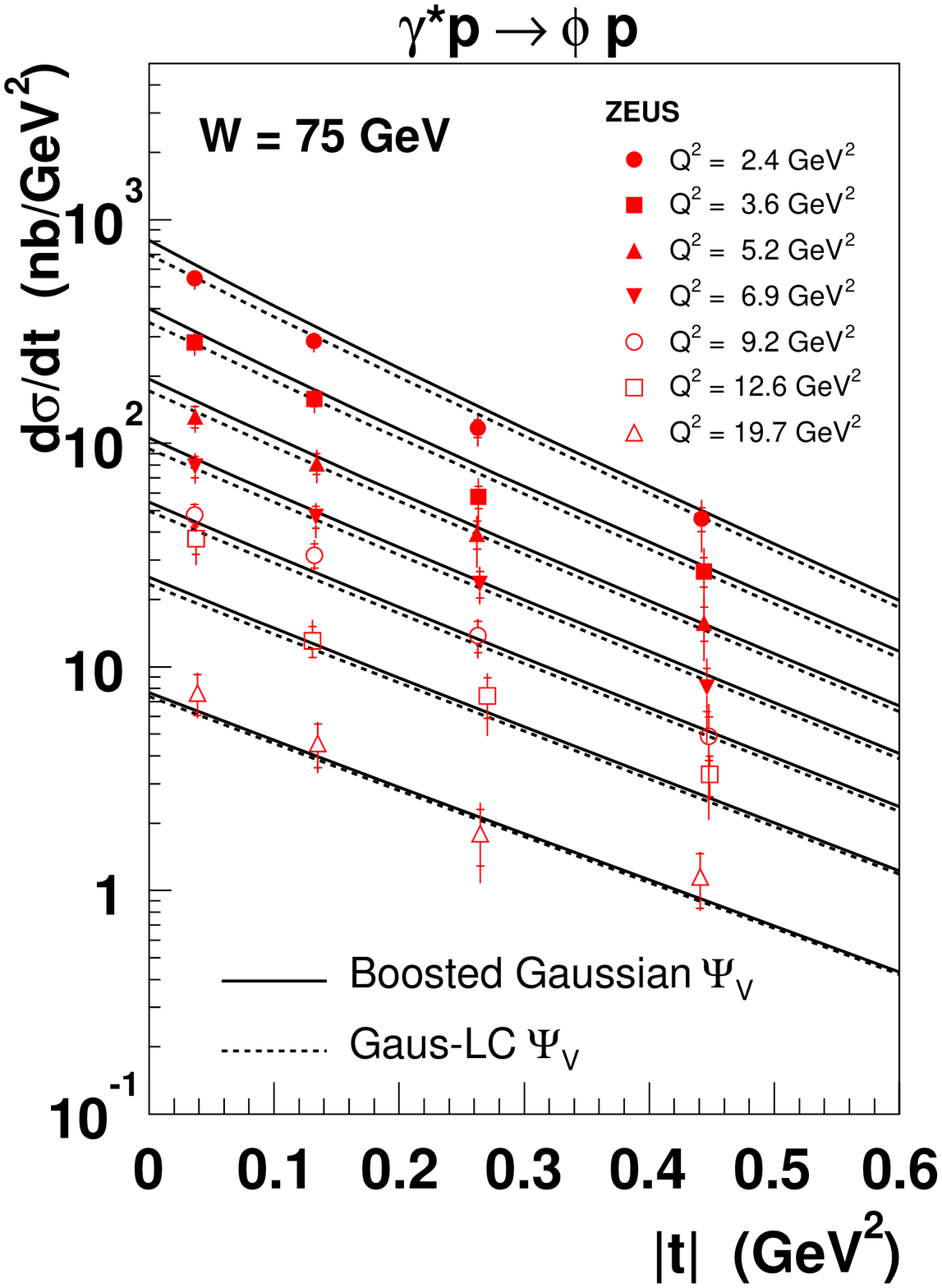}
  \caption{Differential vector meson cross section $\dif{\sigma}/\dif{t}$ vs.~$|t|$ compared to predictions from the b-Sat model using two different vector meson wave functions.  The ZEUS photoproduction ($J/\psi\to\mu^+\mu^-$) data points \cite{Chekanov:2002xi} shown in the second plot show only the statistical errors and are for $W=90$--$110$ GeV with the predictions calculated at $W=100$ GeV.  The ZEUS electroproduction data points \cite{Chekanov:2004mw} shown in the same plot are for $W=90$ GeV.}
  \label{fig:dsdt}
\end{figure}
Fig.~\ref{fig:dsdt} shows the HERA data on $t$-distributions for $J/\psi$  \cite{Chekanov:2002xi,Chekanov:2004mw,Aktas:2005xu} and $\phi$ \cite{Chekanov:2005cq} meson production.
\begin{figure}
  \centering
  \includegraphics[width=0.33\textwidth]{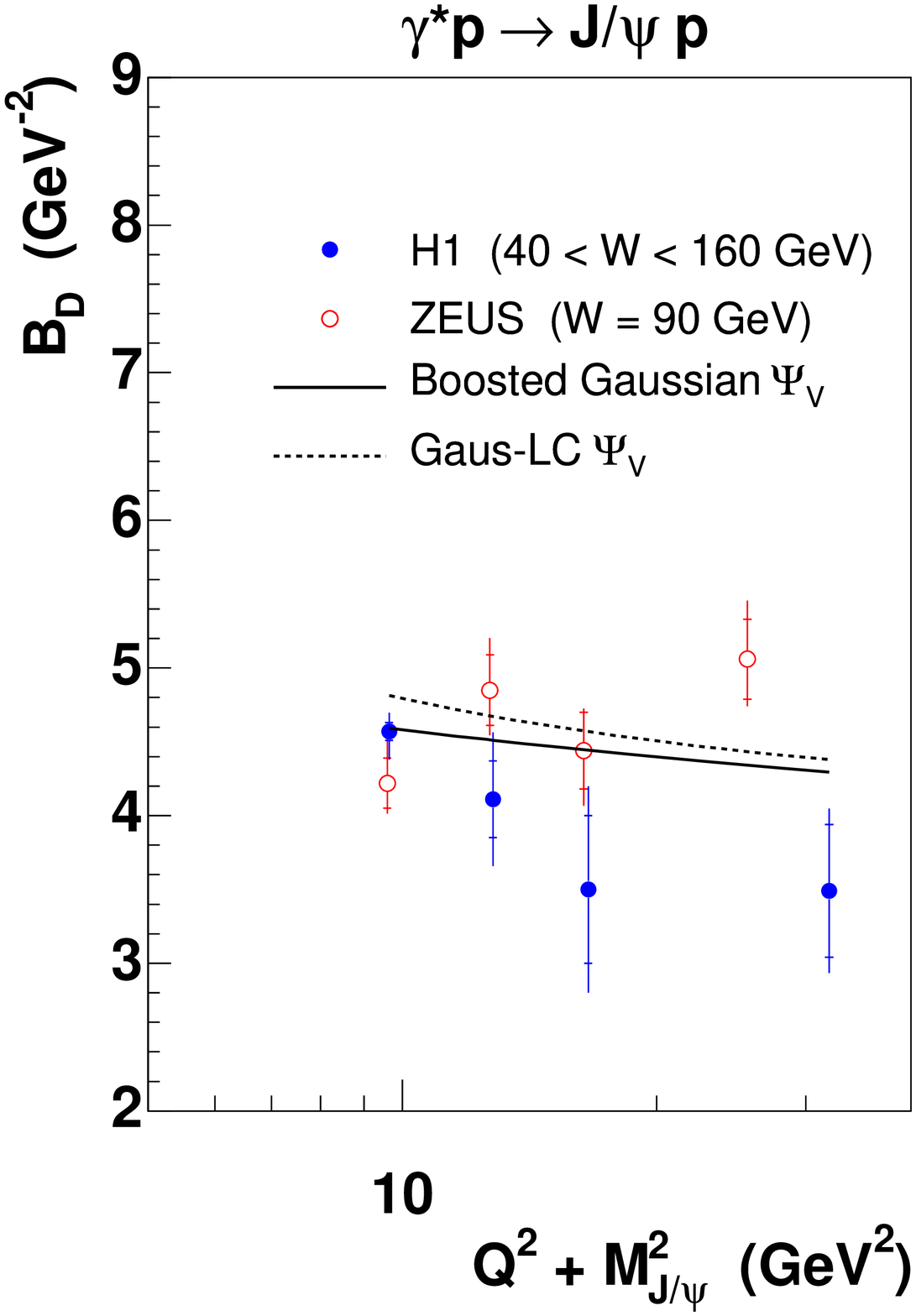}%
  \includegraphics[width=0.33\textwidth]{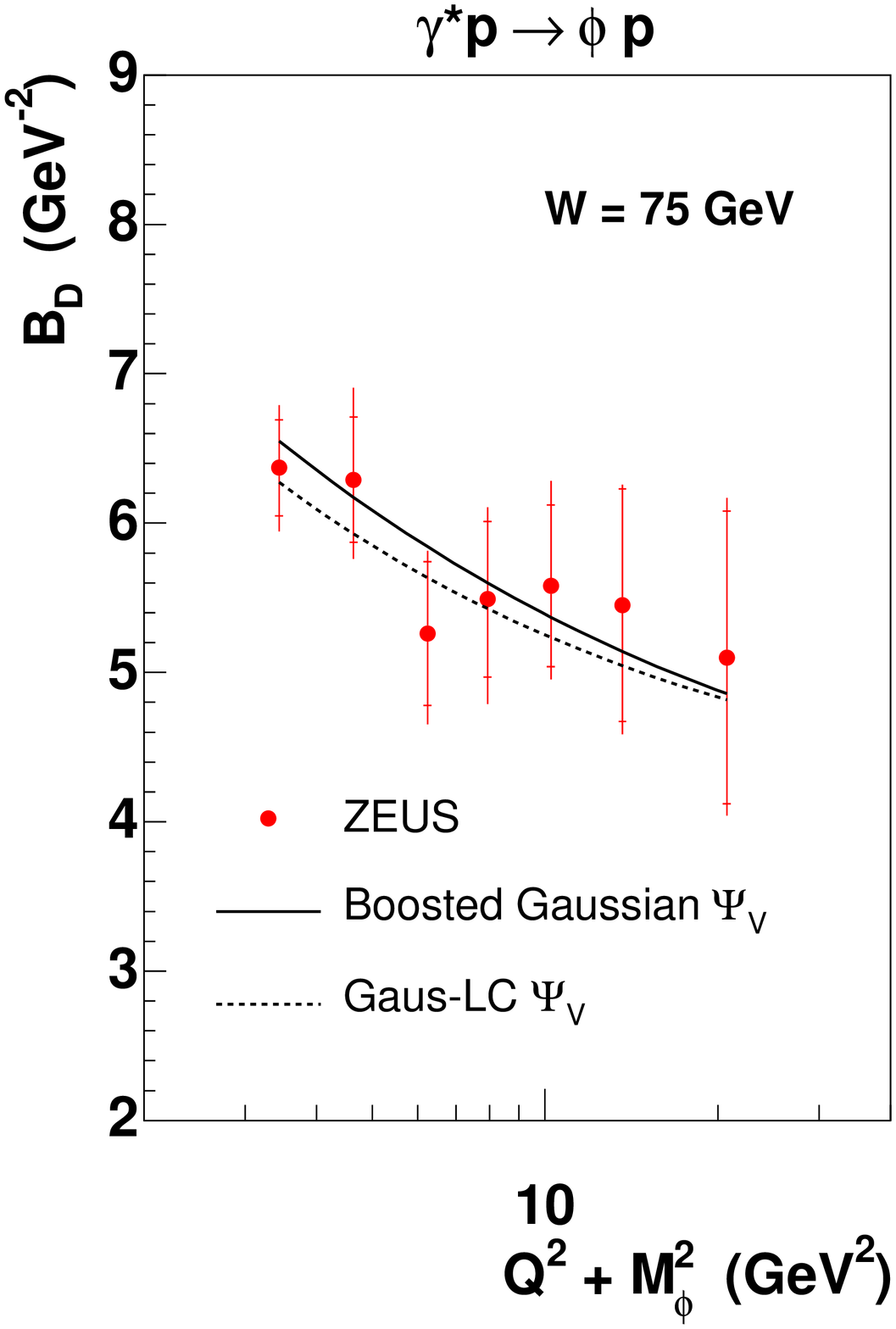}%
  \includegraphics[width=0.33\textwidth]{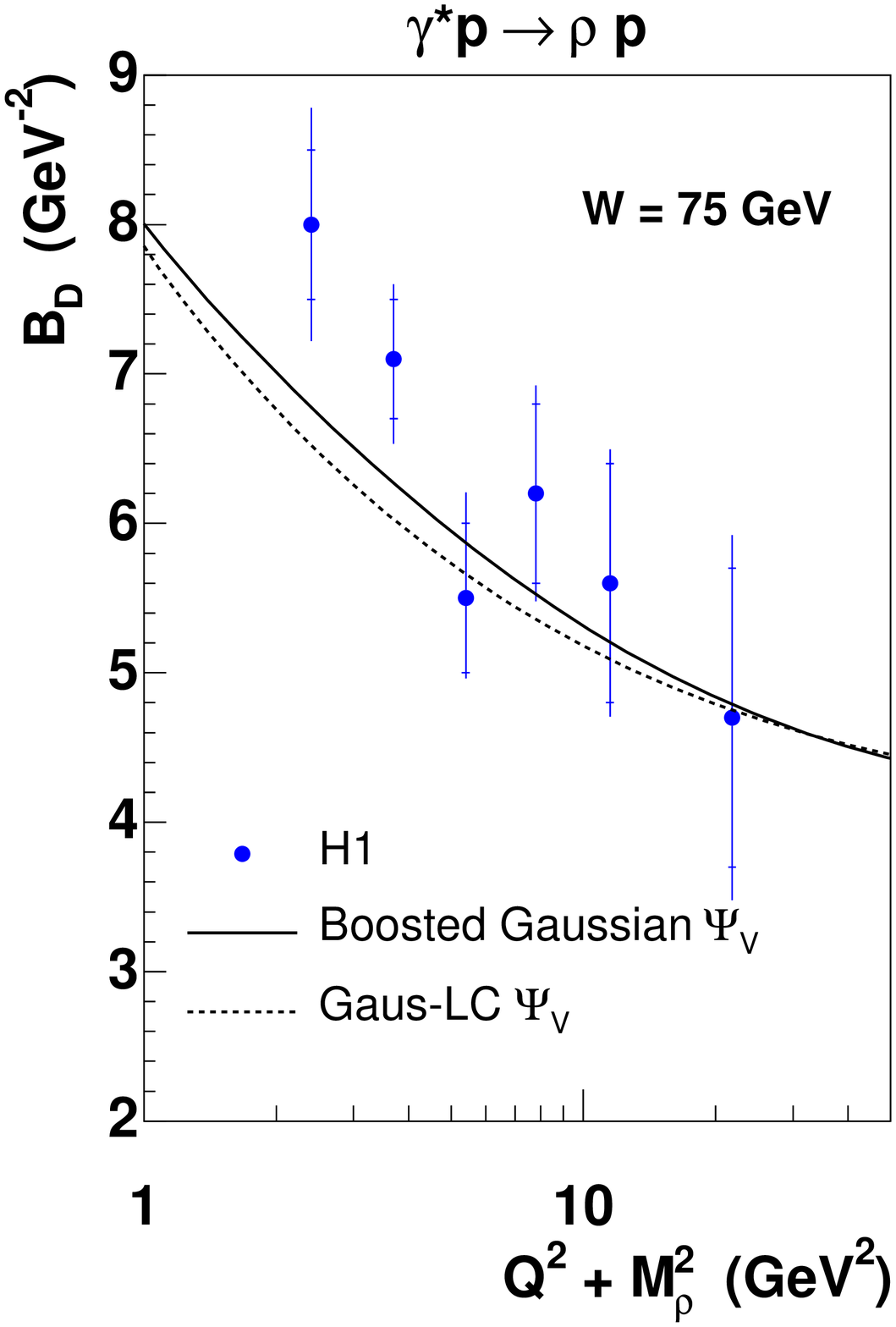}%
  \caption{The $t$-slope parameter $B_D$ vs.~$(Q^2+M_V^2)$, where $B_D$ is defined by fitting $\dif\sigma/\dif t\propto \exp(-B_D|t|)$, compared to predictions from the b-Sat model using two different vector meson wave functions.}
  \label{fig:bd}
\end{figure}
Fig.~\ref{fig:bd} shows the effective slope of the $t$-distribution, the parameter $B_D$, for  $J/\psi$, $\phi$ and $\rho$ \cite{Adloff:1999kg} vector mesons as a function of $(Q^2+M_V^2)$.  The parameter $B_D$ describes the area size of the interaction region and is obtained by making a fit to the observed (or computed in the model) $t$-distributions of the form $\dif\sigma/\dif t\propto \exp(-B_D|t|)$.  The theory predictions for $B_D$ are all obtained by making fits to $\dif\sigma/\dif t$ in the range $|t|<0.5$ GeV$^2$.  Figs.~\ref{fig:dsdt} and \ref{fig:bd} show that the $t$ dependence and the $(Q^2+M_V^2)$ dependence of $B_D$ are well described by the dipole model predictions for all three vector mesons whether using either the ``Gaus-LC'' or the ``boosted Gaussian'' vector meson wave functions.  We note that this good description is obtained with only one value of the width of the proton shape, $B_G$.

The proton shape, in the b-Sat model, is assumed to be purely Gaussian \eqref{eq:GaussianTb}.  The width of the Gaussian, $B_G$, determined by optimising the agreement between the model predictions and data for the $t$-distributions of the vector mesons and their effective slopes $B_D$, is found to be $B_G=4$ GeV$^{-2}$.  This value is mainly determined by the $t$-distributions of $J/\psi$ mesons measured by ZEUS \cite{Chekanov:2002xi,Chekanov:2004mw} and H1 \cite{Aktas:2005xu}.  We note, however, that although the values of the $B_D$ parameters measured by the two experiments are in agreement within errors, the spread of their values is somewhat large; see the first plot of Fig.~\ref{fig:bd}.  We estimate the error on the value of the parameter $B_G$ as being around 0.5 GeV$^{-2}$.

The value of $B_G=4$ GeV$^{-2}$ found in this investigation is slightly smaller than in the KT \cite{Kowalski:2003hm} investigation where $B_G=4.25$ GeV$^{-2}$ was determined using only the ZEUS $J/\psi$ photoproduction data \cite{Chekanov:2002xi}.  Fig.~\ref{fig:bd} shows that the subsequent ZEUS measurements of $J/\psi$ electroproduction \cite{Chekanov:2004mw} exhibit higher values of $B_D$ and therefore require a higher value of $B_G$.  Note that the effect of taking the size of the vector meson into account, that is, including the BGBP \cite{Bartels:2003yj} factor in \eqref{eq:xvecm1} arising from the non-forward wave functions, $\exp\left[\mathrm{i}(1-z)\vec{r}\cdot\vec{\Delta}\right]$, lowers the cross section for non-zero $t$ and therefore lowers the required value of $B_G$; recall that this factor was neglected by KT \cite{Kowalski:2003hm}.

Note also that the obtained values of $B_D$ at the same $(Q^2+M_V^2)$ are larger for light vector mesons than for $J/\psi$, in accordance with the data.  This occurs because the scales $Q^2$ and $m_f^2$ enter the photon wave function in slightly different ways.  We shall illustrate this by comparing $J/\psi$ photoproduction with light vector meson electroproduction at the same value of $(Q^2+M_V^2)$, implying $Q^2 \simeq 4m_c^2$.  The characteristic size of the scattering dipole is set by $1/\epsilon$ with $\epsilon^2 = z(1-z)Q^2+m_f^2$.  For the photoproduction of $J/\psi$, $\epsilon$ has no $z$ dependence, $\epsilon^2 = m_c^2$. In contrast, for light vector mesons $\epsilon^2$ varies with~$z$ from $Q^2/4$ at $z=1/2$ down to $m_{u,d,s}^2$ at $z\to 0$ and $z\to 1$, so that the effective value of $\epsilon^2$ is significantly lower than $Q^2/4+m_{u,d,s}^2 \simeq m_c^2$.  Therefore, for light vector meson production at $Q^2 \simeq 4m_c^2$, the typical dipole size is larger than for photoproduction of $J/\psi$.  This is particularly pronounced at the end-points $z\to 0$ and $z\to 1$ for the transversely polarised light vector mesons.  At sufficiently large values of $Q^2$, however, the longitudinally polarised mesons dominate and the typical dipole size becomes small enough to have a negligible contribution to $B_D$ for both light and heavy mesons.  Hence, at large $(Q^2+M_V^2)$, $B_D$ tends to a universal value determined by the proton shape alone.

\begin{figure}
  \centering
  \includegraphics[width=0.33\textwidth]{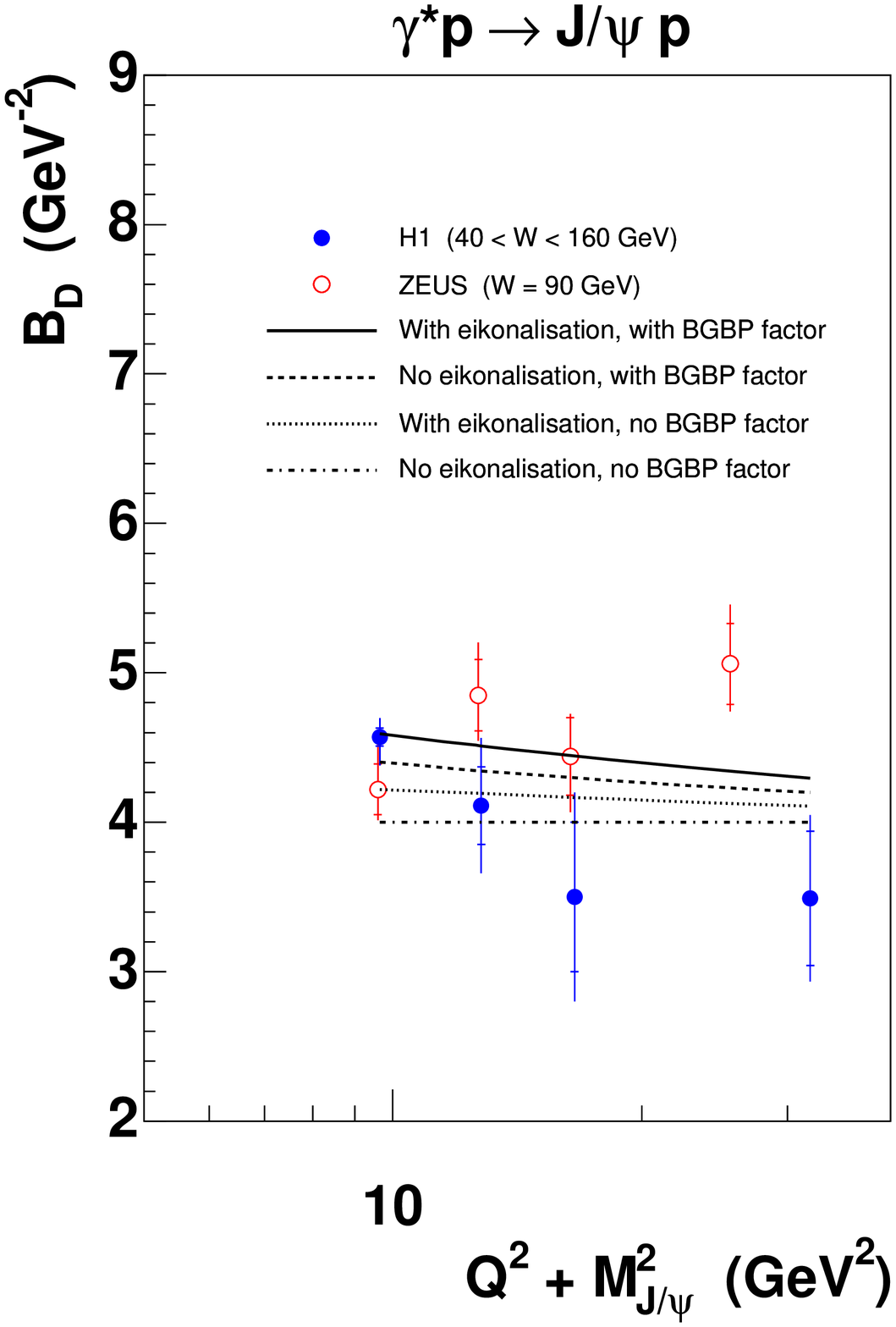}%
  \includegraphics[width=0.33\textwidth]{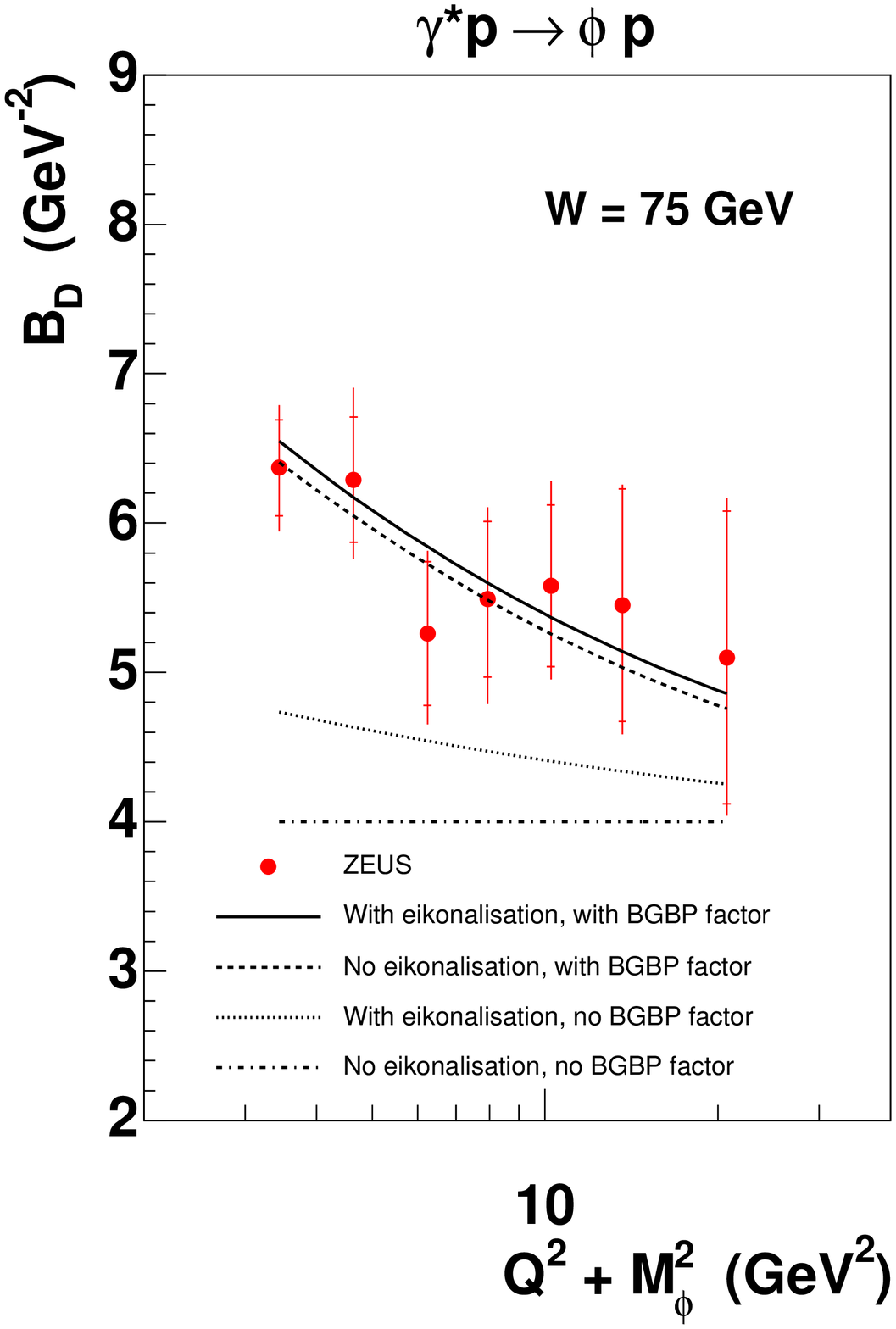}%
  \includegraphics[width=0.33\textwidth]{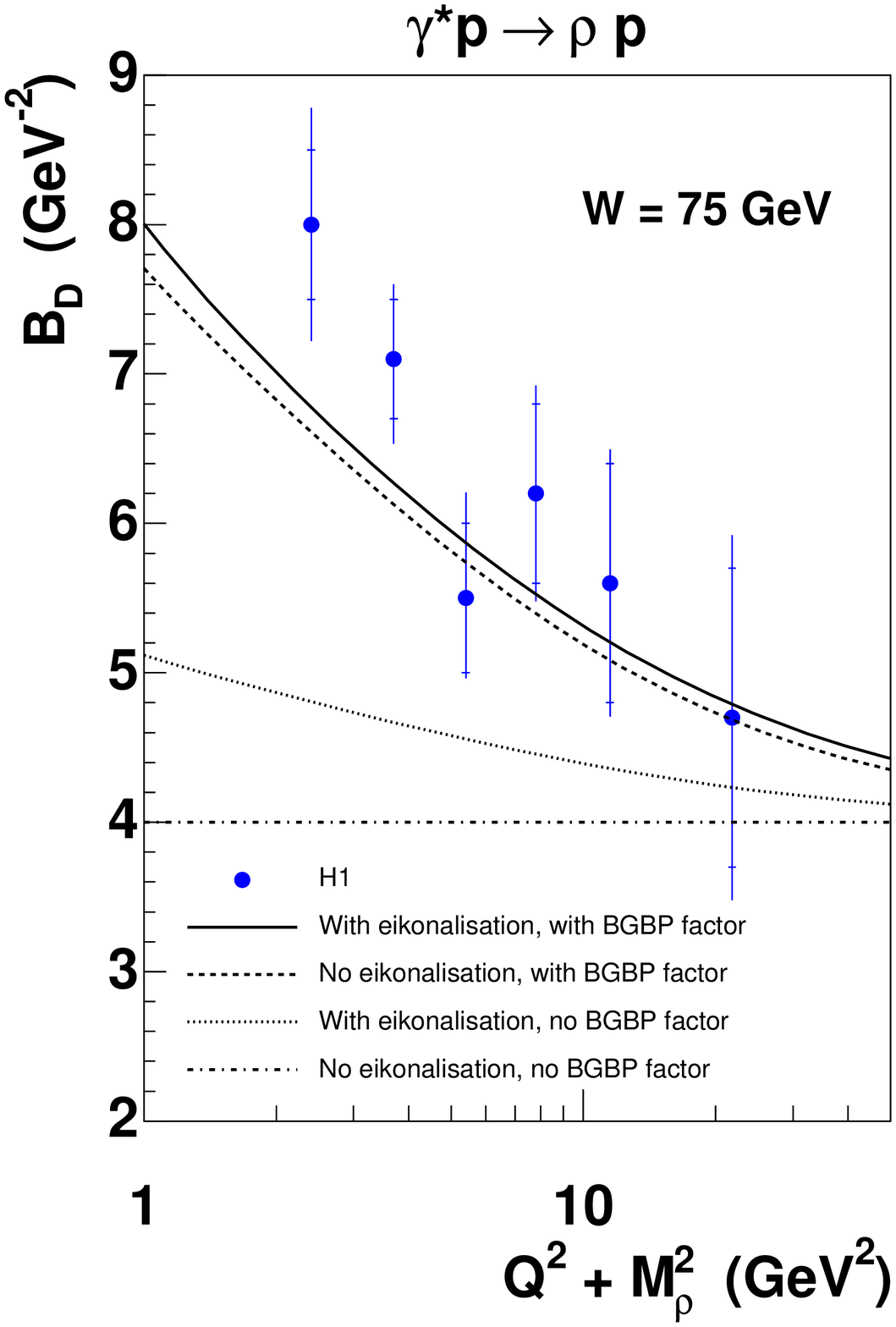}%
  \caption{The $t$-slope parameter $B_D$ vs.~$(Q^2+M_V^2)$ compared to predictions from the b-Sat model using the ``boosted Gaussian'' vector meson wave function.  We show the effect of switching off the eikonalisation in the dipole cross section \eqref{eq:dsigmad2b}, and omitting the BGBP \cite{Bartels:2003yj} factor, $\exp\left[\mathrm{i}(1-z)\vec{r}\cdot\vec{\Delta}\right]$, in \eqref{eq:xvecm1}.}
  \label{fig:bd_noeik_nobgbp}
\end{figure}
It is important to realise that the dependence of $B_D$ on $(Q^2+M_V^2)$ observed for light vector mesons originates from the enlargement of the interaction area due to the dipole transverse extension.  Recall that this effect is taken into account by the BGBP \cite{Bartels:2003yj} prescription of the QCD dipole scattering at $t\neq 0$.  It also partly arises from the saturation effects which play a stronger role for the larger typical dipole sizes at small $(Q^2+M_V^2)$.  We investigate the interplay between these two mechanisms on the value of $B_D$ in Fig.~\ref{fig:bd_noeik_nobgbp}.  We show the effect of switching off the eikonalisation, that is, replacing the dipole cross section \eqref{eq:xdip} by the opacity $\Omega$ \eqref{eq:omega}.  We also show the effect of omitting the BGBP \cite{Bartels:2003yj} factor, $\exp\left[\mathrm{i}(1-z)\vec{r}\cdot\vec{\Delta}\right]$, in \eqref{eq:xvecm1}.  Without these two effects, which diminish with increasing $(Q^2+M_V^2)$, the $t$-slope $B_D$ tends to the universal value of $B_D=B_G=4$ GeV$^{-2}$.  Without the BGBP factor, the eikonalisation has a significant effect for $\phi$ and $\rho$ mesons, but it is not enough to describe the $B_D$ data points.  With the BGBP factor, the eikonalisation has only a small effect and the rise of $B_D$ with decreasing $(Q^2+M_V^2)$ nicely reproduces the rise observed in the data.

\begin{figure}
  \centering
  \includegraphics[width=0.5\textwidth]{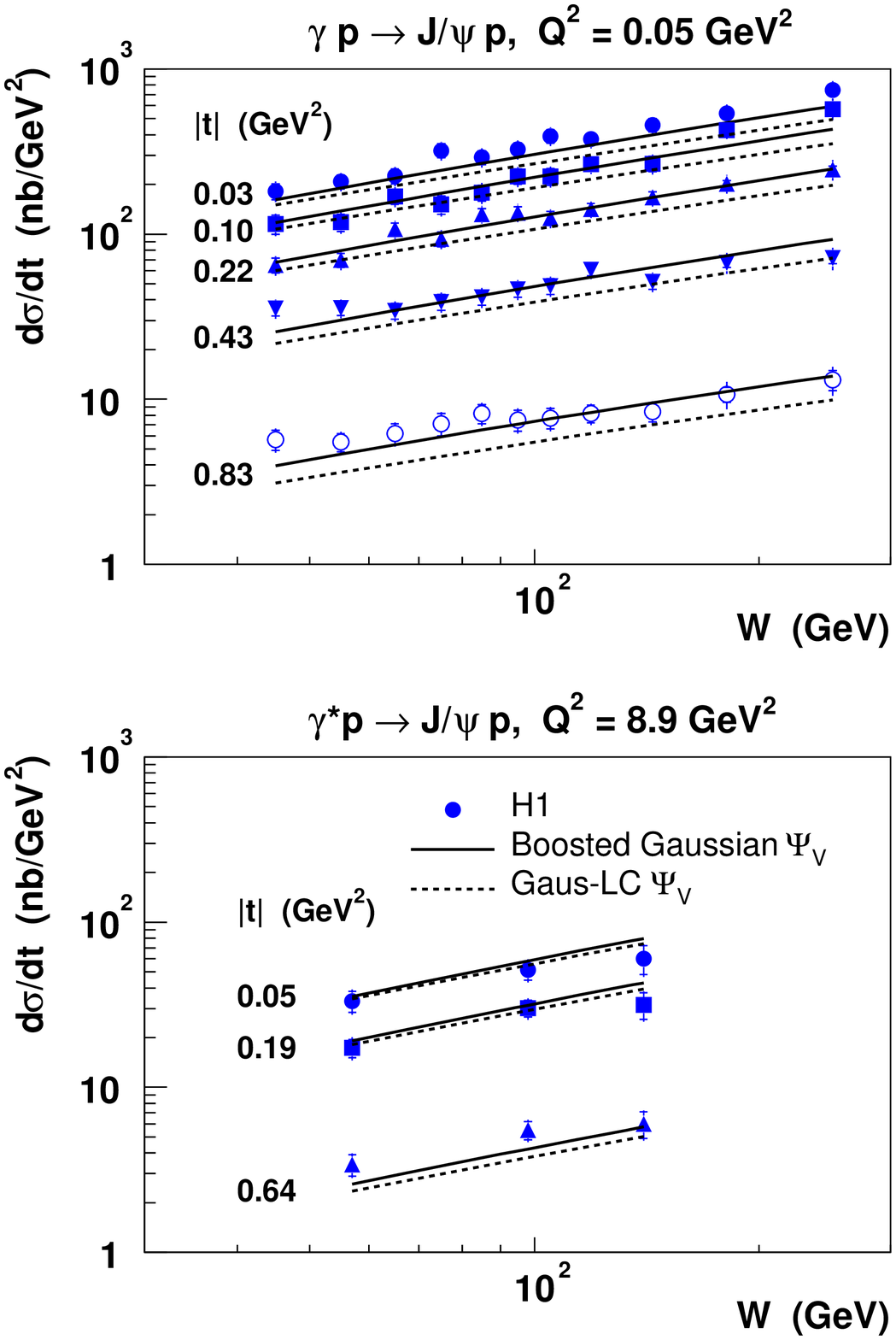}%
  \includegraphics[width=0.5\textwidth]{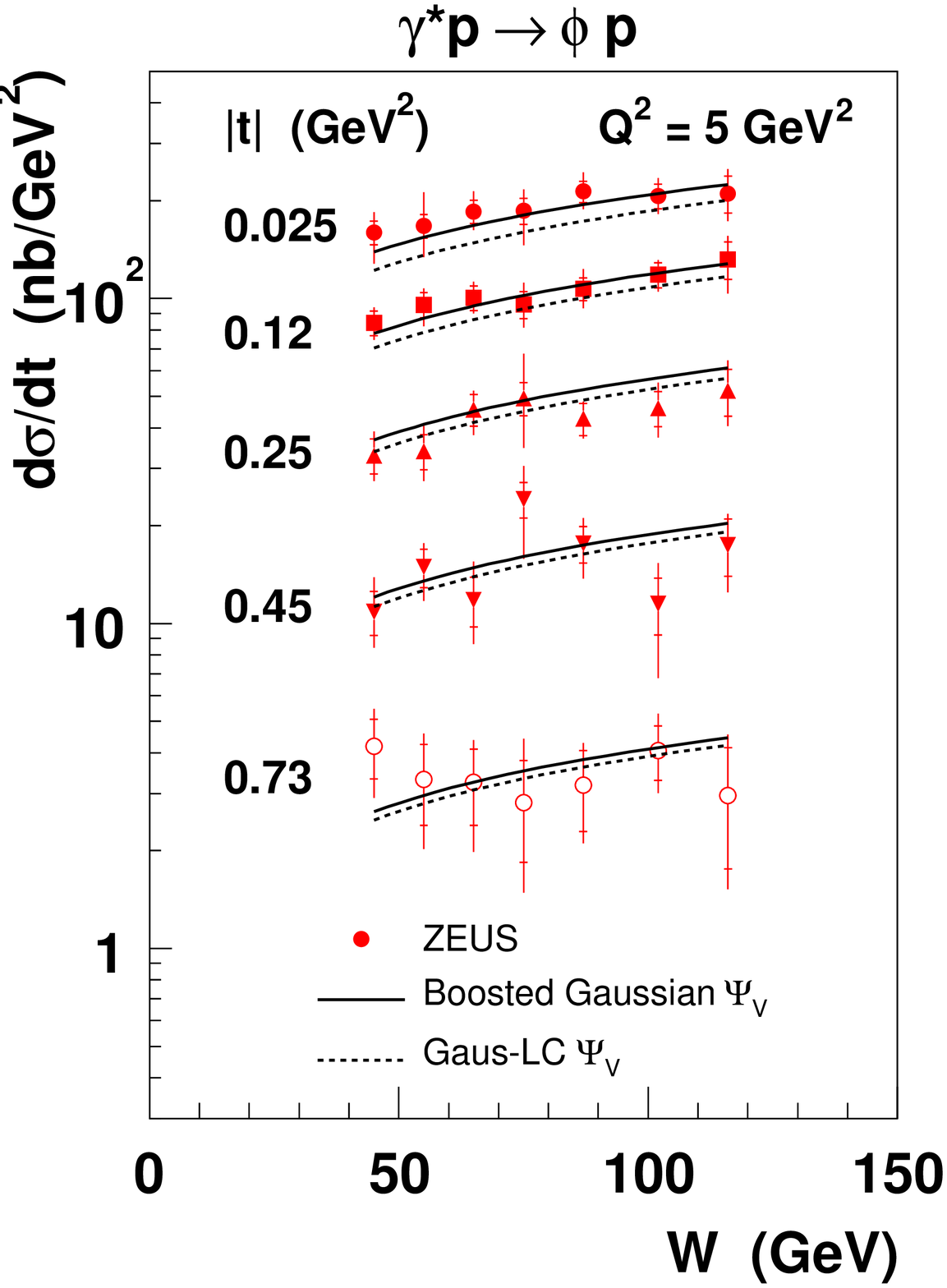}
  \caption{Differential vector meson cross section $\dif{\sigma}/\dif{t}$ vs.~$W$ compared to predictions from the b-Sat model using two different vector meson wave functions.}
  \label{fig:dsdtw}
\end{figure}
\begin{figure}
  \centering
  \includegraphics[width=0.5\textwidth]{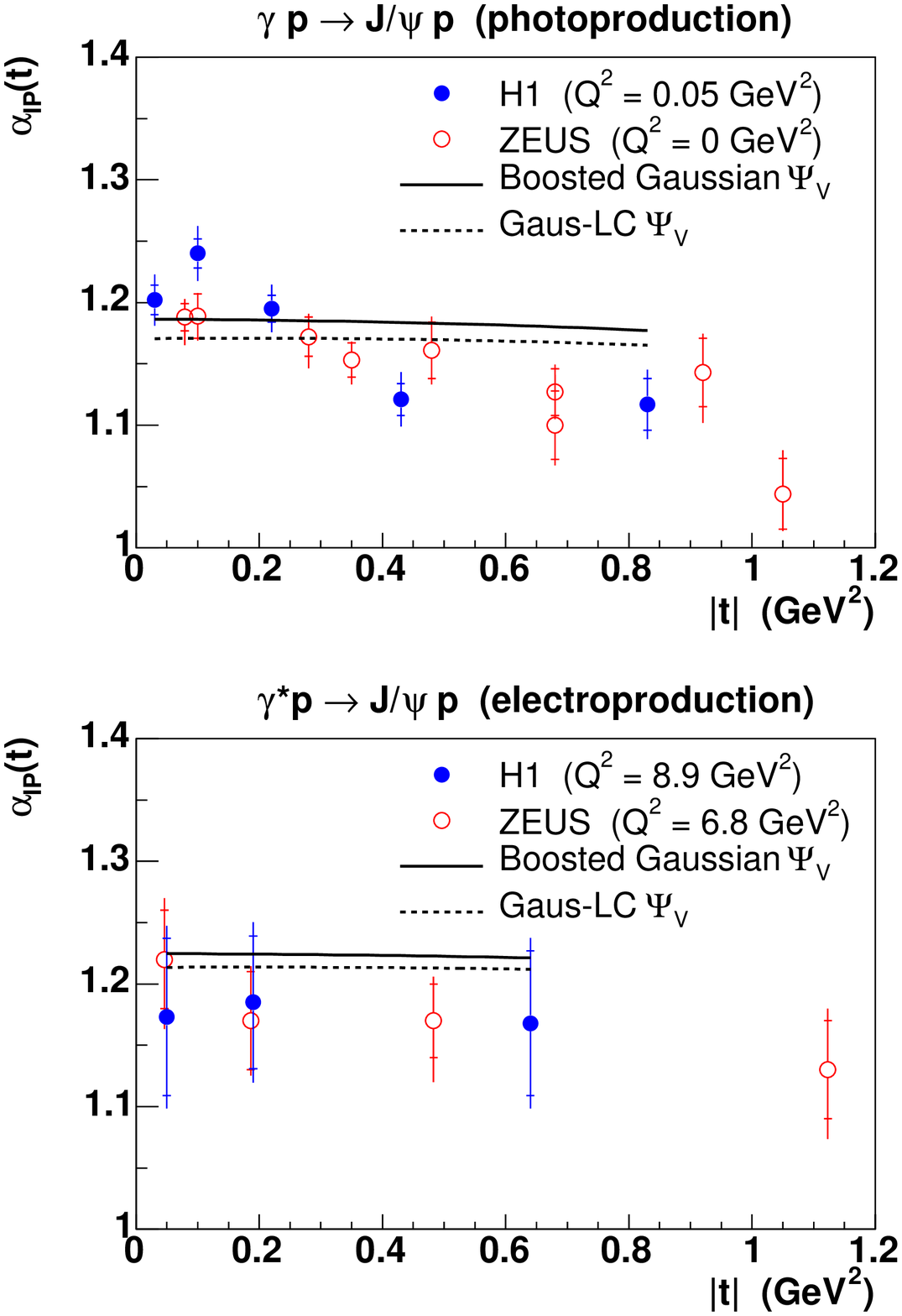}%
  \includegraphics[width=0.5\textwidth]{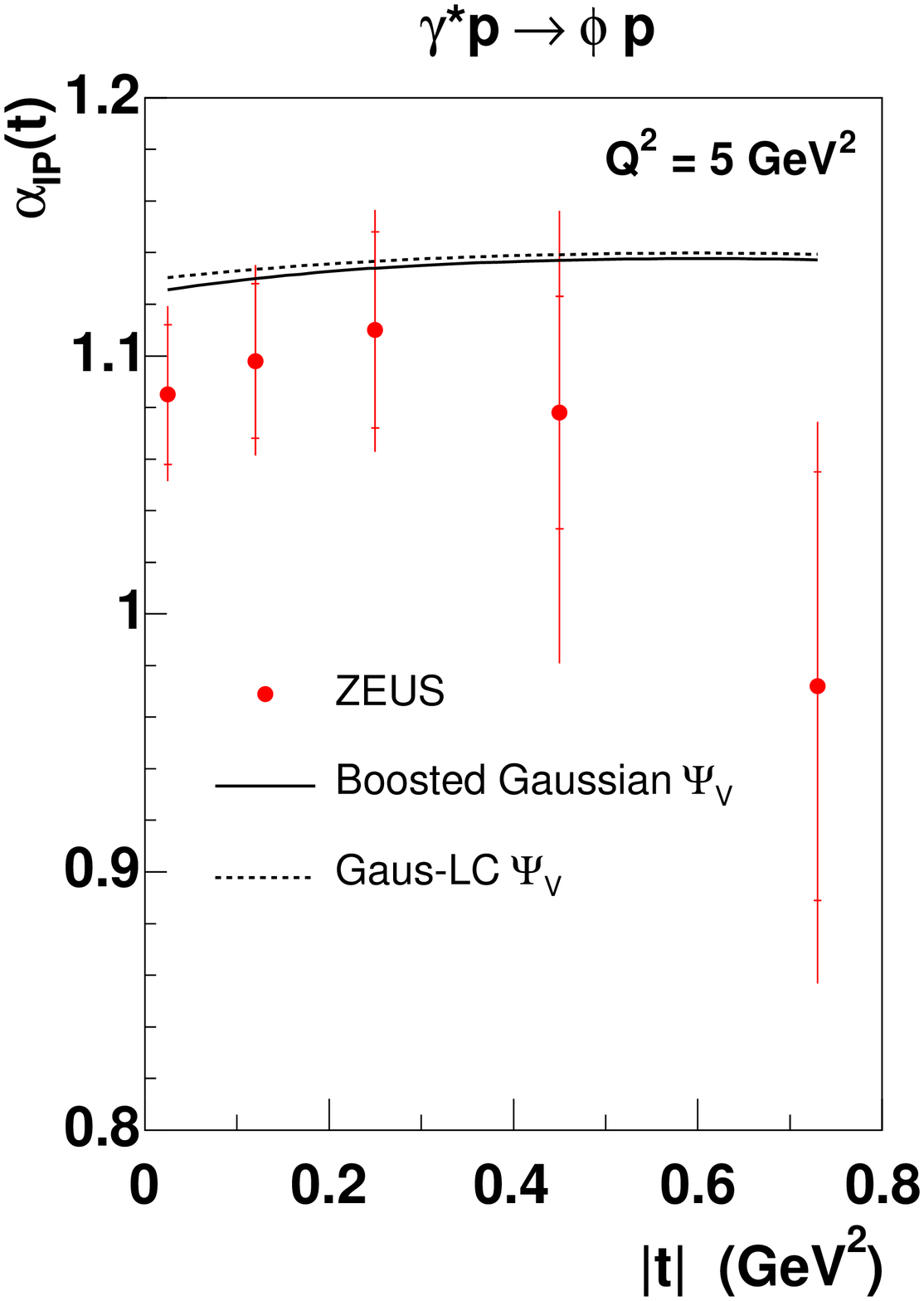}
  \caption{The Pomeron trajectory $\alpha_\Pom(t)$ vs.~$|t|$, where $\alpha_\Pom(t)$ is determined by fitting $\dif\sigma/\dif t\propto W^{4[\alpha_\Pom(t)-1]}$, compared to predictions from the b-Sat model using two different vector meson wave functions.}
  \label{fig:apom}
\end{figure}
\begin{figure}
  \centering
  \includegraphics[width=0.8\textwidth]{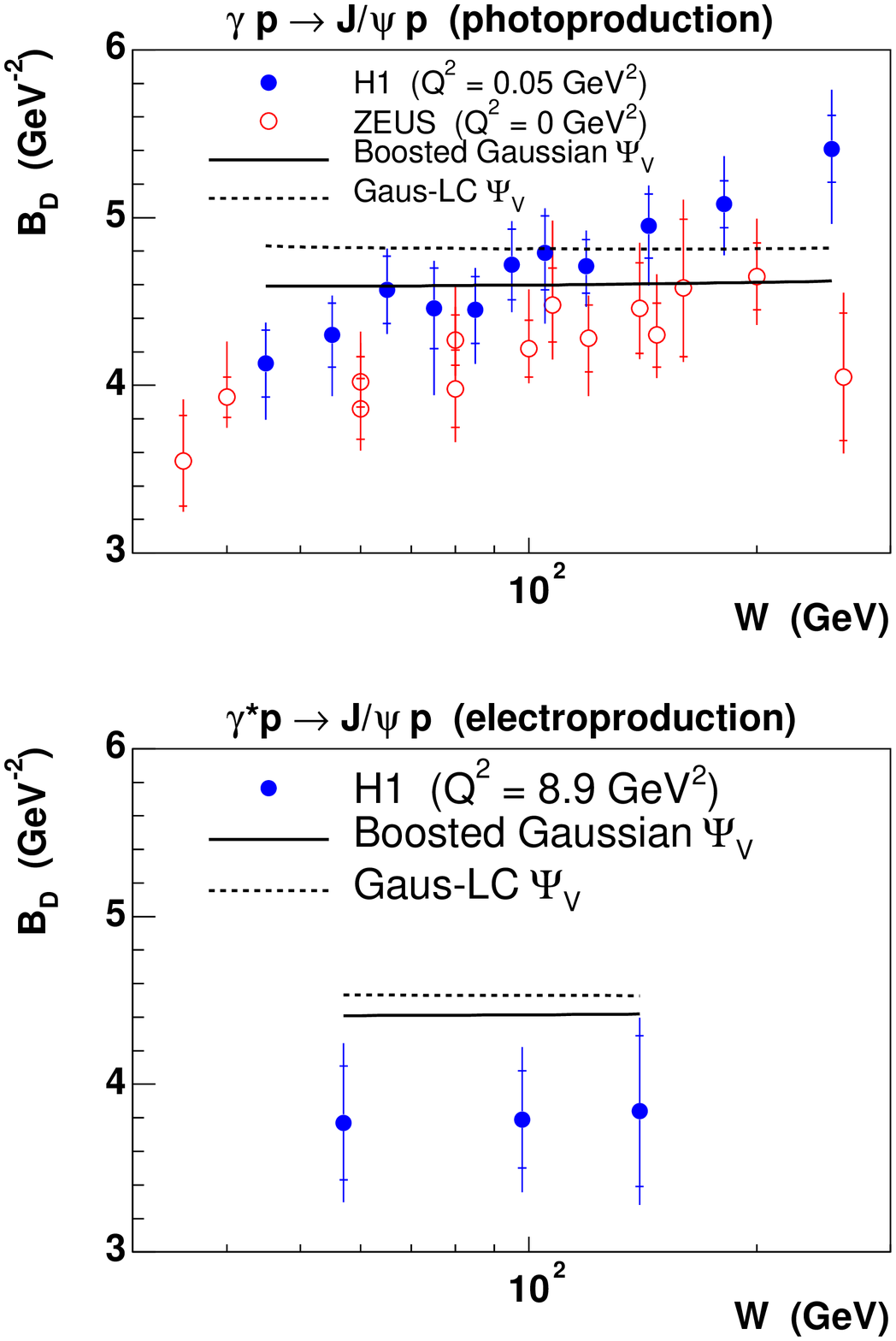}
  \caption{The $t$-slope parameter $B_D$ vs.~$W$, where $B_D$ is defined by fitting $\dif\sigma/\dif t\propto \exp(-B_D|t|)$, compared to predictions from the b-Sat model using two different vector meson wave functions.}
  \label{fig:bdw}
\end{figure}
We also investigated, for completeness, the $W$ dependence of the $t$-distributions.  In Fig.~\ref{fig:dsdtw} we show the $W$ dependence of $\dif\sigma/\dif t$ for fixed values of $|t|$ and $Q^2$.  For each value of $t$, we make a fit of the form $\dif\sigma/\dif t\propto W^{4[\alpha_\Pom(t)-1]}$ and then plot $\alpha_\Pom(t)$ against $|t|$; see Fig.~\ref{fig:apom}.  We also fit the same data to the form $\dif\sigma/\dif t\propto \exp(-B_D|t|)$ for each value of $W$, then we plot $B_D$ against $W$; see Fig.~\ref{fig:bdw}.

\subsection{Deeply virtual Compton scattering} \label{sec:DVCS}
We now compare to the recently published DVCS data from H1 \cite{Aktas:2005ty} and ZEUS \cite{Chekanov:2003ya}.  We use the b-Sat model with a Gaussian $T(b)$ and $B_G=4$ GeV$^{-2}$, and quark masses $m_{u,d,s}=0.14$ GeV and $m_c=1.4$ GeV.
\begin{figure}
  \centering
  \includegraphics[width=0.4\textwidth]{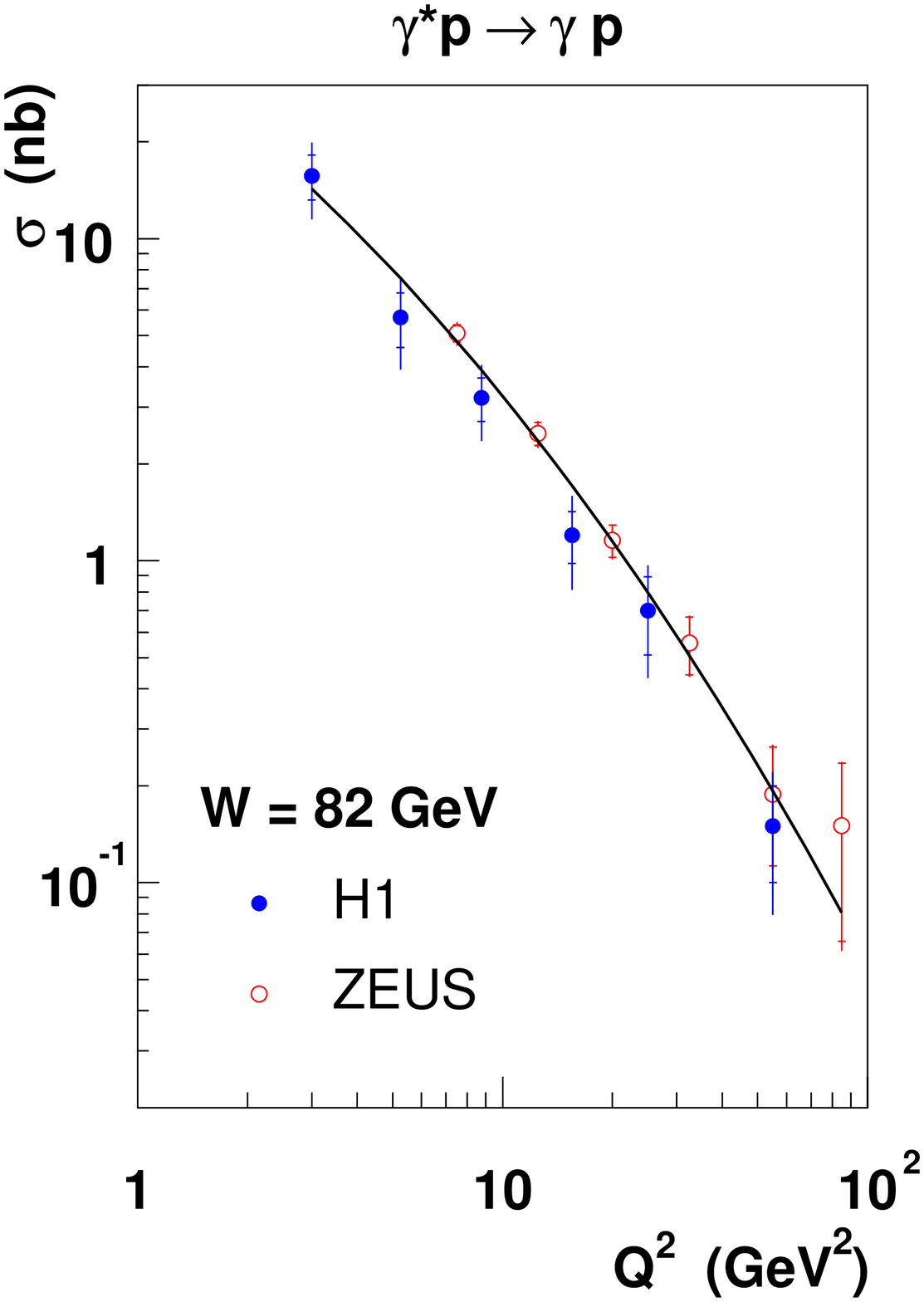}
  \includegraphics[width=0.4\textwidth]{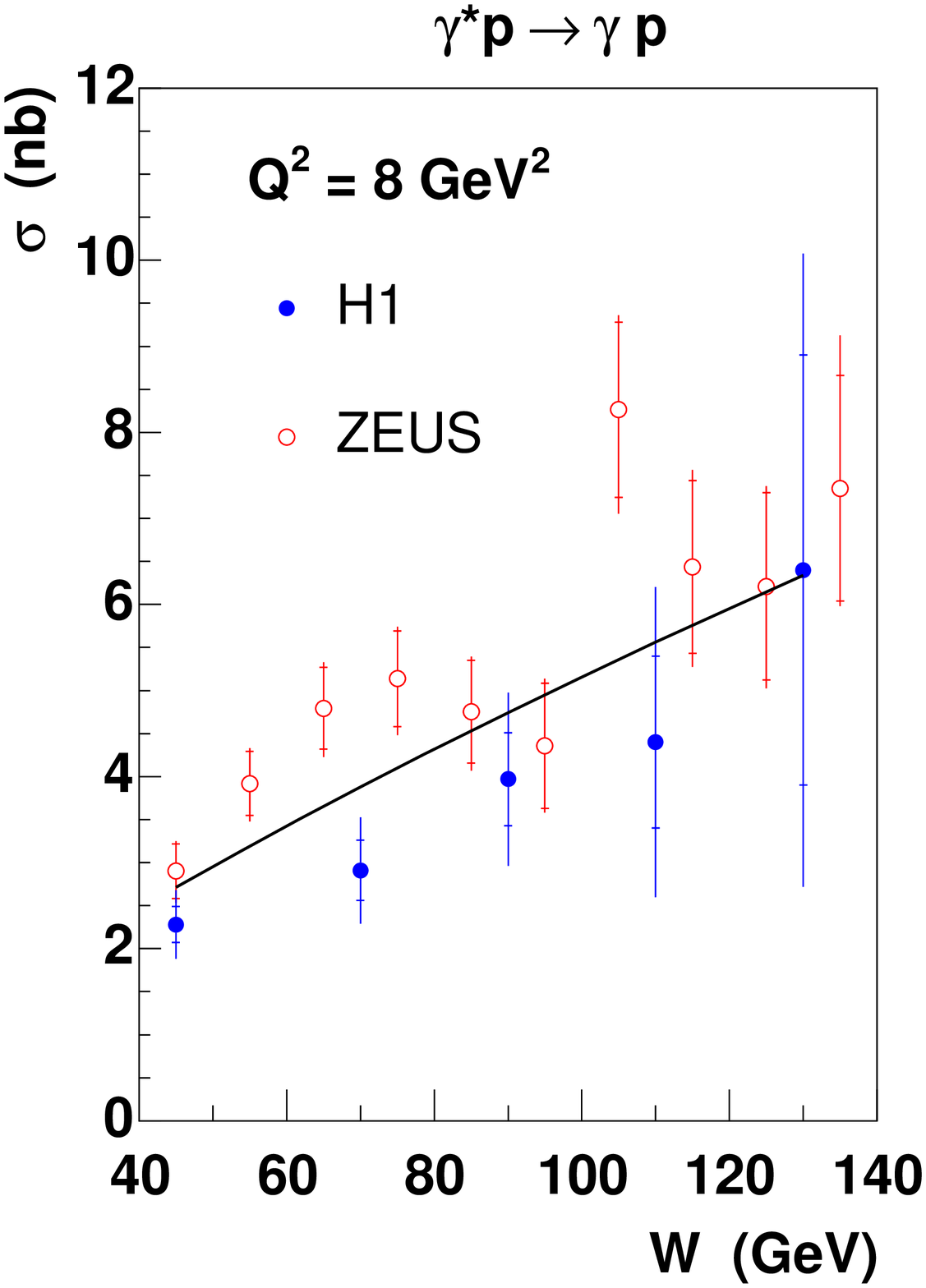}
  \caption{Total DVCS cross sections $\sigma$ vs.~$Q^2$ (left) and $\sigma$ vs.~$W$ (right) compared to predictions from the b-Sat model.}
  \label{fig:crossq_dvcs}
\end{figure}
In Fig.~\ref{fig:crossq_dvcs} (left) we show the $Q^2$ dependence of the cross section integrated over $|t|$ up to 1 GeV$^2$ for $W=82$ GeV compared to the H1 data \cite{Aktas:2005ty}.  We also show the ZEUS data \cite{Chekanov:2003ya} at $W=89$ GeV rescaled to $W=82$ GeV using $\sigma\propto W^\delta$, with $\delta=0.75$ \cite{Chekanov:2003ya}.  In Fig.~\ref{fig:crossq_dvcs} (right) we show the $W$ dependence of the cross section integrated over $|t|$ up to 1 GeV$^2$ for $Q^2=8$ GeV$^2$ compared to the H1 data \cite{Aktas:2005ty}.  We also show the ZEUS data \cite{Chekanov:2003ya} at $Q^2=9.6$ GeV$^2$ rescaled to $Q^2=8$ GeV$^2$ using $\sigma\propto Q^{-2n}$, with $n=1.54$ \cite{Chekanov:2003ya}.  Fitting the theory predictions to the form $\sigma\propto W^\delta$ gives $\delta = 0.80$ to be compared with the experimental value of $0.77\pm0.23\pm0.19$ \cite{Aktas:2005ty}.  We see from Fig.~\ref{fig:crossq_dvcs} that the $Q^2$ and $W$ dependence of the DVCS data, as well as the absolute normalisation, are well described by the b-Sat model.

\begin{figure}
  \centering
  \includegraphics[width=0.45\textwidth]{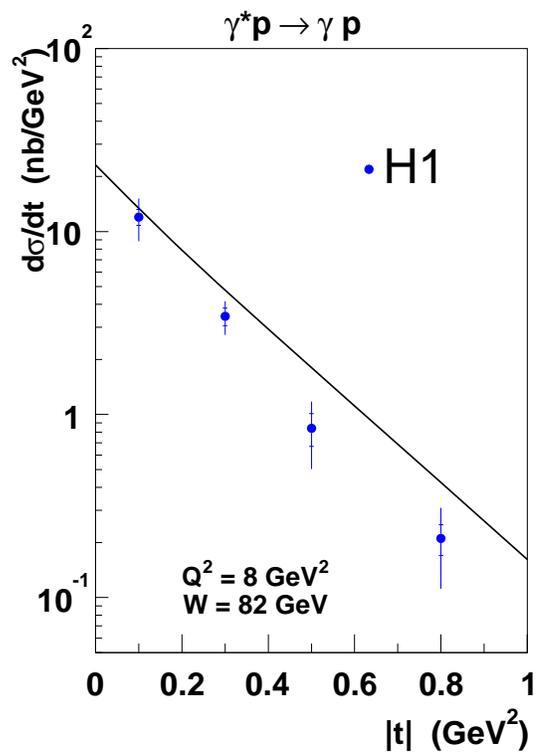}
  \caption{Differential DVCS cross section $\dif{\sigma}/\dif{t}$ vs.~$|t|$ compared to the prediction from the b-Sat model.}
  \label{fig:dsdt_dvcs}
\end{figure}
The $t$-distribution is shown in Fig.~\ref{fig:dsdt_dvcs} for $Q^2=8$ GeV$^2$ and $W=82$ GeV compared to the H1 data \cite{Aktas:2005ty}.  At small $t$ the data are well-described, while at larger $t$ the prediction slightly overestimates the data, due to a $t$-slope which is too small.  Fitting the theory prediction to the form $\dif\sigma/\dif t\propto \exp(-B_D|t|)$ for $|t|< 0.5$ GeV$^2$ gives $B_D=5.29$ GeV$^{-2}$, to be compared with the experimental value of $6.02\pm0.35\pm0.39$ GeV$^{-2}$ \cite{Aktas:2005ty}.  When comparing these values one should bear in mind that the value of the parameter $B_G=4$ GeV$^{-2}$ determined from the $t$-distributions of the vector meson data has a possible uncertainty which could be as large as 0.5 GeV$^{-2}$.

Summarising, we can see that the agreement of the predictions from the b-Sat model with DVCS data is remarkably good, especially if we note that the DVCS data were not used in fixing any parameters of the model.

\section{Impact parameter dependent CGC model}
We have seen that almost all features of the exclusive diffractive HERA processes are well described by the impact parameter dependent saturation (``b-Sat'') model with a Gaussian $T(b)$ of width $B_G=4$ GeV$^{-2}$.  The b-Sat model assumes the validity of DGLAP evolution which may not be appropriate when $x$ approaches the saturation region.  Therefore, we also investigated the impact parameter dependent CGC (``b-CGC'') model, in which the dipole cross section is given by \eqref{eq:bcgc} and \eqref{eq:bcgc1}.  In the b-CGC model the evolution effects are included via an approximate solution to the Balitsky--Kovchegov equation \cite{Balitsky:1995ub,Kovchegov:1999yj,Kovchegov:1999ua}.

\begin{table}
  \centering
  \begin{tabular}{cccc|ccc|c}
    \hline\hline
    Model & $Q^2$/GeV$^2$ & $m_{u,d,s}$/GeV & $m_c$/GeV & $\mathcal{N}_0$ & $x_0$/$10^{-4}$ & $\lambda$ & $\chisq$ \\ \hline
    b-CGC & [0.25,45] & $0.14$ & $1.4$ & 0.417 & $5.95$ & 0.159 & $211.2/130=1.62$ \\ \hline\hline
  \end{tabular}
  \caption{Parameters of the b-CGC model, \eqref{eq:bcgc} and \eqref{eq:bcgc1}, determined from a fit to $F_2$ data \cite{Breitweg:2000yn,Chekanov:2001qu}.}
  \label{tab:bCGC}
\end{table}
Similar to the b-Sat model, the parameter $B_{\rm CGC}=5.5$ GeV$^{-2}$ in \eqref{eq:bcgc1} is determined by requiring a good description of the $t$-slopes of vector meson data, while the three parameters $\mathcal{N}_0$, $\lambda$ and $x_0$ in \eqref{eq:bcgc} and \eqref{eq:bcgc1} are determined by fitting the $F_2$ data \cite{Breitweg:2000yn,Chekanov:2001qu} with $\xB\le 0.01$ and $Q^2\in[0.25,45]$ GeV$^2$.  The results of the fit are shown in Table~\ref{tab:bCGC}.  The fit to the $F_2$ data with the b-CGC model gives a sizably worse description than the b-Sat model as seen from the value of the $\chisq$ in Table~\ref{tab:bCGC} and the comparison with data of the parameter $\lambda_{\rm tot}$ shown in the bottom plot of Fig.~\ref{fig:sigtot}.  The significant deterioration of the fit quality is due to the fact that in the impact parameter dependent description, saturation effects can only be sizable in the core of the proton, see the discussion in Sect.~\ref{sec:Sat}.  The relatively poor quality of the fit is the main reason why we prefer to use a DGLAP-evolved gluon density together with the Glauber--Mueller dipole cross section, that is, the b-Sat model.

\begin{figure}
  \centering
  \includegraphics[width=0.8\textwidth]{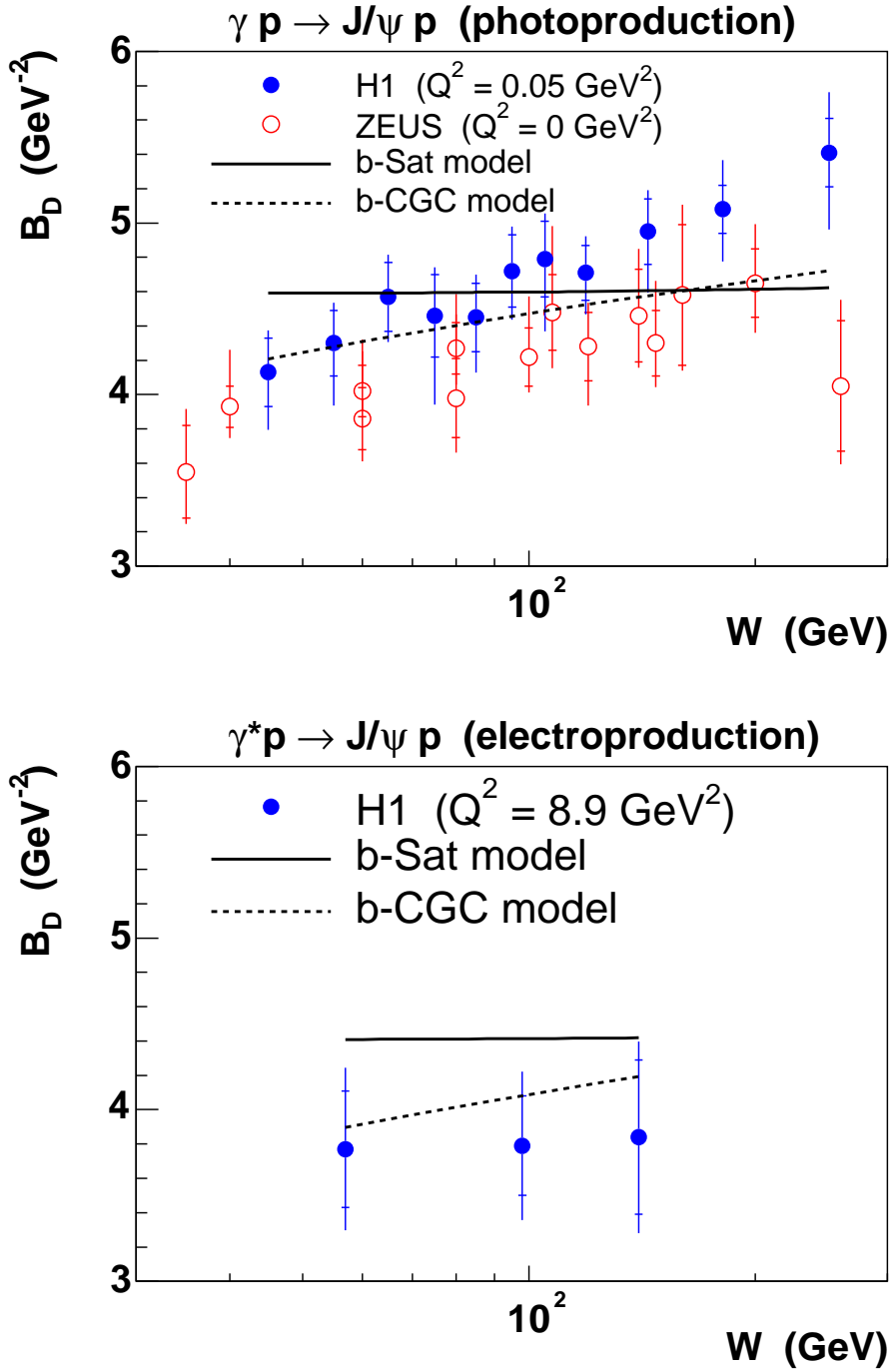}
  \caption{The $t$-slope parameter $B_D$ vs.~$W$ compared to predictions from the b-Sat and b-CGC models using the ``boosted Gaussian'' vector meson wave function.}
  \label{fig:bdw_bcgc}
\end{figure}
Although almost all features of the vector meson and DVCS data are well described by the b-Sat model, there is one exception, namely $\alpha_\Pom^\prime$.  It is predicted to be close to zero, due to the assumed factorisation of $T(b)$ from the gluon distribution $xg(x,\mu^2)$, in some disagreement with the data; see Figs.~\ref{fig:apom} and \ref{fig:bdw}.  In the b-CGC model the $W$ (or $x$) dependence is not factorised from the $b$ dependence.  Therefore, an appreciable $\alpha_\Pom^\prime$ is achievable, as shown in Fig.~\ref{fig:bdw_bcgc}.  Here, we use the ``boosted Gaussian'' vector meson wave function in both cases.  In fact, for photoproduction, a fit to the model predictions of the form $B_D = B_0 +  4\alpha_\Pom^\prime\ln[W/(\text{90 GeV})]$ gives $\alpha_\Pom^\prime=0.075$ for the b-CGC model compared to $\alpha_\Pom^\prime=0.004$ for the b-Sat model.  However, the value of $\alpha_\Pom^\prime$ from the b-CGC model is still slightly low when compared to the values of $0.116\pm0.026\pm^{0.010}_{0.025}$ \cite{Chekanov:2002xi} or $\alpha_\Pom^\prime=0.164\pm0.028\pm0.030$ \cite{Aktas:2005xu} measured by experiment.  We note that, with the exception of $\alpha_\Pom^\prime$, the b-CGC model gives a considerably worse overall description of exclusive processes than the b-Sat model.

\section{Saturation and related topics} \label{sec:Sat}
A frequently asked question, whether or not the HERA data require saturation, is answered in the saturation models like GBW \cite{Golec-Biernat:1998js,Golec-Biernat:1999qd} or CGC \cite{Iancu:2003ge} with a clear yes.  In the impact parameter dependent models, such as the models discussed in this paper, the answer is more involved.  In this section, we will therefore discuss the saturation effects in some detail. 

In the GBW model the effects of saturation are clearly seen, for example, in the change of rate of rise, $\lambda_{\rm tot}$, of the total DIS cross section with $Q^2$, see Fig.~\ref{fig:sigtot} (bottom).  In this model the value of the observed parameter $\lambda_{\rm tot}$ is related to the value of the constant $\lambda_{\rm GBW}\approx 0.3$ modulated by the saturation effects. Since the variation of $\lambda_{\rm tot}$ with $Q^2$ is substantial, saturation has to be an important effect.  Note that, in the GBW model, saturation is the \emph{only} mechanism which can modulate the parameter $\lambda_{\rm tot}$.

\begin{figure}
  \centering
  \includegraphics[width=0.49\textwidth]{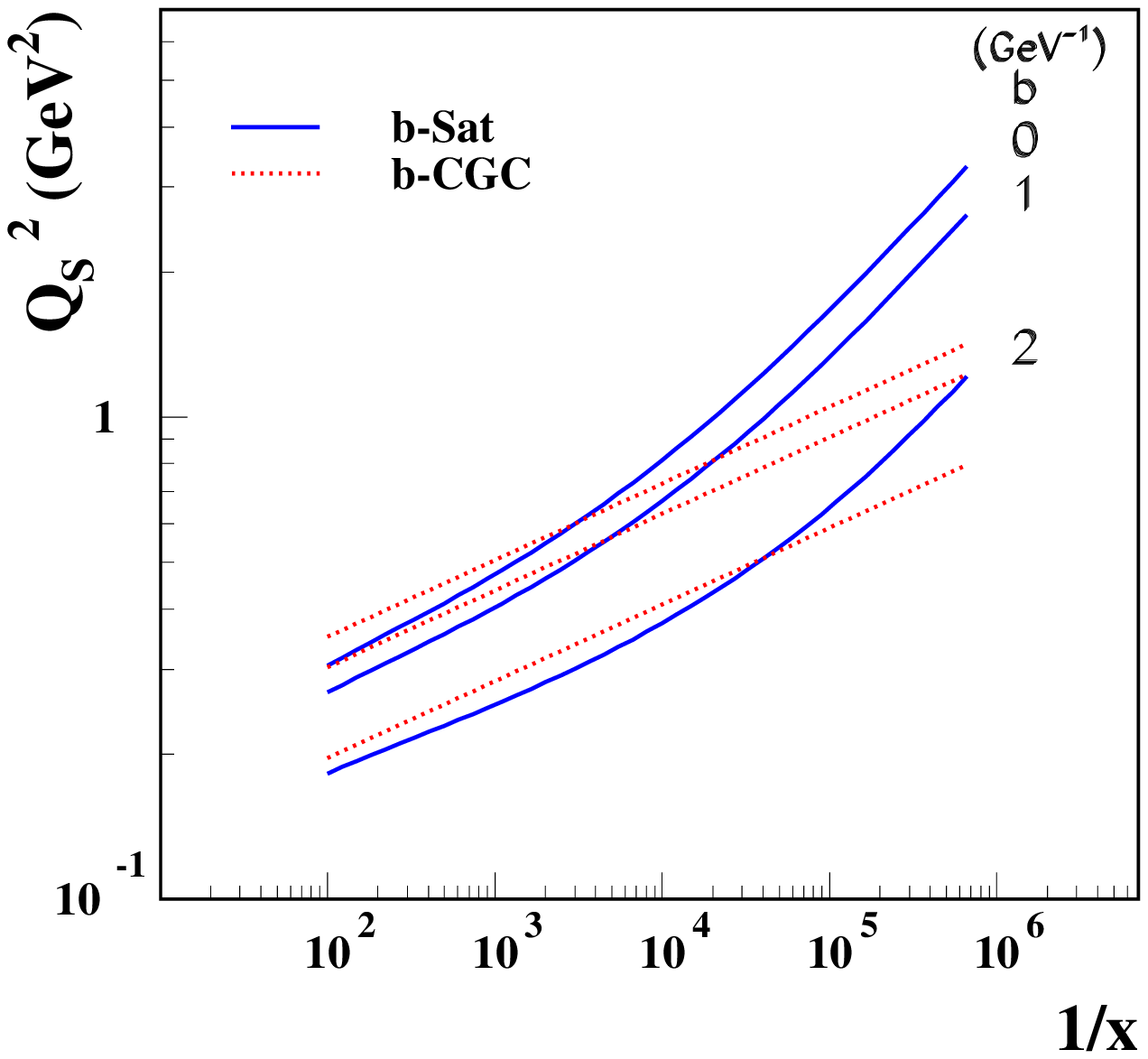}%
  \includegraphics[width=0.49\textwidth]{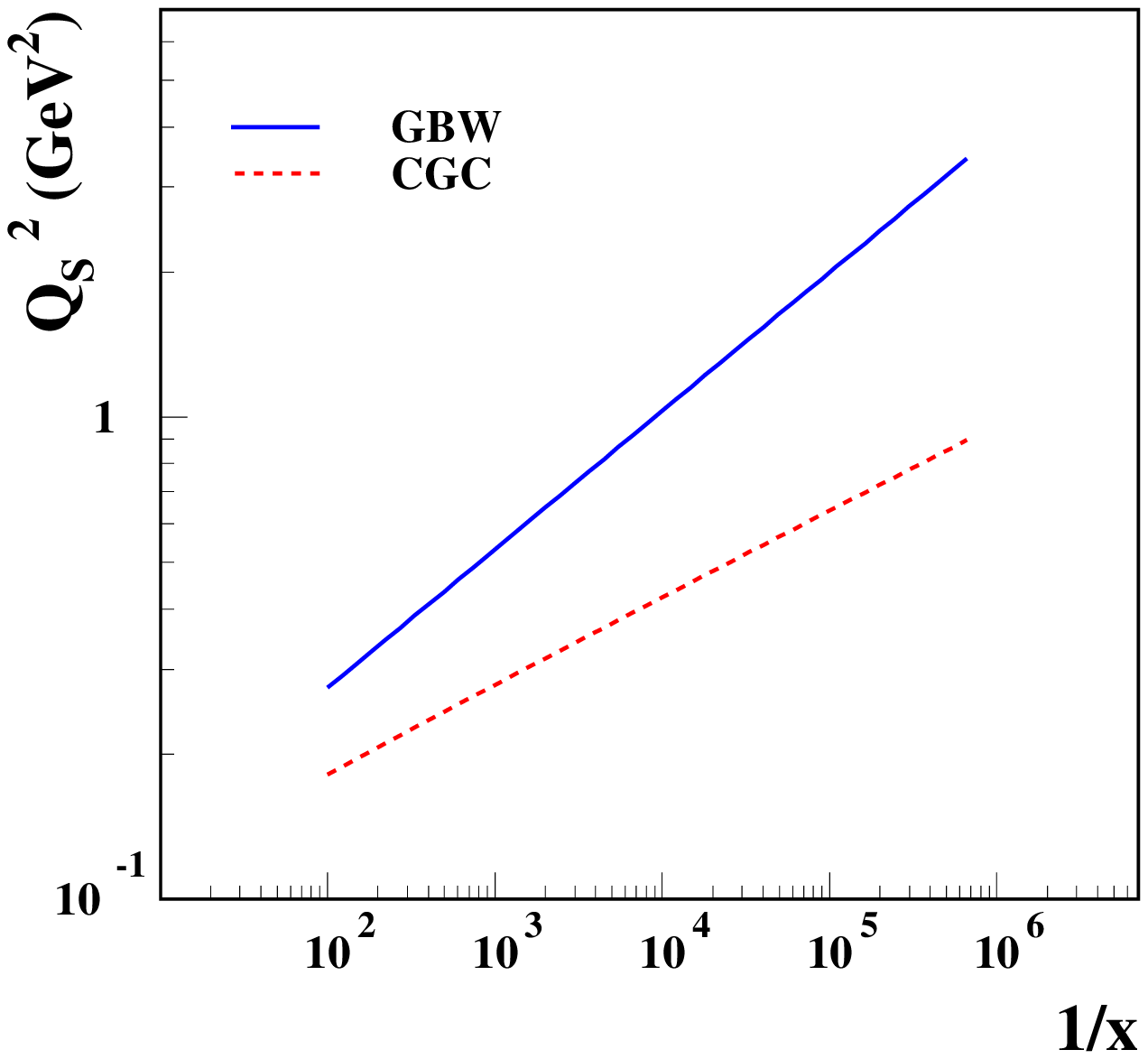}
  \caption{The saturation scale $Q_S^2\equiv 2/r_S^2$, where $r_S$ is defined as the solution of \eqref{eq:satdef}, found in the b-Sat and b-CGC models (left), and in the GBW and CGC models (right).}
  \label{fig:q2sx}
\end{figure}
The saturation effects are best quantified by the value of the saturation scale $Q_S^2\equiv 2/r_S^2$, where the saturation radius $r_S$ is defined as the solution of \eqref{eq:satdef}.  In Fig.~\ref{fig:q2sx} we show the saturation scale for the impact parameter dependent (left) and independent (right) models.  Fig.~\ref{fig:q2sx} (right) shows that the saturation scale in the GBW model is significantly higher than in the CGC model.  (The GBW and CGC fits shown here are described in more detail in Sect.~\ref{AppGBW}.)  This is understandable since in the CGC model the variation of the $\lambda_{\rm tot}$ parameter is partly due to evolution in addition to the saturation effects.  However, even in the CGC model the saturation effects are fairly strong, as discussed by Iancu, Itakura and Munier~\cite{Iancu:2003ge} and by Forshaw and Shaw~\cite{Forshaw:2004vv}.  Fig.~\ref{fig:q2sx} (left) shows that in the b-Sat and b-CGC models the saturation scale is strongly dependent on the impact parameter $b$.  In the centre of the proton ($b\approx 0$), the b-Sat and b-CGC models have a similar saturation scale, comparable to the value in the GBW model.  As $b$ increases the value of the saturation scale drops quickly in both models.  This is again understandable since, in the b-Sat model with a Gaussian proton shape, at larger values of $b$ the gluon density is diluted by the factor $T(b)$ and so the smaller gluon density leads to smaller saturation scales.  In this model, the variation of $\lambda_{\rm tot}$ with $Q^2$ is mostly due to evolution effects, since the gluon density at the initial scale $\mu_0^2$ is characterised by a low value of the parameter $\lambda_g \approx 0$.  The observed large values of $\lambda_{\rm tot}$ can only be generated by evolution, as discussed in detail by KT~\cite{Kowalski:2003hm}.

\begin{figure}
  \centering
  \includegraphics[width=0.6\textwidth]{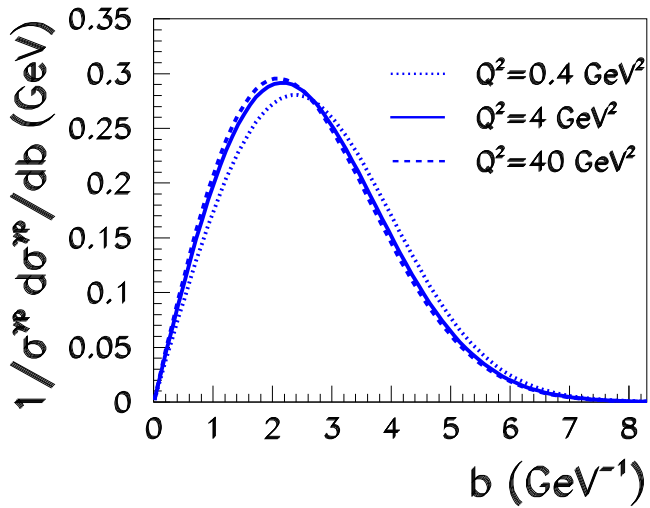}
  \caption{The $b$-dependence of the total cross section, $\sigma_{\rm tot}^{\gamma^*p}$, for $Q^2=0.4$, $4$ and $40$ GeV$^2$ with $x=10^{-4}$, $10^{-3}$ and $10^{-2}$ respectively, using a Gaussian $T(b)$ of width $B_G=4$ GeV$^{-2}$.}
  \label{fig:psitlb}
\end{figure}
In Fig.~\ref{fig:psitlb} we show the $b$-dependence of the total cross section to give a feeling for the relative contributions from the different impact parameters.  The median value of this distribution is around $b=2.6$ GeV$^{-1}$, that is, the majority of the cross section is determined by the dilute gluon region, where the saturation scale is small.

To summarise, in the impact parameter dependent dipole models, evolution plays a greater role than saturation on average.  However, in the centre of the proton ($b\approx 0$), the saturation effects are large in both the b-Sat and b-CGC models.  In the centre of the proton the saturation scale is comparable to the saturation scale found in the original GBW model.

The GBW model is theoretically very attractive since all observables in this model are a function of only one variable, $r^2 Q_s^2(x)$, where $Q_s^2(x) = (x_0/x)^{\lambda_{\rm GBW}}$.  This leads to so-called \emph{geometric scaling} in which $\sigma^{\gamma^* p}$ is only a function of $\tau=Q^2/Q_s^2(x)$, which is confirmed to some accuracy by data \cite{Stasto:2000er}.\footnote{Note, however, that the inclusion of the charm quark contribution violates geometric scaling to a certain extent.}  A similar scaling has recently been observed for ($t$-integrated) diffractive DIS data \cite{Marquet:2006jb}.  The notion of geometric scaling is essential for development of the theoretical approach to saturation.  Indeed, geometric scaling seems to be a universal feature of a wide class of evolution equations with saturation effects, irrespective of the form of the non-linear term \cite{Munier:2003vc,Munier:2003sj,Munier:2004xu}.  In the b-Sat model, approximate geometric scaling is also present, as it is imposed by the fit to the data.  This scaling, however, is not an intrinsic feature of the b-Sat model because of the greater importance of DGLAP evolution compared to saturation effects, and also because of the additional scale introduced by the impact parameter dependence.

\begin{figure}
  \centering
  \includegraphics[width=0.6\textwidth]{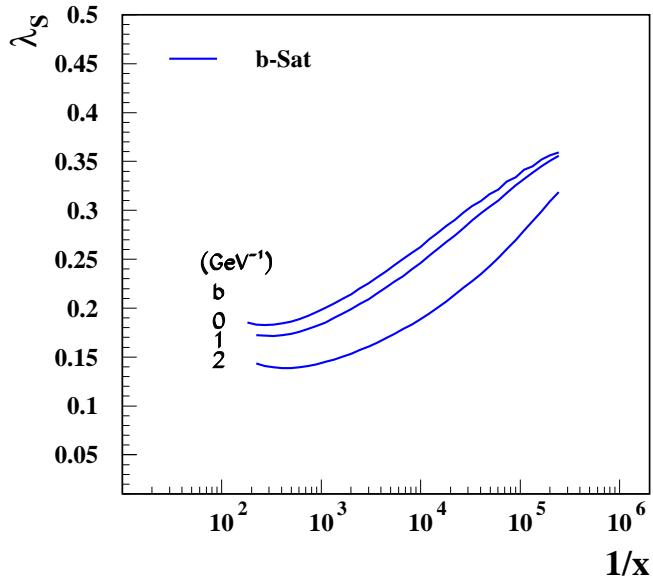}
  \caption{The saturation exponent $\lambda_S$ as a function of $x$ and impact parameter $b$.}
  \label{fig:lambdas}
\end{figure}

The theoretical understanding of saturation phenomena follows from evolution equations obtained within perturbative QCD.  It is therefore interesting to ask the question whether the saturation effects determined in the models from fits to HERA data belong in the perturbative or non-perturbative domain.  As shown in Fig.~\ref{fig:q2sx}, the saturation scale determined in the proton centre in the b-Sat model is around 0.5 GeV$^2$ at $x\approx 10^{-3}$.  This number lies in-between the value of $\Lambda_{\mathrm{QCD}}^2=0.04$ GeV$^2$, being clearly non-perturbative, and the value of around~1~GeV$^2$, considered to be perturbative.  Therefore it is not obvious to what extent the saturation dynamics are driven by the perturbative effects. The models discussed here are, however, by construction perturbative; the renormalisation and factorisation scale $\mu^2 = 4/r^2 + \mu_0^2$, used to evaluate the strong coupling and the gluon density, is bounded from below by $\mu_0^2 \simeq 1$ GeV$^2$ and is around 2 GeV$^2$ if $Q_S^2\equiv 2/r_S^2 \simeq 0.5$ GeV$^{2}$.  Moreover, in the centre of the proton, the value of the saturation exponent
\begin{equation}
  \lambda_S\equiv \frac{\partial\ln(Q_S^2)}{\partial\ln(1/x)}
\end{equation}
varies between $\lambda_S = 0.19$ at $x=10^{-2}$ and $\lambda_S = 0.27$ at $x=10^{-4}$, as shown in Fig.~\ref{fig:lambdas}.  Therefore, the values of this exponent are greater than the value of $\lambda_S \simeq 0.08$ expected for a `soft' process, and are close to the expectations from theoretical studies of perturbative non-linear evolution equations; see, for example, Refs.~\cite{Golec-Biernat:2001if,Mueller:2002zm,Triantafyllopoulos:2002nz}.  This indicates that the saturation phenomena studied in the b-Sat model is outside of the non-perturbative region.

\subsection{Step $T(b)$} \label{step}
We also performed an alternative fit to $F_2$ data using the b-Sat model with the step function $T(b)$ given by \eqref{eq:StepTb}, with the parameter $b_S=4$ GeV$^{-1}$; see the last line of Table~\ref{tab:bSat}.  Recall that this form of $T(b)$ is implicitly used in all $b$-\emph{independent} parameterisations of the dipole cross section.  The fit was of similar quality and gave a slightly larger gluon distribution compared to the corresponding fit with a Gaussian $T(b)$, see Fig.~\ref{fig:gluon}, indicating a slight shift in the balance between evolution and saturation effects.  Note from \eqref{eq:BBstep} that a step $T(b)$ with $b_S=4$ GeV$^{-1}$ corresponds to $\langle b^2\rangle = 8$ GeV$^{-2}$, the same value as for the Gaussian $T(b)$ with $B_G=4$ GeV$^{-2}$ from \eqref{eq:BB}. 

\begin{figure}
  \centering
  \includegraphics[width=0.6\textwidth]{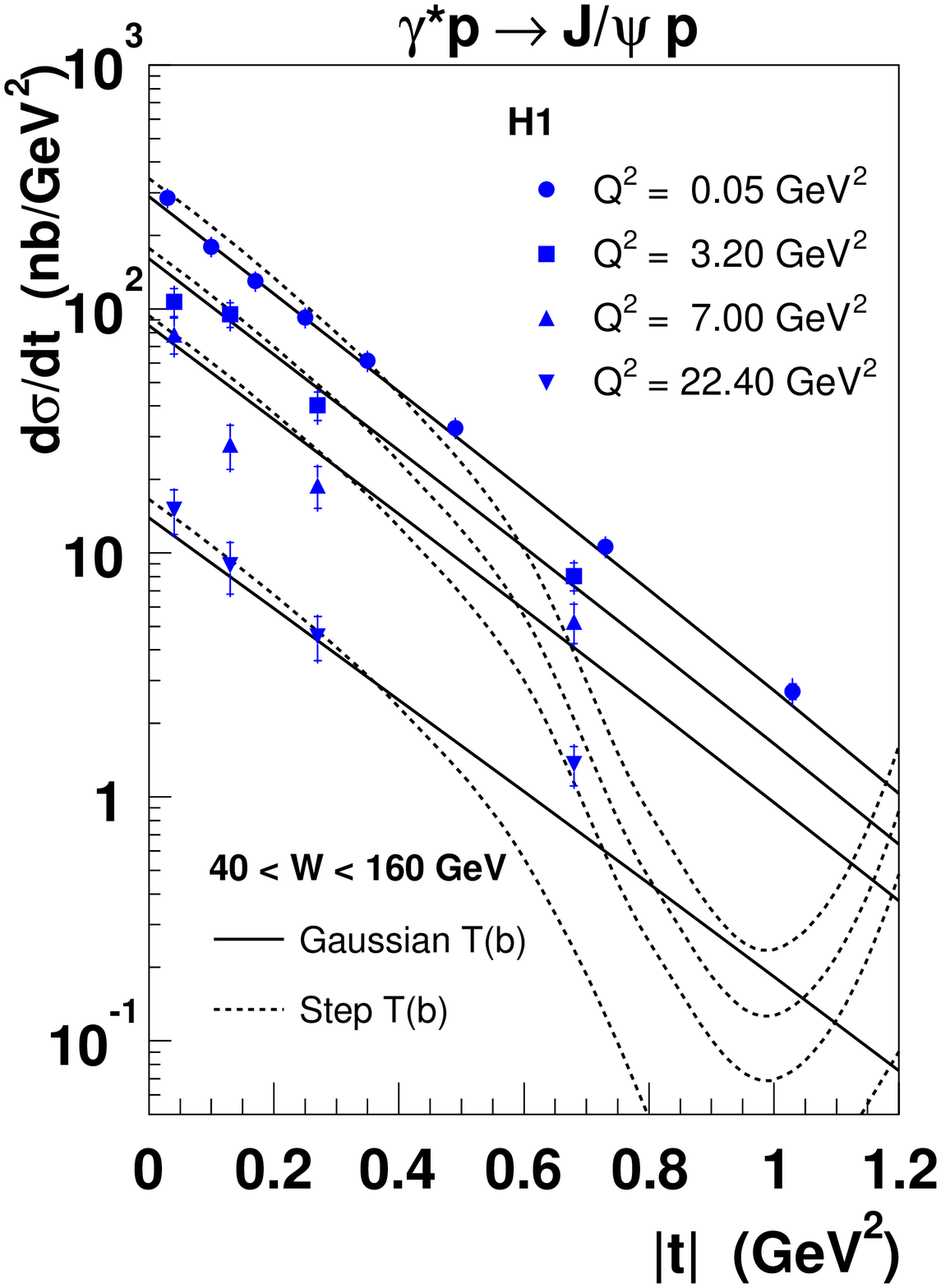}
  \caption{Differential $J/\psi$ meson cross section $\dif{\sigma}/\dif{t}$ vs.~$|t|$ compared to predictions from the b-Sat model using the ``boosted Gaussian'' vector meson wave function with either a Gaussian or a step $T(b)$.}
  \label{fig:dsdt_steptb}
\end{figure}
For small $|t|$, the results with a step $T(b)$ are close to those with a Gaussian $T(b)$, and so the total cross sections for exclusive processes are also similar.  However, for larger values of $|t|$, the step $T(b)$ gives a dip in the $t$-distributions, which is not observed in the data, as seen in Fig.~\ref{fig:dsdt_steptb} for $J/\psi$ production.  Here, we have used the ``boosted Gaussian'' vector meson wave functions in both cases.  The reason for the dip at large $|t|$ can be explained by noticing that the two-dimensional Fourier transform of the step function \eqref{eq:StepTb} gives the Bessel function of the first kind, $2J_1(b_S\Delta)/(b_S\Delta)$, which oscillates through zero, whereas the two-dimensional Fourier transform of a Gaussian is simply another Gaussian.  Although there is some uncertainty in the measured cross section at large $|t|$ due to the treatment of proton dissociation, the uncertainty is not expected to account for the large discrepancy between the predictions with the step $T(b)$ and the data, and so the step $T(b)$ must be ruled out as a model for the proton shape.

\subsection{The GBW and CGC models without impact parameter dependence} \label{AppGBW}
\begin{table}
  \centering
  \begin{tabular}{cccc|ccc|c}
    \hline\hline
    Model & $Q^2$/GeV$^2$ & $m_{u,d,s}$/GeV & $m_c$/GeV & $\sigma_0$/mb & $x_0$/$10^{-4}$ & $\lambda$ & $\chisq$ \\ \hline
    GBW & [0.25,45] & $0.14$ & --- & 20.1 & 5.16 & 0.289 & $216.5/130=1.67$\\
    GBW & [0.25,45] & $0.14$ & $1.4$ & 23.9 & 1.11 & 0.287 & $204.9/130=1.58$ \\
    GBW & [0.25,650] & $0.14$ & $1.4$ & 22.5 & 1.69 & 0.317 & $414.4/160=2.59$ \\ \hline
    CGC & [0.25,45] & $0.14$ & --- & 25.8 & 0.263 & 0.252 & $117.2/130=0.90$ \\
    CGC & [0.25,45] & $0.14$ & $1.4$ & 35.7 & $0.00270$ & 0.177 & $116.8/130=0.90$ \\
    CGC & [0.25,650] & $0.14$ & $1.4$ & 34.5 & $0.00485$ & 0.188 & $173.7/160=1.09$ \\ \hline\hline
  \end{tabular}
  \caption{Parameters of the GBW \eqref{eq:sigGBW} and CGC \eqref{eq:cgc} models determined from fits to $F_2$ data \cite{Breitweg:2000yn,Chekanov:2001qu}.}
  \label{tab:GBW+CGC}
\end{table}
For completeness we give here the results of the fits using the impact parameter \emph{independent} GBW \eqref{eq:sigGBW} and CGC \eqref{eq:cgc} dipole models.  We first make fits to ZEUS $F_2$ data \cite{Breitweg:2000yn,Chekanov:2001qu} with $x\le 0.01$ and $Q^2\in[0.25,45]$ GeV$^2$ using the CGC model \eqref{eq:cgc} with $\mathcal{N}_0=0.7$ (fixed), first without any charm quark contribution as in the original paper \cite{Iancu:2003ge}, then including the charm contribution.  We also show the effect of including the data with higher $Q^2>45$ GeV$^2$.  For comparison, we perform similar fits using the original GBW model \eqref{eq:sigGBW}.  We take $x=\xB$ for light quarks and $x=\xB(1+4m_c^2/Q^2)$ for charm quarks.  The light quark masses are taken to be $m_{u,d,s}=0.14$ GeV, with the charm quark mass $m_c=1.4$ GeV.  The results of these fits are shown in Table~\ref{tab:GBW+CGC}.

We note that the description of the data by the CGC model is sizably better than by the GBW model.  This is presumably due to the lack of evolution effects in the GBW model and can be seen from the fact that the worsening of the $\chi^2$ value when the data points with $Q^2>45$ GeV$^2$ are included is more prominent for the GBW model than the CGC model; see the right-hand column of Table~\ref{tab:GBW+CGC}.

\begin{figure}
  \centering
  \includegraphics[width=0.49\textwidth]{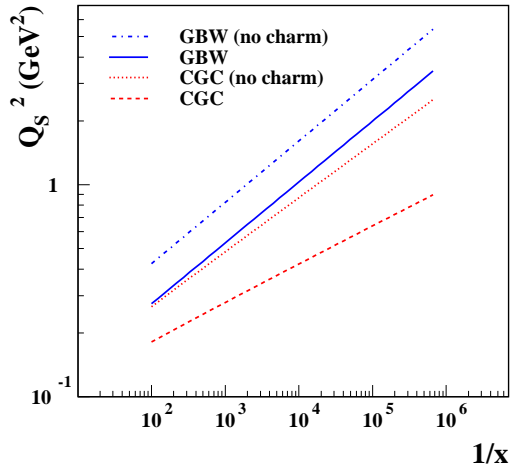}
  \caption{The saturation scale $Q_S^2\equiv 2/r_S^2$, where $r_S$ is defined as the solution of \eqref{eq:satdef}, found in the GBW and CGC models with and without charm quarks.  The presence of charm quarks dramatically lowers the saturation scale, especially for the CGC model.}
  \label{fig:q2sxnoch}
\end{figure}
Notice also that the saturation scale in the CGC fit is dramatically lowered with the introduction of charm quarks, as shown in Fig.~\ref{fig:q2sxnoch}.  The fact that saturation effects are very sensitive to the presence of the charm contribution was first noticed in the original GBW paper \cite{Golec-Biernat:1998js} and also in the KT \cite{Kowalski:2003hm} impact parameter dependent analysis.  In particular, Thorne \cite{Thorne:2005kj} has emphasised the importance of the charm contribution, which has been omitted in some analyses of the saturation scale at HERA.

\begin{figure}
  \centering
  \includegraphics[width=0.6\textwidth]{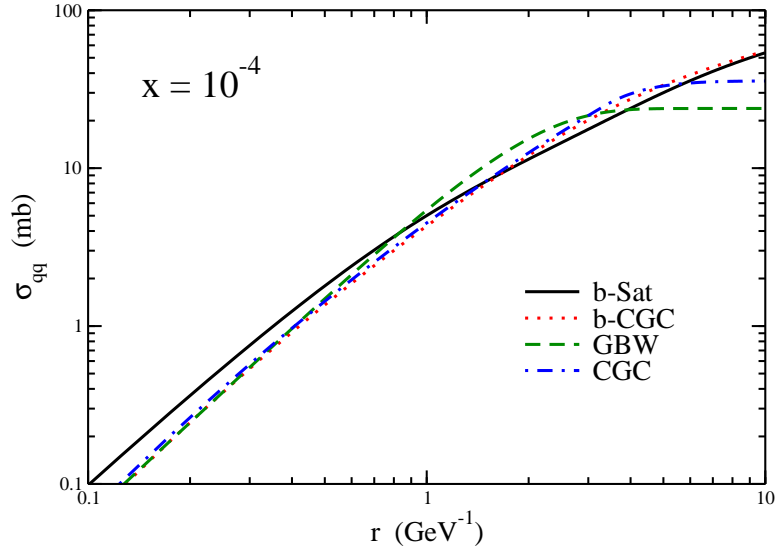}
  \caption{The dipole cross section $\sigma_{q\bar{q}}(x,r)$ at fixed $x=10^{-4}$, integrated over the impact parameter $\vec{b}$, obtained in the b-Sat, b-CGC, GBW and CGC models.}
  \label{fig:sighati}
\end{figure}
\begin{figure}
  \centering
  \includegraphics[width=0.6\textwidth]{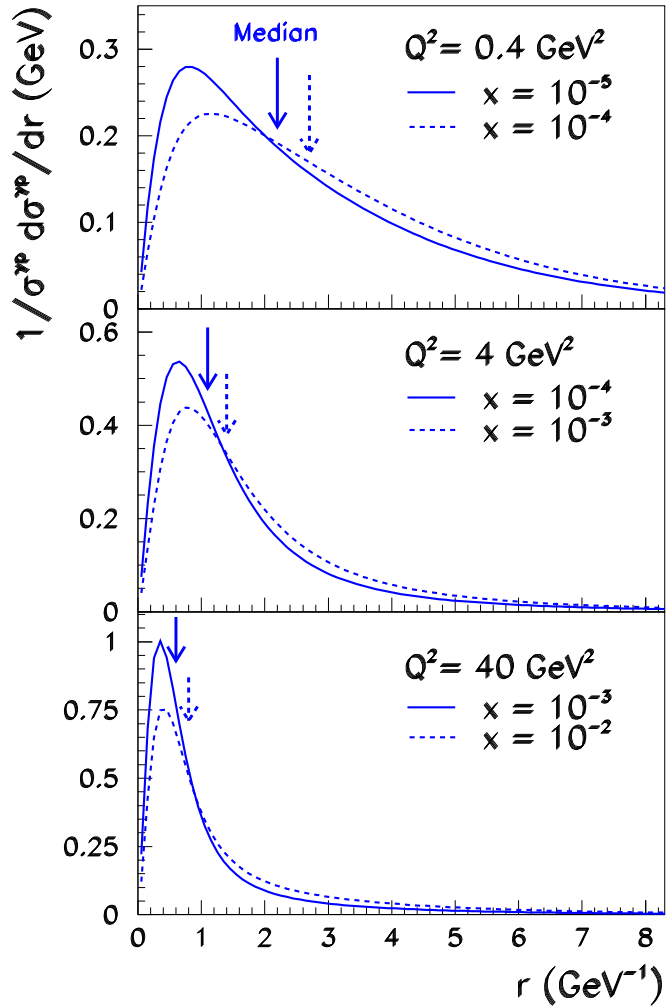}
  \caption{The distribution of dipole sizes $r$ contributing to the total inclusive DIS cross section in the b-Sat model for various  virtualities, $Q^2$,  of the photon.  The median values are indicated by vertical arrows.}
  \label{fig:psitlr}
\end{figure}

In Fig.~\ref{fig:sighati} we show the dipole cross section $\sigma_{q\bar{q}}(x,r)$ at fixed $x=10^{-4}$, integrated over the impact parameter $\vec{b}$, obtained in the fits using the b-Sat, b-CGC, GBW and CGC models with $m_{u,d,s}=0.14$ GeV and $m_c=1.4$ GeV.  At smaller values of $r$ the b-Sat model has a slightly larger dipole cross section than the other models due to the presence of DGLAP evolution.  At larger $r$ the GBW and CGC models tend to a constant value of $\sigma_0$, while the b-Sat and b-CGC models continue to increase with increasing $r$; see also the discussion in Ref.~\cite{Kowalski:2003hm}.  However, as discussed in Sect.~\ref{sec:wavefunction0}, the contribution to the total cross section from large dipole sizes is generally suppressed by the photon wave functions, as is clearly seen in Fig.~\ref{fig:psitlr}.

\section{Summary and outlook}
We have presented an analysis of exclusive diffractive vector meson and DVCS data measured at HERA within an impact parameter dependent saturated dipole (``b-Sat'') model.  Various cross sections measured as a function of $Q^2$, $W$ and $t$ can be described by a model with a minimal number of free parameters, namely the parameters $\mu_0^2$, $A_g$ and $\lambda_g$ of the initial gluon distribution, $xg(x,\mu_0^2) = A_g\,x^{-\lambda_g}\,(1-x)^{5.6}$, and the proton width $B_G$.  The wave functions of the virtual photon are known from QED, while the vector meson wave functions are assumed to have a Gaussian shape.  The variable which fluctuates in the Gaussian is, of course, not known precisely.  However, we have shown that the observed distributions are fairly insensitive to the particular assumptions, with possible exception of the $\sigma_L/\sigma_T$ ratio for the $\rho$ meson.  A more precise measurement of this distribution and of the spin density matrix elements would allow better constraints to be made on the form of the vector meson wave functions.

An important finding of this investigation is that, although the vector meson wave functions are not fully known, one obtains a good description of the measured data.  The model parameters, which were fixed by the fit to the total inclusive DIS cross section and the vector meson $t$-distributions, describe the measured $Q^2$ and $W$ dependence of vector meson production and DVCS very well, together with the absolute normalisation.  The measured DVCS $t$-distribution agrees with the model expectation within the measurement error.  We expect that the high luminosity achieved by HERA will allow the $t$-distributions of vector mesons and DVCS to be measured more precisely.  They provide important information about the proton size and the transverse dynamics of the evolution process.

The b-Sat model, which gives the best description of data, uses the Glauber--Mueller dipole cross section \eqref{eq:dsigmad2b} with DGLAP evolution of the gluon density.  Although the overall description of exclusive processes is very good, this approach has some limitations, seen most clearly in the lack of $W$ dependence of $B_D$ in $J/\psi$ photoproduction, Fig.~\ref{fig:bdw}.  Although this is a delicate effect, the measurement precision is sufficient to show that there is a coupling between the transverse and longitudinal evolution variables, that is, $\alpha_\Pom^\prime \ne 0$.  We therefore introduced impact parameter dependence into the CGC model, the ``b-CGC'' model, which leads to a considerably poorer fit to $F_2$ than the b-Sat model and a worse overall description of exclusive processes, but a better description of the $\alpha_\Pom^\prime$ parameter.  The saturation scale $Q_S^2$ evaluated in this investigation does not depend sizably on the adopted evolution scheme and is consistent with the results of Ref.~\cite{Kowalski:2003hm}.

An important finding of this investigation is that the $t$-dependences of all three vector mesons and the DVCS process can be simultaneously described with one universal shape of the proton.  The parameter characterising the size of the proton, $B_G=4$ GeV$^{-2}$, determined in this investigation, corresponds to a root-mean-square impact parameter $\sqrt{\langle b^2\rangle}$, given by \eqref{eq:BB}, of 0.56 fm.  This is rather smaller than the proton charge radius of $0.870\pm0.008$ fm \cite{Eidelman:2004wy}.\footnote{The proton charge radius was first measured by Hofstadter \cite{Mcallister:1956ng} to be $0.74\pm0.24$ fm.}  This leads to a rather surprising result that gluons are more concentrated in the centre of the proton than quarks.  DVCS measurements planned at JLab should help clarify this somewhat puzzling picture (see, for example, \cite{JLab}).

The investigation presented here demonstrates that a wide class of high-energy scattering processes measured at HERA may be understood within a simple and unified framework.  The key ingredient is the gluon density which is probed in the longitudinal and transverse directions.  The success of the description indicates the universality of the emerging gluon distribution.

Let us finish with a general remark that vector meson and DVCS processes may be used to probe the properties of nuclear matter in a new way.  In measurements with polarised beams it is possible to achieve precision which would allow a holographic picture of protons and nuclei to be obtained \cite{kow_tsu,Brodsky:2006in,Muller:2006pm}.  Such a measurement could be performed at an $ep$ collider with roughly a third of the HERA centre-of-mass energy, similar to the one described in the eRHIC proposal \cite{eRHIC,Deshpande:2005wd}.

\section*{Acknowledgments}
We are grateful to Al Mueller for discussions and comments. We thank 
Markus Diehl for his suggestion to investigate DVCS.
L.M.\ gratefully acknowledges the support of the grant 
of the Polish State Committee for Scientific Research 
No.\ 1~P03B~028~28. 
 
\appendix

\section{Connection to the KT paper} \label{sec:appendix}
In the preceding analysis \cite{Kowalski:2003hm} of $J/\psi$ photoproduction in the impact parameter dependent dipole saturation model, Kowalski and Teaney (KT) used a somewhat different convention to define the wave functions and to calculate the decay constants and the overlaps.

KT \cite{Kowalski:2003hm} defined the overlap functions between the vector meson and the photon wave functions in the following way:
\begin{align}
  (\Psi_V^*\Psi)_{T} &= \hat{e}_f e\frac{\sqrt{2N_c}}{2\pi m_f}\left\{m_f^2 K_0(\epsilon r)\bar\phi_T(r,z) - \left[z^2+(1-z)^2\right]\epsilon K_1(\epsilon r) \partial_r \bar\phi_T(r,z)\right\},\label{overV1}\\
  (\Psi_V^*\Psi)_{L} &= \hat{e}_f e\frac{\sqrt{2N_c}}{2\pi}2Q K_0(\epsilon r)z(1-z)\bar\phi_L(r,z),
  \label{overV2}
\end{align}
where the scalar ``Gaus-LC'' wave functions $\bar\phi_{T,L}(r,z)$ were defined as the Fourier transforms of factorised wave functions given in the momentum space by
\begin{align}
  \tilde{\phi}_{T,L}(k,z) = \bar N_{T,L} z(1-z)\exp(-k^2R_{T,L}^2/2),
  \label{eq:Gaus-LCKT}
\end{align}
leading to
\begin{align}
  \bar\phi_{T,L}(r,z) &= \int\!\frac{\dif^2\vec{k}}{(2\pi)^2}\;\exp\left(\mathrm{i}\vec{k}\cdot\vec{r}\right)\tilde{\phi}_{T,L}(k,z) \notag \\
  &= \bar N_{T,L} z(1-z)\int\!\frac{\dif^2\vec{k}}{(2\pi)^2}\;\exp\left(\mathrm{i}\vec{k}\cdot\vec{r}\right)\exp(-k^2R_{T,L}^2/2) \notag \\
  &= \frac{\bar N_{T,L}}{2\pi R_{T,L}^2}z(1-z)\exp\left(-\frac{r^2}{2R_{T,L}^2}
  \right).
\end{align}
In that representation the normalisation conditions were given by
\begin{gather}
  1 = \int\!\frac{\dif^2\vec{k}}{(2\pi)^2}\int_0^1\!\frac{\dif{z}}{4\pi}\;\left\{\left[z^2+(1-z)^2\right]\frac{k^2}{m_f^2}+1\right\}\left\rvert\tilde{\phi}_T(k,z)\right\rvert^2, \label{oldnormt}\\
  1 = \int\!\frac{\dif^2\vec{k}}{(2\pi)^2}\int_0^1\!\frac{\dif{z}}{4\pi}\;\left\lvert\tilde{\phi}_L(k,z)\right\rvert^2,
  \label{oldnorml}
\end{gather}
and the decay constants read,
\begin{gather}
  f_{V,T} = \hat{e}_f\sqrt{2N_c}\frac{m_f}{M_V}\int\!\frac{\dif^2\vec{k}}{(2\pi)^2}\int_0^1\!\frac{\dif{z}}{4\pi z(1-z)}\;\left\{\left[z^2+(1-z)^2\right]\frac{k^2}{m_f^2}+1\right\}\tilde{\phi}_T(k,z),
  \label{oldft}\\
  f_{V,L} = \hat{e}_f\sqrt{2N_c}2\int\!\frac{\dif^2\vec{k}}{(2\pi)^2}\int_0^1\!\frac{\dif{z}}{4\pi}\;\tilde{\phi}_L(k,z).
  \label{oldfl}
\end{gather}

It is straightforward to observe that the KT formulae (\ref{overV1},\ref{overV2}), (\ref{oldnormt},\ref{oldnorml}) and (\ref{oldft},\ref{oldfl}) may be obtained from the formulae of the present paper (\ref{eq:overt},\ref{eq:overl}), (\ref{eq:nnz_normt},\ref{eq:nnz_norml}) and (\ref{eq:nnz_fvt},\ref{eq:nnz_fvl}) if $\delta=0$ and the previously used wave functions $\,\bar\phi_T\,$ and $\,\bar\phi_L\,$ are expressed in terms of the wave functions $\,\phi_T\,$ and $\,\phi_L\,$ written in the conventions of this paper:
\begin{align}
  \bar\phi_T(r,z) = \frac{\sqrt{2N_c}}{z(1-z)}\, m_f \, \phi_T(r,z), \\
  \bar\phi_L(r,z) = \sqrt{2N_c} \, M_V \, \phi_L(r,z).
\end{align}
Note the modification of the $z$-dependent part of $\phi_T(r,z)$.  Of course, the radius parameters $R_{T,L}$ are the same in both conventions. The normalisation factors are, however, transformed according to
\begin{align}
  \bar N_T = \sqrt{2N_c}\, m_f \, 2\pi R_T ^2 \; N_T, \label{eq:NbarT} \\
  \bar N_L = \sqrt{2N_c}\, M_V \, 2\pi R_L ^2 \; N_L. \label{eq:NbarL}
\end{align}
The parameters of the ``Gaus-LC'' wave functions in its initial formulation are given in Table~\ref{tab:GLCparams_old2}. 
\begin{table}
  \centering
  \begin{tabular}{cccc|cccc}
    \hline\hline
    Meson & $M_V$/GeV & $f_V$ & $m_f$/GeV & $\bar N_T$/GeV$^{-1}$ & $R_T^2$/GeV$^{-2}$ & $\bar N_L$/GeV$^{-1}$ & $R_L^2$/GeV$^{-2}$ \\ \hline
    $J/\psi$ & 3.097 & 0.274 & 1.4 & 171 & 6.5 & 119 & 3.0 \\
    $\phi$ & 1.019 & 0.076 & 0.14 & 164 & 16.0 & 214 & 9.7 \\
    $\rho$ & 0.776 & 0.156 & 0.14 & 211 & 21.9 & 222 & 10.4 \\
    \hline\hline
  \end{tabular}
  \caption{Parameters of the ``Gaus-LC'' vector meson wave functions.  These are identical to those in Table~\ref{tab:GLCparams}, but using $\bar{N}_{T,L}$ instead of $N_{T,L}$; see \eqref{eq:NbarT} and \eqref{eq:NbarL}.}
  \label{tab:GLCparams_old2}
\end{table}

\end{document}